\documentclass[preprint]{aastex}
\usepackage{graphicx,natbib,amsmath,multirow}

\shorttitle{Helioseismic data inclusion in solar dynamo models} \shortauthors{Mu\~noz-Jaramillo,
Nandy and Martens}

\begin{document}

\title{Helioseismic data inclusion in solar dynamo models}

\author{Andr\'es Mu\~noz-Jaramillo}
\affil{Department of Physics, Montana State University, Bozeman, MT 59717, USA}
\email{munoz@solar.physics.montana.edu}

\and

\author{Dibyendu Nandy}
\affil{Indian Institute for Science Education and Research-Kolkata, WB 741252, India}
\email{dnandi@iiserkol.ac.in}

\and

\author{Petrus C. H. Martens}
\affil{Harvard-Smithsonian Center for Astrophysics, Cambridge, MA 02138, USA}
\email{pmartens@cfa.harvard.edu}

\begin{abstract}
An essential ingredient in kinematic dynamo models of the solar cycle is the internal velocity
field within the simulation domain -- the solar convection zone. In the last decade or so, the
field of helioseismology has revolutionized our understanding of this velocity field. In
particular, the internal differential rotation of the Sun is now fairly well constrained by
helioseismic observations almost throughout the solar convection zone. Helioseismology also gives
us some information about the depth-dependence of the meridional circulation in the near-surface
layers of the Sun. The typical velocity inputs used in solar dynamo models, however, continue to
be an analytic fit to the observed differential rotation profile and a theoretically constructed
meridional circulation profile that is made to match the flow speed only at the solar surface.
Here we take the first steps towards the use of more accurate velocity fields in solar dynamo
models by presenting methodologies for constructing differential rotation and meridional
circulation profiles that more closely conform to the best observational constraints currently
available. We also present kinematic dynamo simulations driven by direct helioseismic measurements for the rotation  and
four plausible profiles for the internal meridional circulation -- all of which are made to match
the helioseismically inferred near-surface depth-dependence, but whose magnitudes are made to
vary. We discuss how the results from these dynamo simulations compare with those that are driven
by purely analytic fits to the velocity field. Our results and analysis indicate that the
latitudinal shear in the rotation in the bulk of the solar convection zone plays a more important
role, than either the tachocline or surface radial shear, in the induction of the toroidal field.
We also find that it is the speed of the equatorward counterflow in the meridional circulation
right at the base of the solar convection zone, and not how far into the radiative interior it
penetrates, that primarily determines the dynamo cycle period. Improved helioseismic constraints
are expected to be available from future space missions such as the Solar Dynamics Observatory
and through analysis of more long-term continuous data sets from ground-based instruments such as
the Global Oscillation Network Group. Our analysis lays the basis for the assimilation of these
helioseismic data within dynamo models to make future solar cycle simulations more realistic.
\end{abstract}

\keywords{Sun: magnetic fields, Sun: interior, Sun: rotation, Sun: helioseismology, Sun: activity }

\section{Introduction}

The dynamic nature of solar activity can often be traced back to the presence and evolution of
magnetic fields in the Sun. The more intense magnetic fields on the order of $1000$ Gauss (G) are
observed to be concentrated within regions known as sunspots, which often appear in pairs of
opposite magnetic polarities (Hale 1908\nocite{hale1908}). Sunspots have been observed regularly
now for about four centuries starting with the telescopic observations of Galileo Galilei in the
early 1600s. These observations point out that the number of sunspots on the solar surface varies
in a cyclic fashion with an average periodicity of 11 years (Schwabe 1844)\nocite{schwabe1844},
although there are variations both in the amplitude and period of this cycle. At the beginning of a
cycle sunspots appear at about mid-latitudes in both hemispheres (with opposite polarity
orientation across the hemispheres) and then progressively appear at lower and lower latitudes as
the cycle progresses (Carrington 1858)\nocite{carrington1858} until no sunspots are seen (i.e.,
solar minimum). In the next cycle, the same pattern repeats, but the new cycle spots have their
bipolar magnetic orientation reversed relative to the previous cycle (in both hemispheres). So
considering sign as well as amplitude, the solar cycle has a period of $22$ years.

There is a weaker, more diffuse component of the magnetic field outside of sunspots which is seen
to have a somewhat different evolution. This field -- whose radial component has been
observable at the solar surface since the advent of the magnetograph -- was believed to be on the order of $10$
Gauss. It is found that this field is concentrated in unipolar patches, which originate at
sunspot latitudes at the time of sunspot maxima, and then moves poleward with the progress of the
sunspot cycle (Babcock 1959\nocite{babcock59}; Bumba and Howard 1965\nocite{bumba-howard65}; Howard
and LaBonte 1981\nocite{howard-labonte81}). The sign of this radial field of any given cycle
is opposite to the old cycle polar field, which it cancels and reverses upon reaching the poles.
The amplitude of this radial field achieves a maximum (at the poles) at the time of sunspot
minimum (i.e. with a $90^{\circ}$ phase difference relative to the sunspots). However, the
periodicity of the cycle of this field matches the sunspot cycle period, underscoring that
they are related. Recent observations by Hinode indicate that this radial field gets concentrated within
unipolar flux tubes with field strength on the order of $10^3$ Gauss (Tsuneta et al.\ 2008\nocite{tsuneta-etal08}).

Explanations of this observations of the solar magnetic cycle rely on the field of
magnetohydrodynamic dynamo theory, which seeks to address the generation and evolution of magnetic
fields as a complex non-linear process involving interactions between the magnetic field and plasma
flows within the solar interior. In particular, it is now believed that the solar cycle involves
the generation and recycling (feeding on the energy available in plasma motions) of two
components of the magnetic field -- the toroidal component and the poloidal component. In an
axisymmetric spherical coordinate system, the magnetic and velocity fields can be expressed as
\begin{eqnarray}
  {\bf B} &=& B_{\phi} {\bf \hat{e}}_{\phi} + \nabla \times (A {\bf \hat{e}}_{\phi}) \\
  {\bf v} &=& r\sin(\theta)\Omega{\bf \hat{e}}_{\phi} + {\bf v}_p
\end{eqnarray}
where the first term on the right hand side (R.H.S.) of Equations $1$ and $2$ is the toroidal
component (in the case of the velocity field this corresponds to the differential rotation) and the
second term is the poloidal component of the field (in the case of the velocity field this
corresponds to the meridional circulation). The toroidal component of the magnetic field is thought
to be produced by stretching of an initially poloidal field by the differential rotation of the Sun
(the dynamo $\Omega$-effect); subsequently, strong toroidal flux loops rise up due to magnetic
buoyancy emerging through the solar surface as sunspots (Parker 1955a\nocite{parker55a}). To
complete the dynamo cycle, the poloidal field (whose radial component is manifested as the observed vertical field on the solar surface) has to be regenerated from this toroidal field in a process that
is traditionally called the dynamo $\alpha$-effect. The first explanation of this $\alpha$-effect
was due to Parker (1955b\nocite{parker55b}) who suggested that helical turbulent convection in the
solar convection zone (SCZ) would twist rising toroidal flux tubes into the poloidal plane,
recreating the poloidal component of the magnetic field. Much has changed since this pioneering
description of the first solar dynamo model by Parker, although the basic notion of the recycling
of the toroidal and poloidal components remain the same.

First of all, simulations of the buoyant rise of toroidal flux tubes point out that to match the
observed properties of sunspots at the solar surface, the strength of these flux tubes at the
base of the SCZ has to be much more than the equipartition field strength of $10^4$ G (D'Silva \&
Choudhuri 1993\nocite{dsilva-choudhuri93}; Fan, Fisher \& DeLuca
1993\nocite{fan-fisher-deluca93}). The classical dynamo $\alpha$-effect due to helical turbulence
is expected to be quenched for super-equipartition field strengths and therefore other physical
processes have to be invoked as a regeneration mechanism for the poloidal field. One of the
alternatives is an idea originally due to Babcock (1961)\nocite{babcock61} and Leighton
(1969)\nocite{leighton69}. The Babcock and Leighton (hereby BL) model proposes that the decay and
dispersal of tilted bipolar sunspot pairs at the near-surface layers, mediated by diffusion,
differential rotation, and meridional circulation, can regenerate the poloidal field. This
mechanism is actually observed and is simulated as a surface flux transport process that can
reproduce the solar polar field reversals (Wang, Nash \& Sheeley
1989\nocite{wang-nash-sheeley89}). Therefore a synthesis of Parker's original description along
with the BL mechanism for poloidal field generation is now widely accepted as a leading contender
for explaining the solar dynamo mechanism (Choudhuri, Sch\"ussler \& Dikpati 1995; Durney 1997;
Dikpati \& Charbonneau 1999\nocite{dikpati-charbonneau99}; Nandy \& Choudhuri
2001\nocite{nandy-choudhuri01}; Nandy 2003\nocite{nandy03}), although there are other alternative
suggestions as well. A description of all of those is beyond the scope of this paper and
interested readers are referred to the review by Charbonneau (2005)\nocite{charbonneau05}.

Second, helioseismology has now mapped the solar internal rotation profile (Schou et al.\ 1998\nocite{schou-etal98}; Charbonneau et al.\ 1999\nocite{charbonneau-etal99}), which is observed
to be vary mainly in the latitudinal direction in the main body of the SCZ. Helioseismology has
also discovered the tachocline -- a region of strong radial and latitudinal shear beneath the base
of the SCZ which is expected to play an important role in the generation and storage of strong
toroidal flux tubes.

Third, more is now known about the meridional circulation, which is observed to be polewards at
the surface (Hathaway 1996\nocite{hathaway96}). To conserve mass, this circulation should turn
equatorwards in the solar interior. This circulation is deemed to be important for the dynamics
of the solar cycle (see Hathaway et al.\ 2003\nocite{hathaway-etal03} and the review by Nandy 2004) but the profile of
this in the solar interior remains poorly constrained. Nandy and Choudhuri
(2002)\nocite{nandy-choudhuri02} proposed a deep equatorward counterflow in the circulation
(penetrating into the radiative interior beneath the SCZ) to better reproduce in dynamo
simulations the latitudinal distribution of sunspots (because equatorward advective transport and
storage of the deep seated toroidal field is more efficient at these depths where turbulence is
greatly reduced). However, Gilman and Miesch (2004)\nocite{gilman-miesch04}, based on a laminar
analysis, argued that the penetration of the circulation would be limited to a shallow Ekman
layer close to the base of SCZ. A recent and more detailed analysis of the problem by Garaud and
Brummell (2008)\nocite{garaud-brummell08} suggests that the circulation can in fact penetrate
deeper down into the radiative interior. At this point there is no consensus on the profile and
nature of the meridional circulation in the solar interior. Helioseismic data does provide some
information about the depth-dependence of this circulation at near-surface layers (Braun \& Fan
1998\nocite{braun-fan98}, Giles 2000\nocite{giles00}, Chou \& Ladenkov 2005
\nocite{chou-ladenkov05}, Gonz{\'a}lez-Hern{\'a}ndez et al.\ 2006\nocite{gonzalezhernandez-etal06}), which, in conjunction with reasonable theoretical
arguments, can be used to construct some plausible interior profiles of this flow.

Numerous kinematic dynamo models have been constructed in recent years (see Charbonneau
2005\nocite{charbonneau05} for a review) incorporating these large scale flows (differential
rotation and meridional circulation) as drivers of the magnetic evolution. More recently such
dynamo models (based on the BL idea of poloidal field generation) have also been utilized to make
predictions for the upcoming cycle (Dikpati, de Toma \& Gilman
2006\nocite{dikpati-detoma-gilman06}; Choudhuri, Chatterjee \& Jiang
2007\nocite{choudhuri-chatterjee-jiang07}). At present, all these kinematic dynamo models
incorporate the information on large-scale flows as analytic fits to the differential rotation
profile and a theoretically constructed meridional circulation profile that is subject to mass
conservation but matches the flow speed only at the solar surface (i.e., without incorporating
the depth-dependent information that is available). However, these large-scale flows are crucial
to the generation and transport of magnetic fields; the differential rotation is the primary
source of the toroidal field that creates solar active regions, and the meridional flow is
thought to play a crucial role in coupling the two source regions for the poloidal and toroidal
field through advective flux transport. Given this, it is obvious that the next step in
constructing more sophisticated dynamo models of the solar cycle is to move towards a more rigorous use of
helioseismic data to constrain these models in a way such that they conform more closely to the best
available observational constraints; that is the goal of this study.

In Section~2, we describe the basic features of the kinematic dynamo model based on the BL idea
that we use for our study; in this model, we use fairly standard parametrization (commonly used
in the community) of various processes such as the diffusivity, dynamo $\alpha$-effect and
magnetic buoyancy. In Sections $3.1$ and $3.2$ we present the methodologies for using the
helioseismically observed solar differential rotation and constraining the meridional circulation profiles within
this dynamo model and describe how they improve upon the commonly used analytic profiles. In
Section $4$ we present results from dynamo simulations using these improved helioseismic
constraints and conclude in Section $5$ with a summary of our main results and their contextual
relevance.


\section{Our Model}

We substitute Equations $1$ and $2$ into the magnetic induction equation
\begin{equation}\label{IE}
    \frac{\partial \textbf{B}}{\partial t} = \nabla\times\left( \textbf{v}\times\textbf{B} - \eta\nabla\times\textbf{B}\right)
\end{equation}
and add the phenomenological BL poloidal field source $\alpha$ to obtain the axisymmetric dynamo
equations:
\begin{eqnarray}
  \frac{\partial A}{\partial t} + \frac{1}{s}\left[ \textbf{v}_p \cdot \nabla (sA) \right] &=& \eta\left( \nabla^2 - \frac{1}{s^2}  \right)A + \alpha(r,\theta)F(B_{av})B_{av},\\
  & & \nonumber\\
  \frac{\partial B}{\partial t}  + s\left[ \textbf{v}_p \cdot \nabla\left(\frac{B}{s} \right) \right] + (\nabla \cdot \textbf{v}_p)B&=& \eta\left( \nabla^2 - \frac{1}{s^2}  \right)B + s\left(\left[ \nabla \times (A\bf \hat{e}_\phi) \right]\cdot \nabla \Omega\right)  \\
  & & + \frac{1}{s}\frac{\partial (sB)}{\partial r}\frac{\partial \eta}{\partial
  r}, \nonumber
\end{eqnarray}
where $s = r\sin(\theta)$.  The terms on the LHS of both equations with the poloidal velocity
($\textbf{v}_p$) correspond to the advection and deformation of the magnetic field by the meridional flow.  The
first term on the RHS of both equations corresponds to the diffusion of the magnetic field.  The
second term on the RHS of both equations is the source of that type of magnetic field (BL
mechanism for $A$ and rotational shear for $B$).  Finally, the third term on the RHS of
Equation $5$ corresponds to the advection of toroidal field due to a turbulent diffusivity
gradient.

\begin{figure}[c]
\begin{tabular}{c}
  \begin{tabular}{cc}
  \includegraphics[scale=0.5]{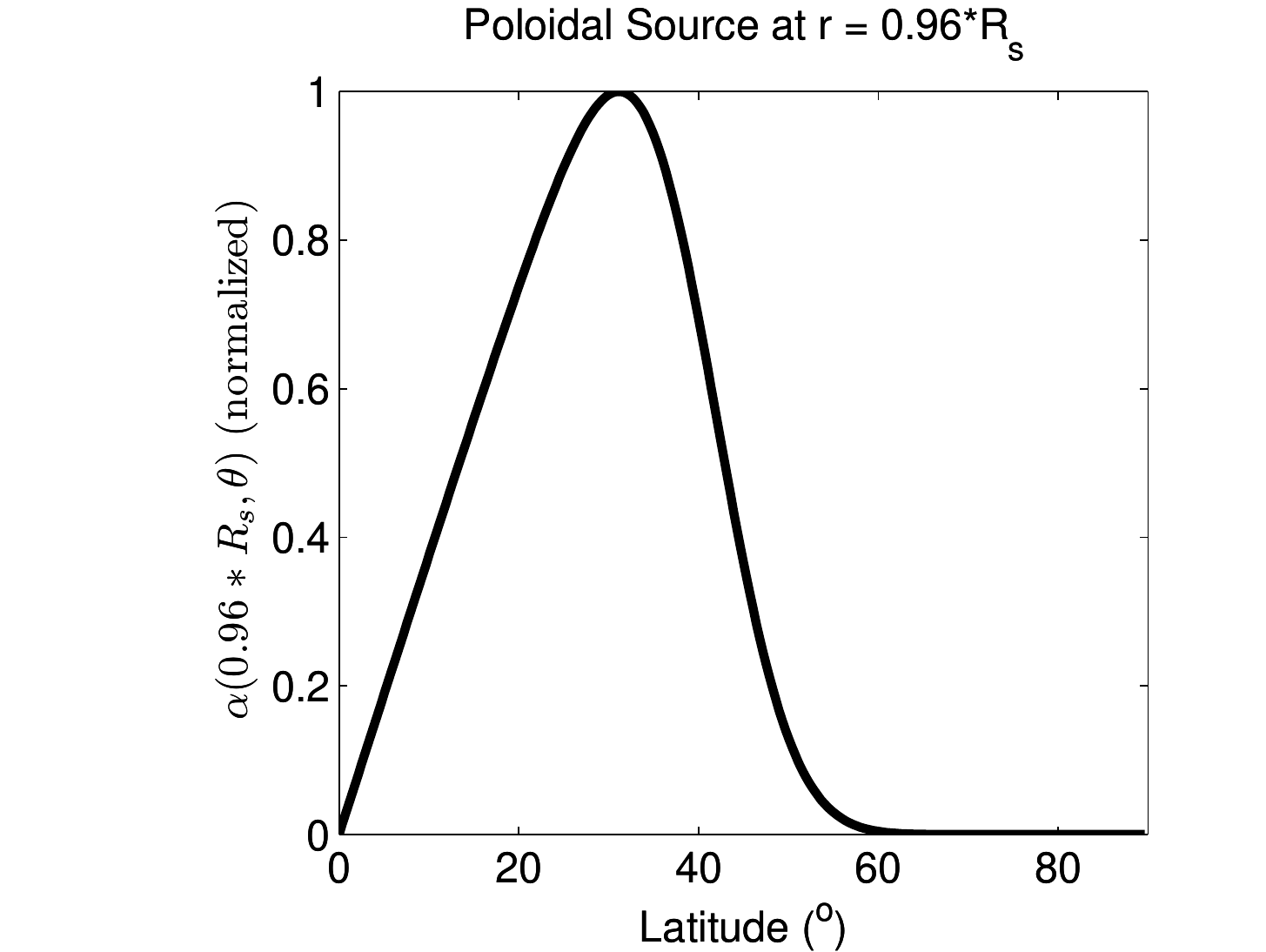} & \includegraphics[scale=0.5]{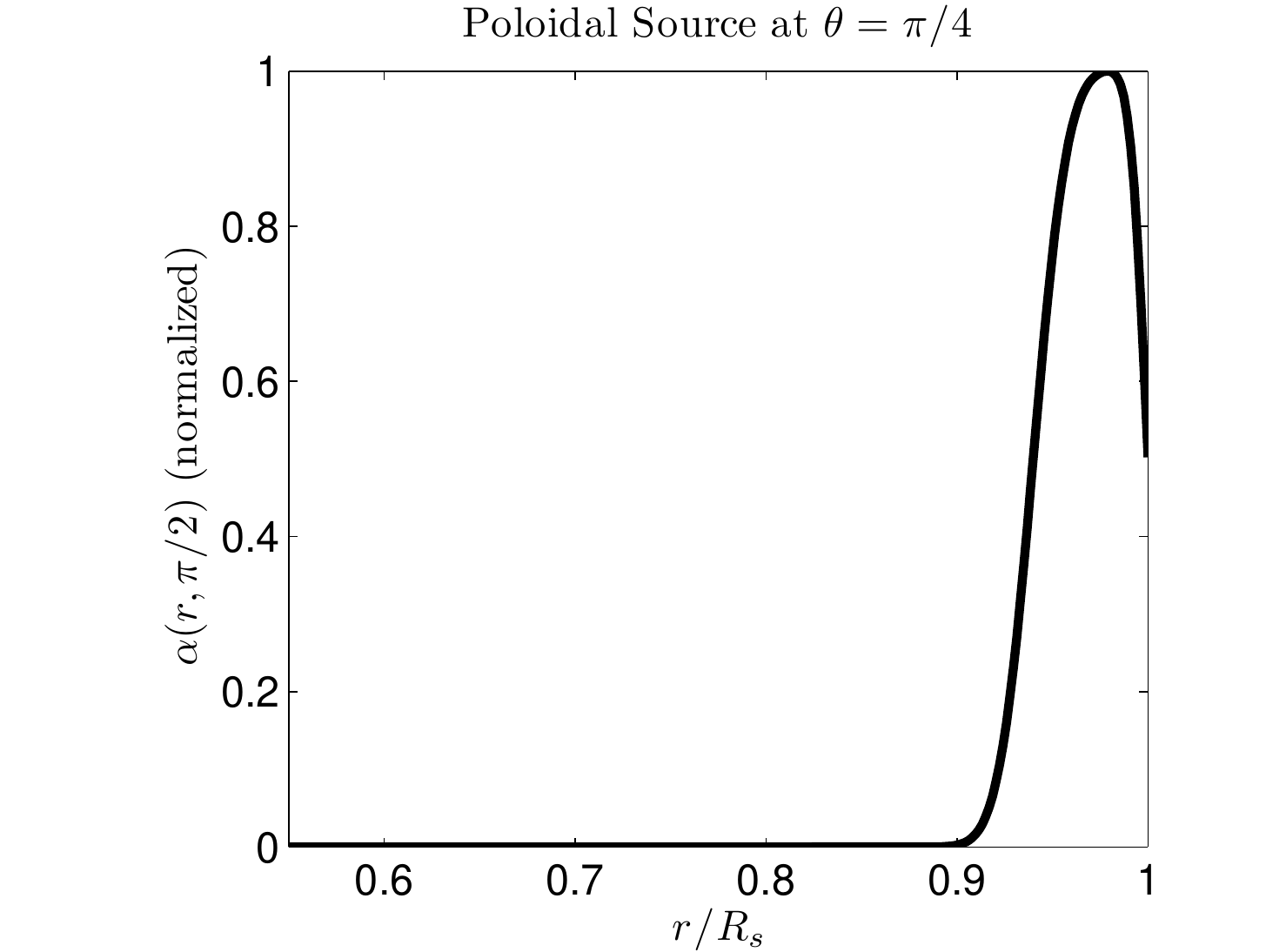}\\
    $(a)$ & $(b)$\\
  \end{tabular} \\
  \includegraphics[scale=0.5]{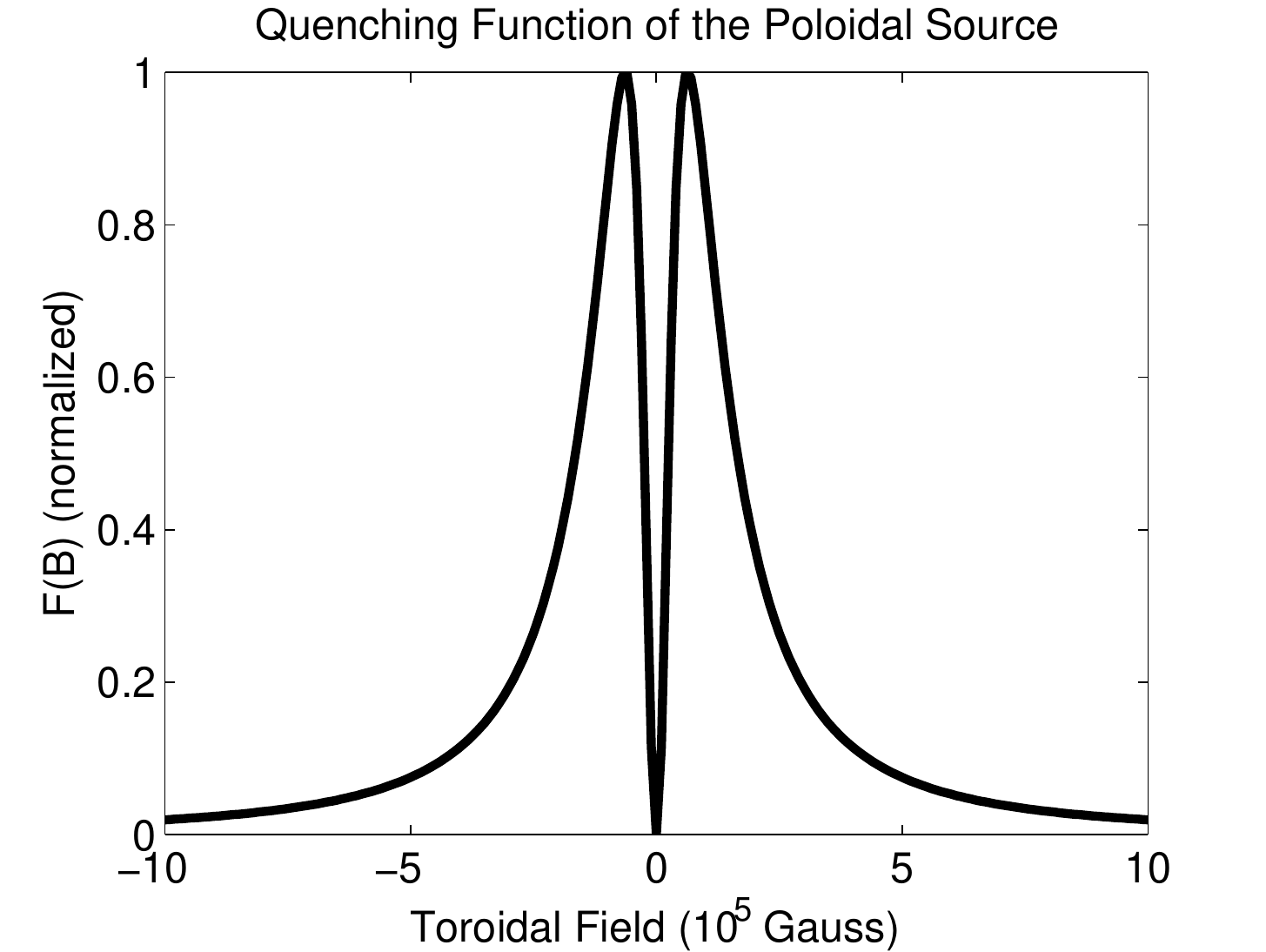} \\
  (c) \\
\end{tabular}
\caption{ Latitudinal(a) and radial(b) dependence of the poloidal source as quantified by the
dynamo $\alpha$-term. (c) Magnetic quenching of the poloidal source term.}\label{AE}
\end{figure}

As mentioned before, the generation of poloidal field near the surface due to the decay of active
regions is modeled through the inclusion of a source term,
which acts as a source for the vector potential $A$. It is localized both in radius and latitude
matching observations of active region emergence patterns and depends on the average toroidal
field at the bottom of the convection zone (Dikpati \& Charbonneau
1999)\nocite{dikpati-charbonneau99}. The radial and latitudinal dependence of the source is the
following:
\begin{equation}\label{AE_rt}
    \begin{split}
    \alpha(r,\theta) = \frac{\alpha_0}{16}\cos(\theta) &  \left( 1 + \operatorname{erf}\left( \frac{\theta - (90^o - \beta)}{\gamma} \right) \right)\left( 1 - \operatorname{erf}\left( \frac{\theta - (90^o + \beta)}{\gamma} \right) \right) \\
        & *\left( 1 + \operatorname{erf}\left( \frac{r - r_{al}}{d_{al}}\right) \right)\left( 1 - \operatorname{erf}\left(\frac{r - r_{ah}}{d_{ah}} \right) \right)
    \end{split}
\end{equation}
where $\alpha_0$ sets the strength of the source term and we set it to the value in which our
system starts having oscillating solutions. The parameters $\beta = 40^o$ and $\gamma = 10^o$ characterize the
latitudes in which sunspots appear. On the other hand, $r_{al}=0.94R_\odot$, $d_{al}=0.04R_\odot$, $r_{ah}=R_\odot$
and $d_{ah}=0.01R_\odot$ characterize the radial extent of the region in which the poloidal field
is deposited (see Figure \ref{AE}-a and b).  Besides radial and latitudinal dependence, we also
introduce lower and upper operating thresholds on the poloidal source that is dependent on
toroidal field amplitude -- which is a more realistic representation of the physics involving
magnetic buoyancy (as argued in Nandy \& Choudhuri 2001; Nandy 2002; see also Charbonneau, St-Jean \& Zacharias 2005). The presence of a lower threshold is in response to the fact that the plasma
density inside weak flux tubes is not low enough (compared to the density of the surrounding
plasma) to make them unstable to buoyancy (Caligari, Moreno-Insertis \& Schussler
1995\nocite{caligari-morenoinsertis-schussler95}) and those that manage to rise have very long
rising times (Fan, Fisher and De Luca 1993\nocite{fan-fisher-deluca93}).  On the other hand, if
flux tubes are too strong they are not tilted enough when they reach the surface to contribute to
poloidal field generation (D'Silva and Choudhuri 1993\nocite{dsilva-choudhuri93}; Fan, Fisher and
Deluca 1993\nocite{fan-fisher-deluca93}).  The dependence of the poloidal source on magnetic
field is
\begin{equation}\label{FB}
    F(B_{av}) = \frac{K_{ae}}{1 + \left( B_{av}/B_h \right)^2}\left( 1 - \frac{1}{1 + \left( B_{av}/B_l \right)^2}
    \right),
\end{equation}
where $K_{ae} = 1/\max(F(B_{av}))$ is a normalization constant and $B_h=1.5\times10^5G$ and
$B_l=4\times10^4G$ are the operating thresholds (see Figure \ref{AE}-c).

\begin{figure}[c]
\begin{center}
  \includegraphics[scale=0.5]{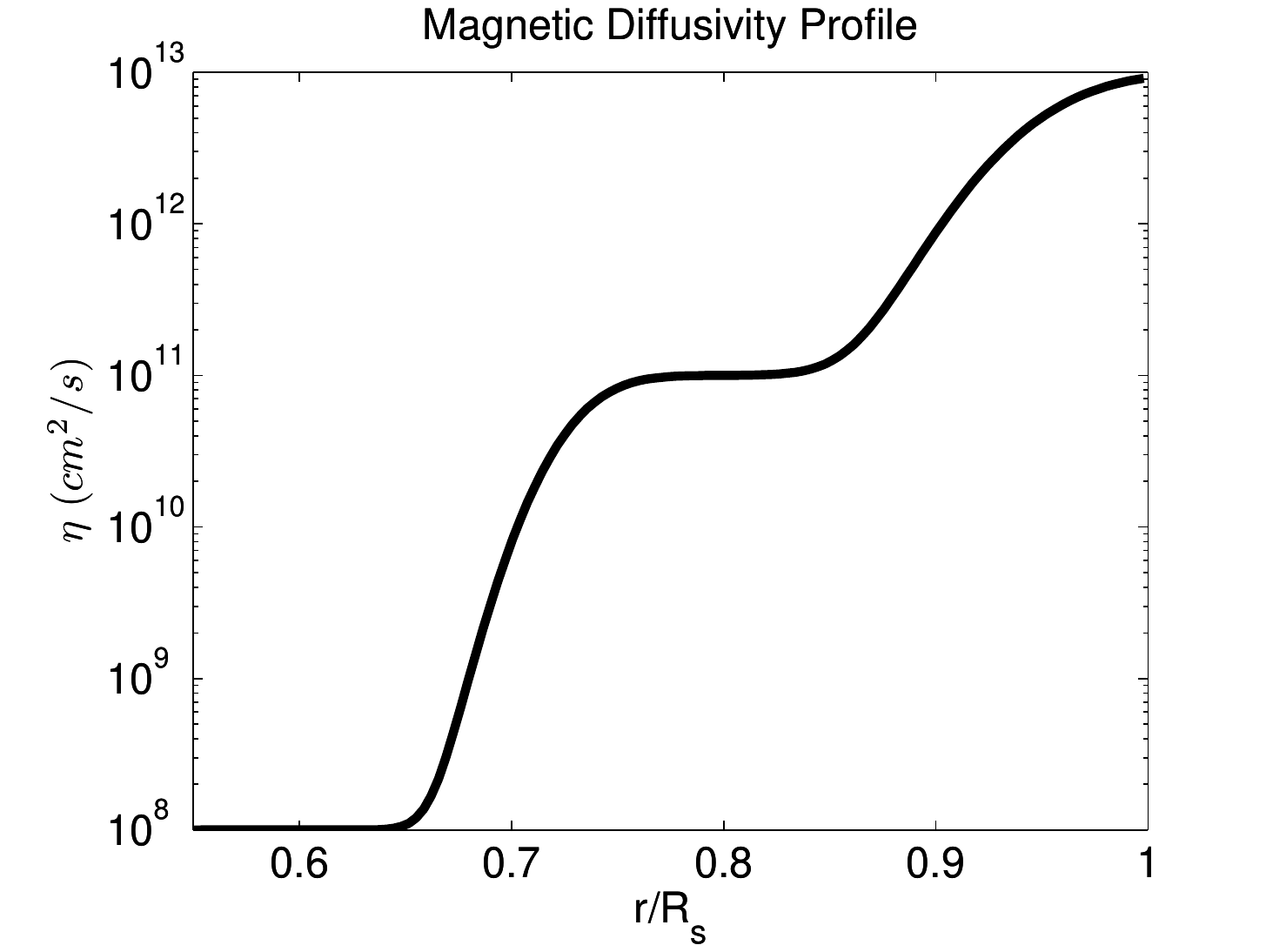}
\end{center}
\caption{Radial dependence of the turbulent magnetic diffusivity.}\label{Etapl}
\end{figure}

Another ingredient of this model is a radially dependent magnetic diffusivity; in this work we use
a double-step profile (see Figure \ref{Etapl}) given by
\begin{equation}\label{Eta}
      \eta(r) = \eta_{bcd} + \frac{\eta_{cz} - \eta_{bcd}}{2}\left( 1 + \operatorname{erf}\left( \frac{r - r_{cz}}{d_{cz}}  \right)
      \right) + \frac{\eta_{sg} - \eta_{cz} - \eta_{bcd}}{2}\left( 1 + \operatorname{erf}\left( \frac{r - r_{sg}}{d_{sg}},  \right) \right)
\end{equation}
where $\eta_{bcd} = 10^8 cm^2/s$ corresponds to the diffusivity at the bottom of the
computational domain,$\eta_{cz} = 10^{11} cm^2/s$ corresponds to the diffusivity in the
convection zone, $\eta_{sg} = 10^{13} cm^2/s$ corresponds to the supergranular diffusivity and
$r_{cz} = 0.73R_\odot$, $d_{cz} = 0.03R_\odot$, $r_{sg} = 0.95R_\odot$ and $d_{sg} = 0.05R_\odot$
characterize the transitions from one value of diffusivity to the other. Although it is common
these days to use these sort of profile, a point we note in passing is that the exact depth
dependence of turbulent diffusivity within the SCZ is poorly, if at all, constrained.  Besides
the value of the supergranular diffusivity $\eta_{sg}$, which can be estimated by observations
(Hathaway \& Choudhary 2008\nocite{hathaway-choudhary08}), the others parameters are not
necessarily well constrained.

Once all ingredients are defined(see Section \ref{Assimilation} for the flow field), we solve the
dynamo equations using a recently developed and novel numerical technique called exponential
propagation (see Appendix). Our computational domain is defined in a $250\times250$ grid
comprising only one hemisphere. Since we run our simulations only in one hemisphere our
latitudinal boundary conditions at the equator ($\theta=\pi/2$) are $\partial A /
\partial \theta = 0$ and $B=0$.  Furthermore, since the equations we are solving are axisymmetric, both the
potential vector and the toroidal field need to be zero ($A = 0$ and $B=0$) at the pole
($\theta=0$), to avoid singularities in spherical coordinates.  For the lower boundary condition
($r = 0.55R_\odot$), we assume a perfectly conducting core, such that both the poloidal field and
the toroidal field vanish there (i.e., $A, B = 0$ at the lower boundary). For the upper boundary condition we assume that the magnetic field has only a radial component ($B = 0$ and $\partial (r A)/\partial r = 0$); this condition has been found necessary for stress balance between subsurface and coronal magnetic fields (for more details refer to van Ballegooijen and Mackay 2007\nocite{vanBallegooijen-mackay07}). As initial conditions we set $A = 0$
throughput our computational domain and $B \propto \sin(2\theta)*\sin( \pi*( ( r - 0.55R_\odot)/(
R_\odot - 0.55R_\odot) ) )$. After a few cycles, all transients related to the initial conditions
typically disappear and the dynamo settles into regular oscillatory solutions whose properties
are determined by the parameters in the dynamo equations.


\section{Constraining the Flow Fields}\label{Assimilation}

The last two ingredients of the dynamo model are the velocity fields (differential rotation and
meridional circulation), which are the focus of this work.  The differential
rotation is probably the best constrained of all dynamo ingredients but the actual helioseismology
data has never before been used directly in dynamo model, only an analytical fit to it.  We discuss
below how the actual rotation data can be used directly within dynamo models through the use of a
weighting function to filter out the observational data in the region where it cannot be trusted. On the other hand, the
meridional circulation is one of the most loosely constrained ingredients of the dynamo.
Traditionally only the peak surface flow speed is used to constrain the analytical functions that
are used to parameterize it, in conjunction with mass conservation. In this work we take advantage
of the properties of such functions and make a fit to the helioseismic data on the meridional flow
that constrains the location and extent of the polar downflow and equatorial upflow, as well as the
radial dependence of the meridional flow near the surface -- thereby taking steps towards better
constrained flow profiles.

\subsection{Differential Rotation}

As opposed to meridional circulation, there are helioseismic measurements of the differential
rotation for most of the convective envelope which can be used directly in our simulations. Here we
use data from the Global Oscillation Network Group (GONG) (courtesy Dr. Rachel Howe) obtained using
the RLS inversion mapped onto a $51\times51$ grid (see Figure \ref{DR}-a).  However, these
observations cannot be trusted fully in the region within $0.3R_\odot$ of the rotation axis
(specifically at high latitudes), because the inversion kernels have very
little amplitude there.  Below we outline a method to
deal with this suspect data by creating a composite rotation profile that replaces these data
at high latitudes with plausible synthetic data, that smoothly matches to the observations in the
region of trust.


\subsubsection{Adaptation of the Data to the Model}\label{DR_Comp}

\begin{figure}[c]
  \begin{tabular}{cc}
  \includegraphics[scale=0.5]{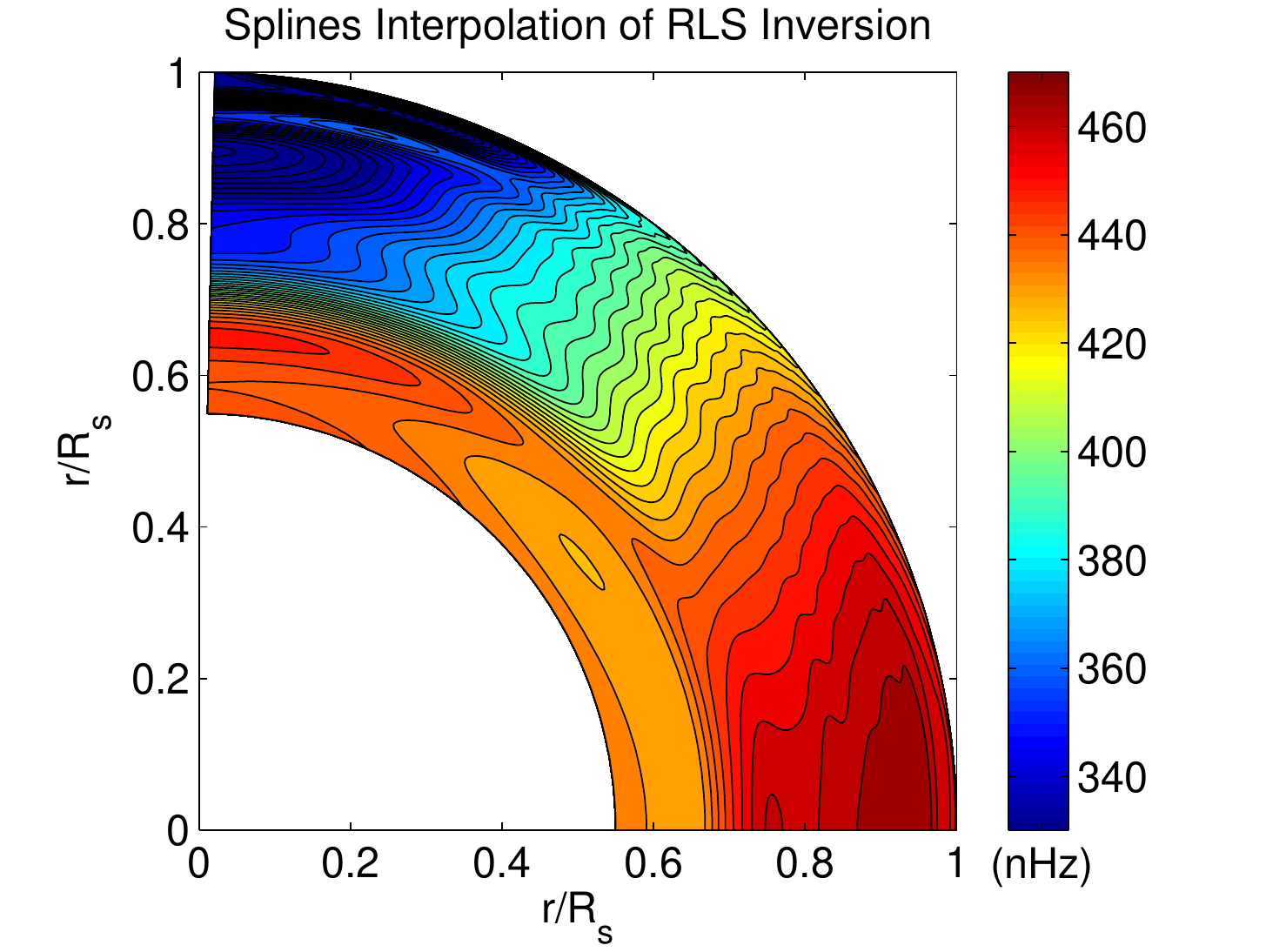} & \includegraphics[scale=0.5]{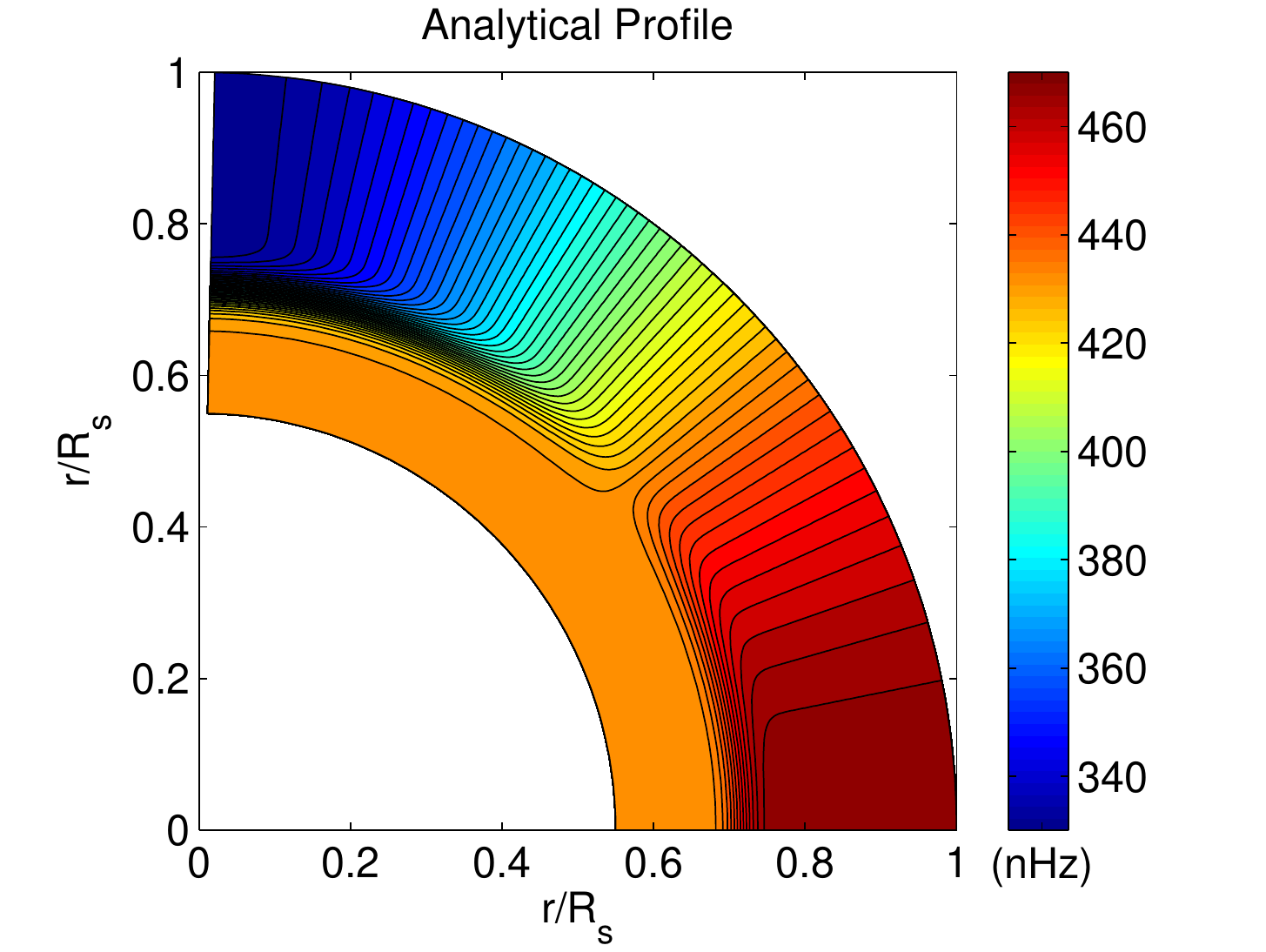}\\
    $(a)$ & $(b)$\\
  \includegraphics[scale=0.5]{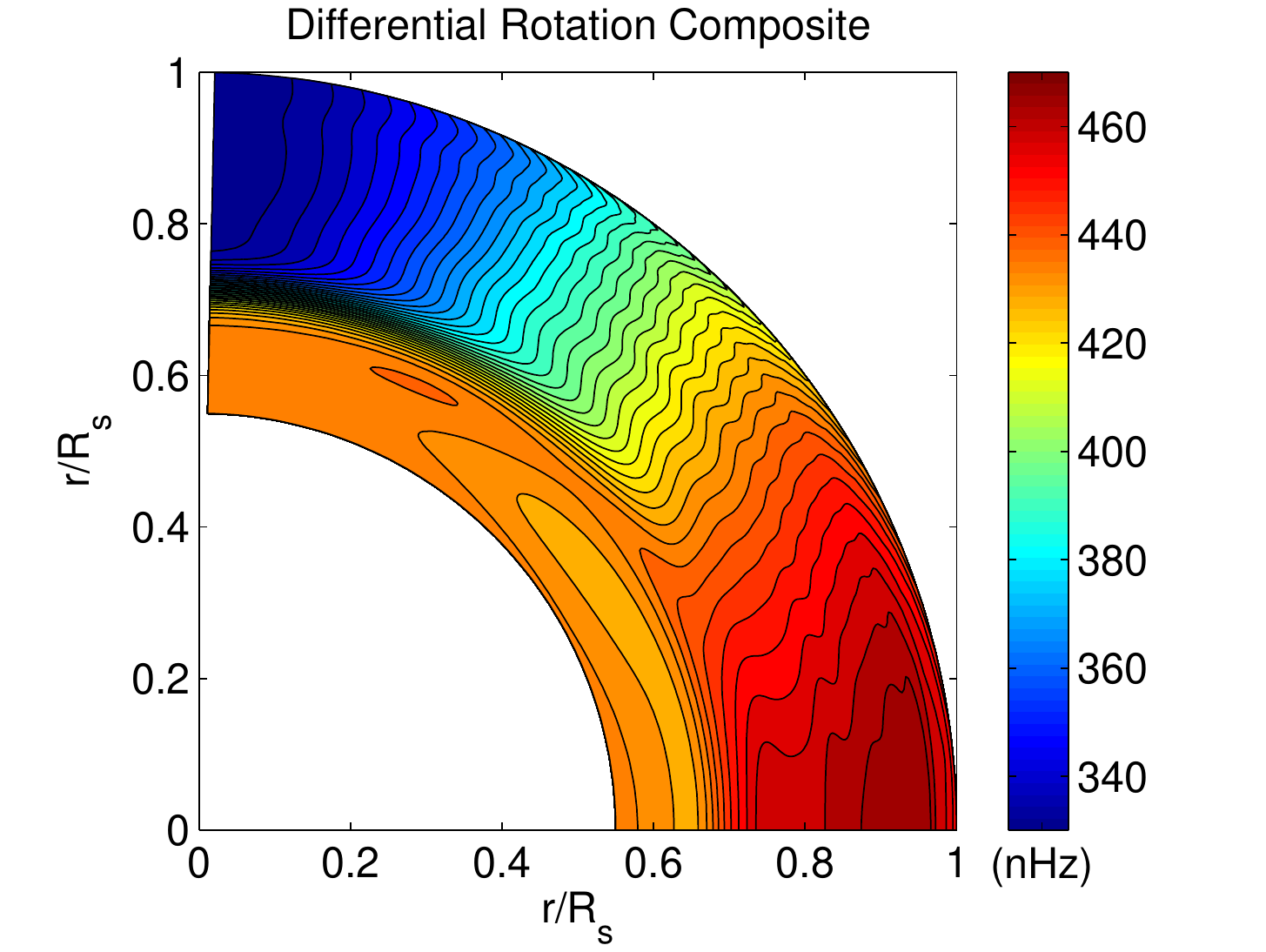} & \includegraphics[scale=0.5]{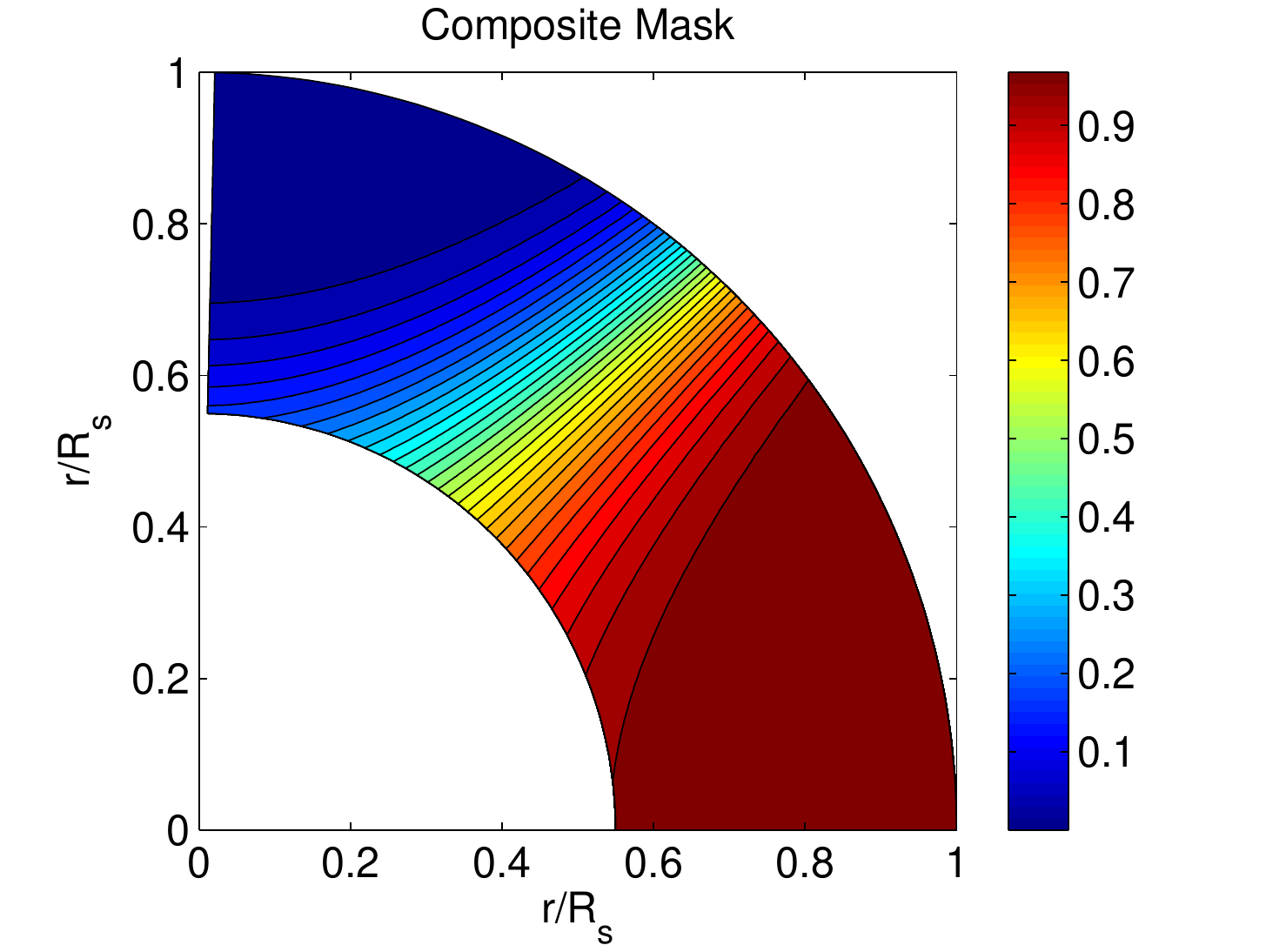}\\
    $(c)$ & $(d)$\\
  \end{tabular} \\
\caption{(a) Spline interpolation of the RLS inversion.  (b) Analytical profile of Charbonneau et al.\ (1999). (c) Differential rotation composite used in our simulations. (d) weighting function used to create a
composite between the RLS inversion and the analytical profile of Charbonneau et al.\ 1999.
 For all figures the red denotes the highest and blue the lowest value and the units are nHz with the exception of the weighting function.}\label{DR}
\end{figure}

In the first step we use a splines interpolation in order to map the data to the resolution of our
simulation (a grid of $250\times250$ see Figure \ref{DR}-a).  The next step is to make a composite
with the data and the analytical form of Charbonneau et al.\ (1999;\nocite{charbonneau-etal99} see
Figure \ref{DR}-b).  The analytical form is defined as:

\begin{equation}\label{DRan}
    \begin{array}{c}
      \Omega_A(r,\theta) = 2\pi\left[\Omega_{c} + \frac{1}{2}\left( 1 - \texttt{erf}\left( \frac{r - r_{tc}}{w_{tc}}  \right)\right)\left( \Omega_{e} - \Omega_{c} + ( \Omega_{p} - \Omega_{e} )\Omega_S(\theta) \right)\right]\\
      \\
      \Omega_S(\theta) = a\cos^2(\theta) + (1-a)\cos^4(\theta), \\
    \end{array}
\end{equation}
where $\Omega_c = 432$ nHz is the rotation frequency of the core, $\Omega_e = 470$ nHz is the
rotation frequency of the equator, $\Omega_p = 330$ nHz is the rotation frequency of the pole, $a
= 0.483$ is the strength of the $\cos^2(\theta)$ with respect to the $\cos^4(\theta)$ term,
$r_{tc} = 0.716$ the location of the tachocline and $w_{tc} = 0.03$.  We use the parameters
defining the tachocline's location and thickness as reported by Charbonneau et al.\ (1999) for a
latitude of $60^o$. This is because they match the data better at high latitudes (which is the
place where the data merges with the analytical profile) than those reported for the equator.
This composite replaces the suspect data at high latitudes within $0.3R_\odot$ of the rotation
axis with that of the analytic profile. However, it is important to note that at low latitudes,
within the convection zone, the actual helioseismic data is utilized.


In order to make the composite we create a weighting function $m(r,\theta)$ with values between 0 and 1 for each
grid-point defining how much information will come from the RLS data and how much from the
analytical form (see Figure \ref{DR}-d).  We define the weighting function in the following way:

\begin{equation}\label{Mask}
    m(r,\theta) = 1 - \frac{1}{2}\left(1 + \texttt{erf}\left(\frac{r^2\cos(2\theta) -
    r_{m}^2}{d_{m}^2}\right)\right),
\end{equation}
where $r_m = 0.5R_\odot$ is a parameter that controls the center of the transition and $d_m =
0.6R_\odot$ controls the thickness.  The resultant differential rotation profile, which can be seen in
Figure \ref{DR}-c, is then calculated using the following expression:

\begin{equation}\label{DR_Cmp}
      \Omega(r,\theta) = m(r,\theta)\Omega_{RLS}(r,\theta) + [1 - m(r,\theta)]\Omega_A(r,\theta)
\end{equation}

\subsubsection{Differences Between the Analytical Profile and the Composite Data}

\begin{figure}[c]
  \begin{tabular}{cc}
  \includegraphics[scale=0.5]{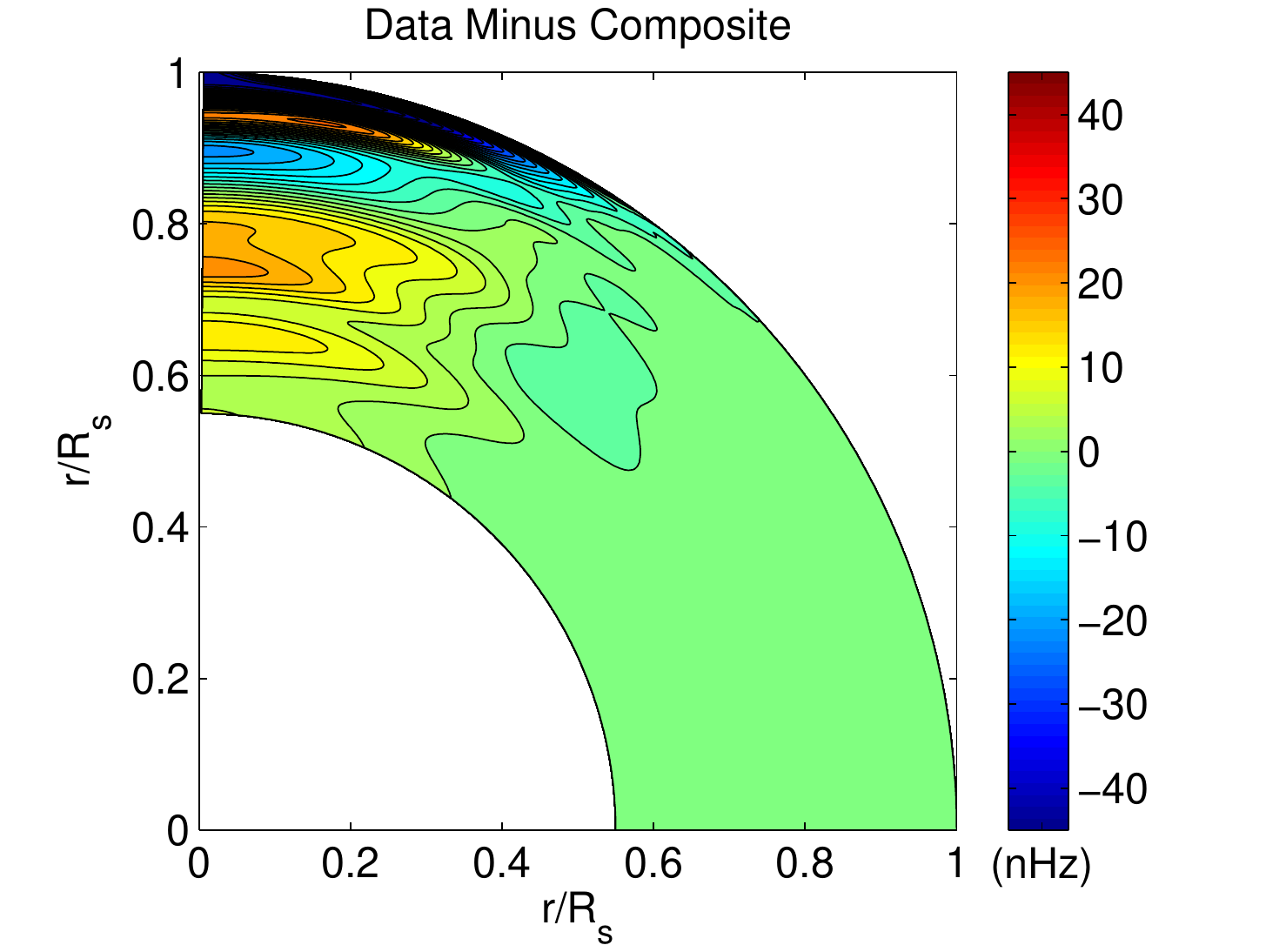} & \includegraphics[scale=0.5]{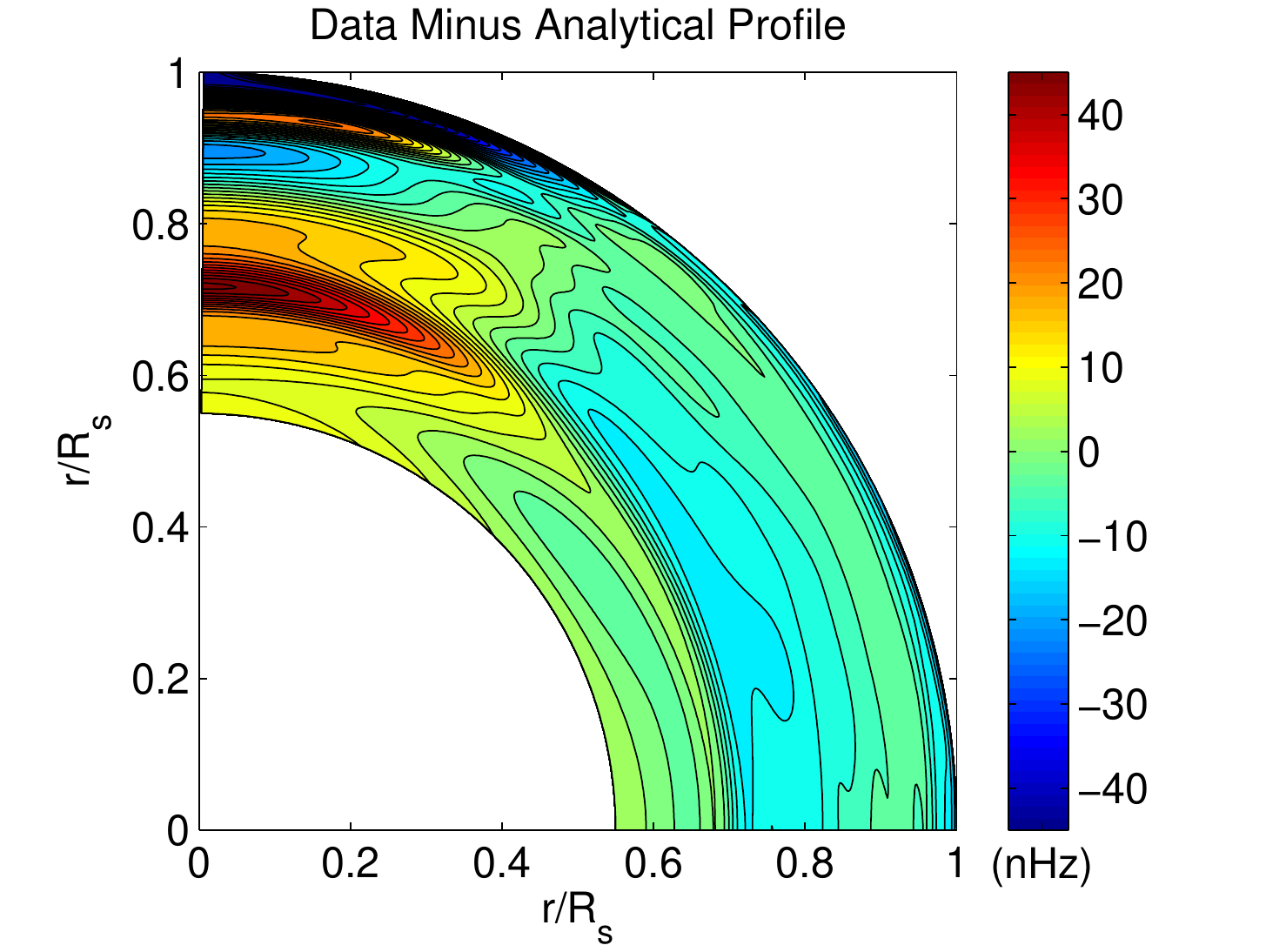}\\
    $(a)$ & $(b)$\\
  \end{tabular}
\caption{(a) Residual of subtracting the composite used in this work to the RLS inversion. (b)
Residual of subtracting the analytical profile commonly used by the community to the RLS inversion.
 Red color corresponds to the hightest value and blue to the lowest.  Graphs in units of nHz}\label{DR_Res}
\end{figure}

It is instructive to compare the analytical and composite profiles with the actual
helioseismology data. In Figure \ref{DR_Res}-a we present the residual error of subtracting the
analytical profile of Charbonneau et al.\ (1999), with $r_{tc} = 0.7$ and $w_{tc} = 0.025$, from the
RLS data. In Figure \ref{DR_Res}-b we present the residual error of subtracting our composite from
the RLS data. As is expected, there is no difference between the composite and raw data at low
latitude, but the residual increases as we approach the rotation axis -- where the RLS data cannot
be trusted.  For the analytical profile, the residuals errors are more significant, even at low
latitudes. This demonstrates the ability of our methodology to usefully integrate the helioseismic
data for differential rotation.


\subsection{Meridional Circulation}

The meridional flow profile remains rather poorly constrained in the solar interior even though
the available helioseismic data can be used to constrain the analytic flow profiles that are in
use currently. Here we present ways to betters constrain this profile with helioseismic data. In
order to do that, we use data from GONG (Courtesy Dr. Irene Gonz\'alez-Hern\'andez) obtained
using the ring-diagrams technique. This data, which we can see in Figure \ref{VthFit}-a,
corresponds to a time average of the meridional flow between $2001$ and $2006$; and it comprises
19 values of $r$ from $R_\odot$ down to a depth of $0.97R_\odot$, and 15 different latitudes
between $-52.5^o$ and $52.5^o$. It is important to note that our work relies heavily in the
assumption that the meridional flow is adequately described by a stream function with separable
variables.  This is consistent with the assumption present implicitly in all work on axisymmetric
solar dynamo models up to this date. Below we use this property of our stream function, along
with weighted latitudinal and radial averages of the data, to completely constrain its
latitudinal dependence, as well as the topmost ten percent of its radial dependence. As this data
currently does not constrain the depth of penetration of the flow in the deep solar interior, we
explore two different plausible penetration depths of the circulation. For reasons described
later, we choose to perform simulations with two different peak meridional flow speeds, therefore
exploring four plausible meridional flow profiles altogether.

\subsubsection{Constraining the Latitudinal Dependence of the Meridional Flow}

Meridional circulation has been typically implemented in these type of dynamo models by using a
stream function in combination with mass conservation, i.e.:
\begin{equation}\label{Vel_F}
    \overrightarrow{\textbf{v}}_p\left(r,\theta\right) =
    \frac{1}{\rho(r)}\vec{\nabla}\times\left(\Psi(r,\theta)\widehat{\textbf{e}}_{\phi}\right).
\end{equation}
The two stream functions that are commonly used were proposed by van Ballegooijen and Choudhuri
(1988)\nocite{vanBallegooijen-Choudhuri88} and by Dikpati and Choudhuri
(1995)\nocite{Dikpati-Choudhuri95}.  They have in common the separability of variables and thus
can be written in the following way:

\begin{equation}\label{Str_F}
    \Psi(r,\theta)=v_0F(r)G(\theta),
\end{equation}
where $v_0$ is a constant which controls the amplitude of the meridional flow.

\begin{figure}[c]
  \includegraphics[scale=0.6]{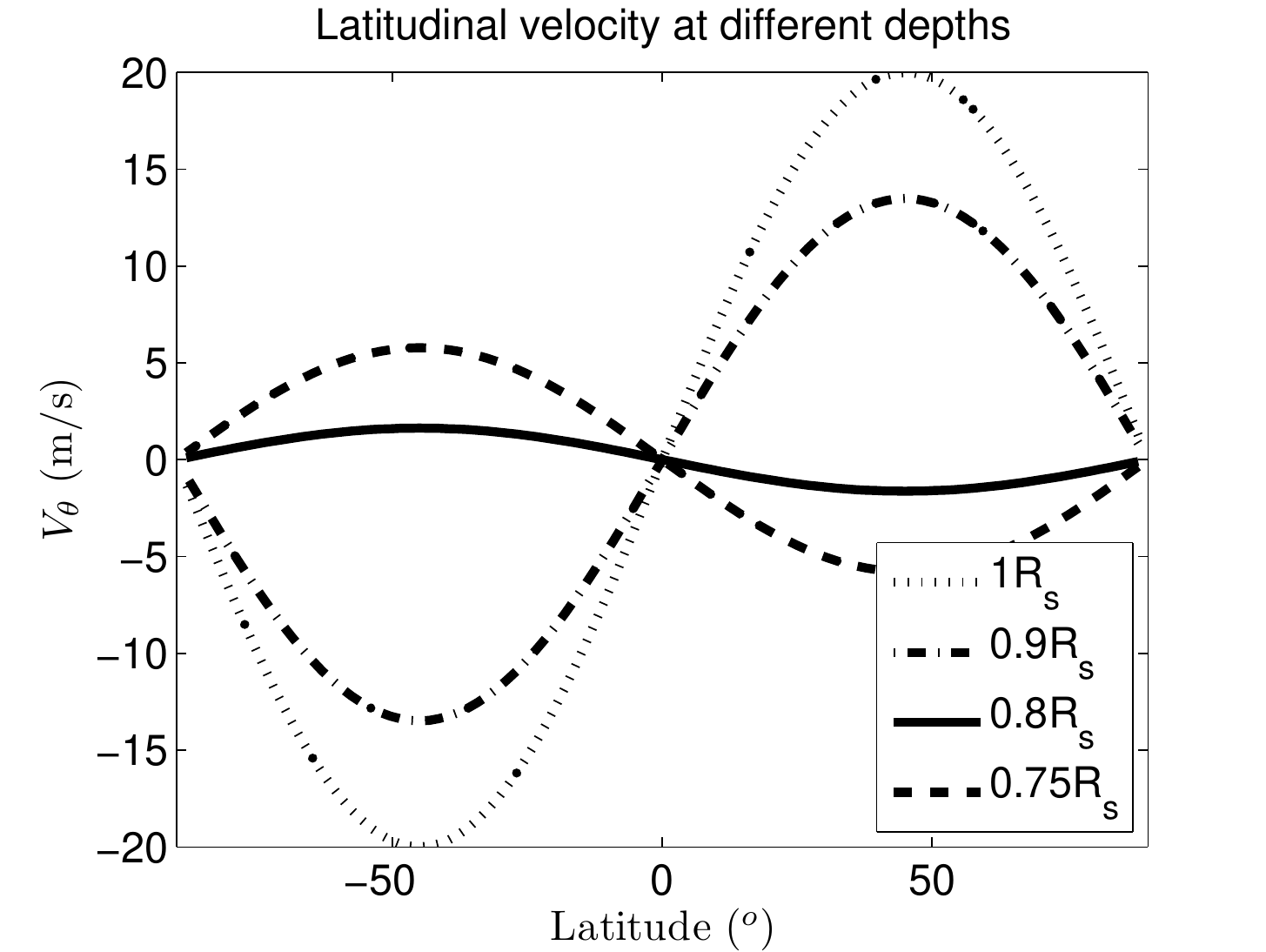}\\
  \caption{Latitudinal velocity as a function of $\theta$ used by van Ballegooijen and Choudhuri 1988 for different depths.  Notice that the curves differ from each other only on their amplitude.}\label{Vth_r}
\end{figure}

Using such a stream function the components of the meridional flow become:
\begin{equation}\label{Vel_Fr}
    v_r(r,\theta) =
    v_0\frac{F(r)}{r\rho(r)}\frac{1}{\sin(\theta)}\frac{\partial}{\partial\theta}\left(\sin(\theta)G(\theta)\right),
\end{equation}
\begin{equation}\label{Vel_Ft}
    v_{\theta}(r,\theta) = -v_0\frac{1}{r\rho(r)}\frac{\partial}{\partial r}(r F(r))G(\theta),
\end{equation}
which can themselves be separated into the multiplication of exclusively radially and latitudinally
dependent functions.  This property allows us to constrain the entire latitudinal dependence of
this family of functions by using the available helioseismology data for $v_{\theta}$ at the
surface.  This can be done because the latitudinal dependence of $v_{\theta}$ is exactly the same
as that of the stream function and only the amplitude of this functional form changes with depth,
see for example Figure \ref{Vth_r} for the latitudinal velocity at different depths used by van
Ballegooijen and Choudhuri (1988)\nocite{vanBallegooijen-Choudhuri88}. In this work we assume a
latitudinal dependence like the one they used, i.e.,
\begin{equation}\label{Psi_th}
    G(\theta) = \sin^{(q+1)}(\theta)\cos(\theta).
\end{equation}

\begin{figure}[c]
\begin{tabular}{c}
  \begin{tabular}{cc}
  \includegraphics[scale=0.5]{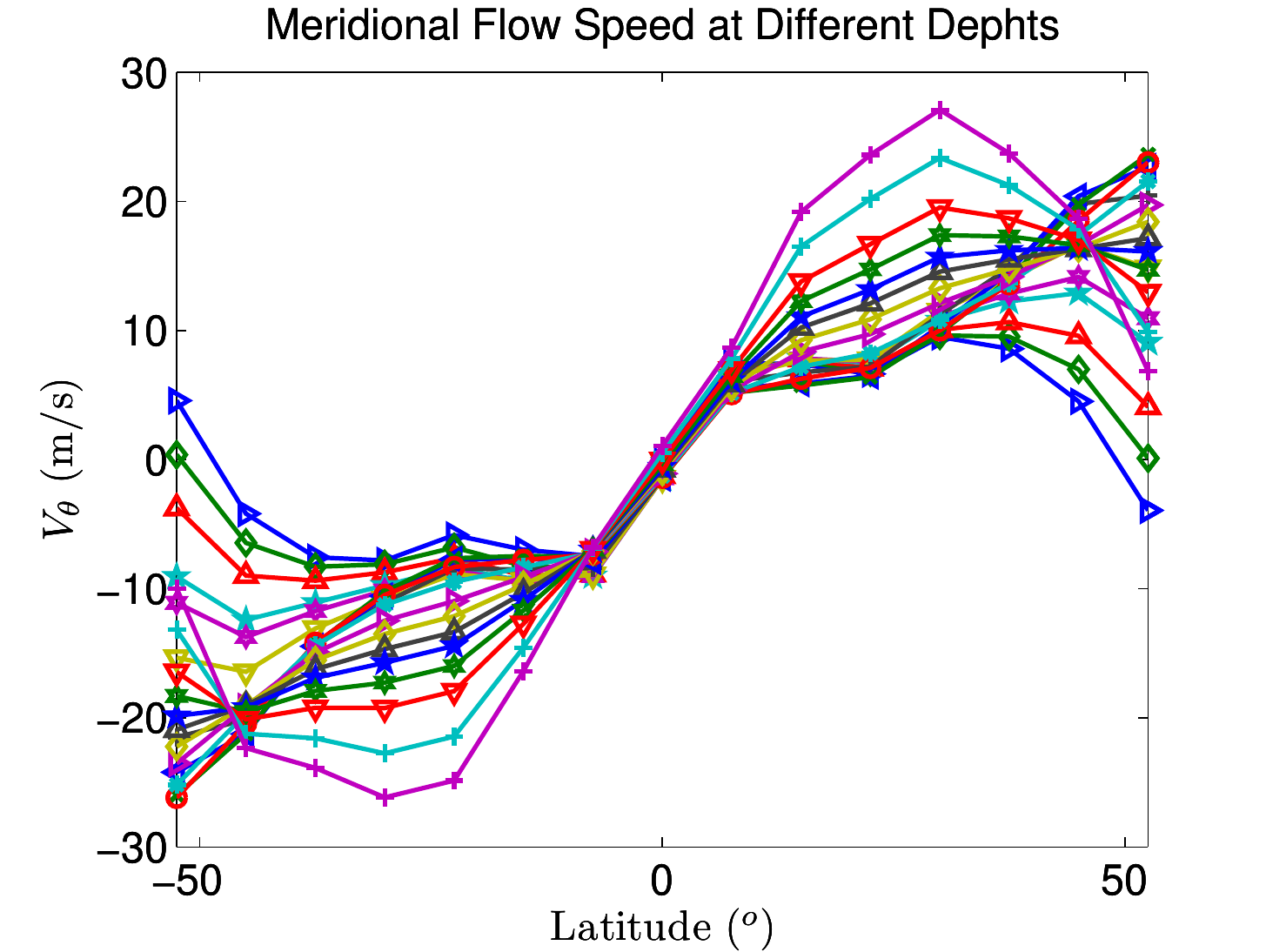} & \includegraphics[scale=0.5]{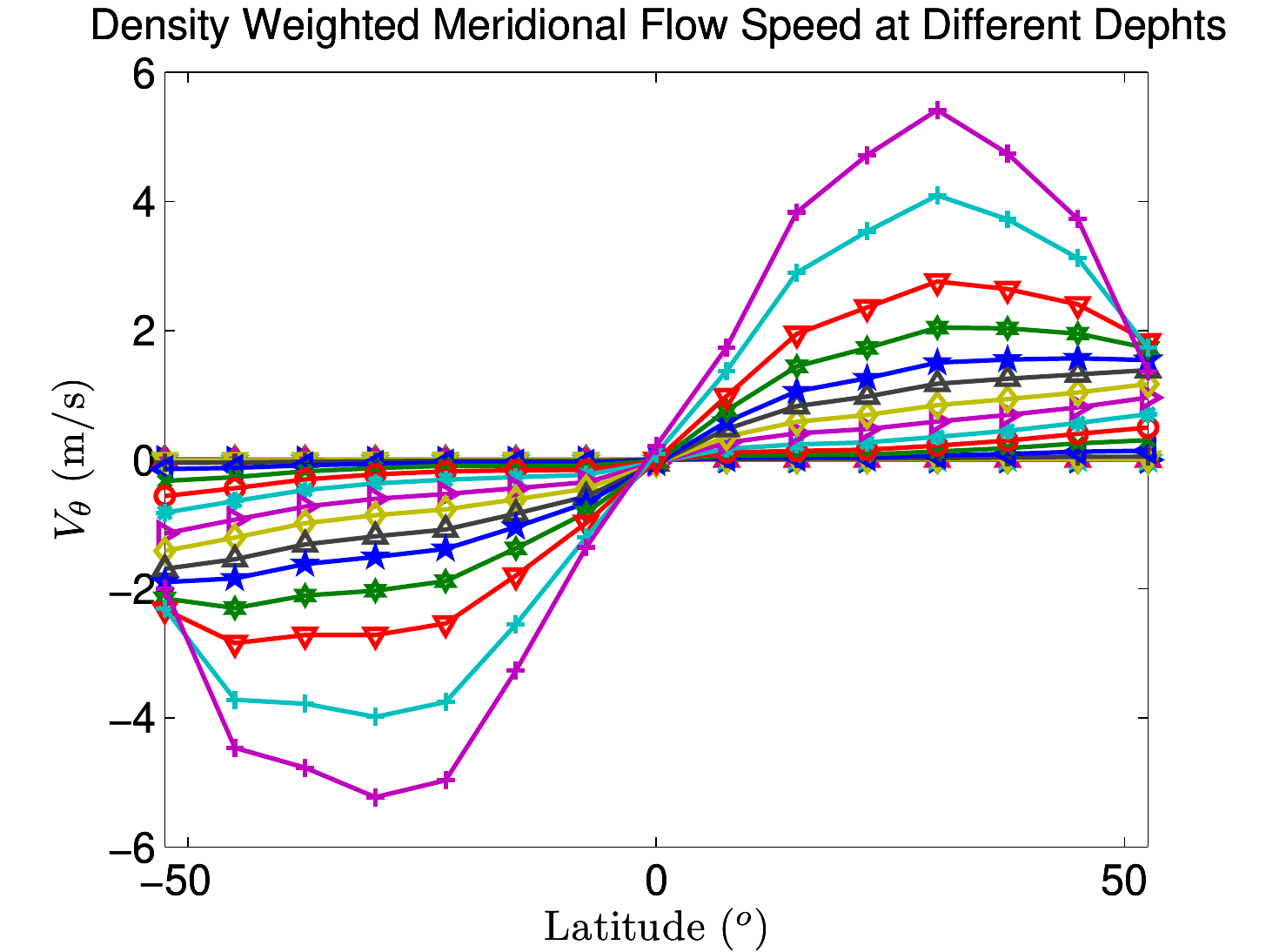}\\
    $(a)$ & $(b)$\\
  \end{tabular} \\
  \includegraphics[scale=0.6]{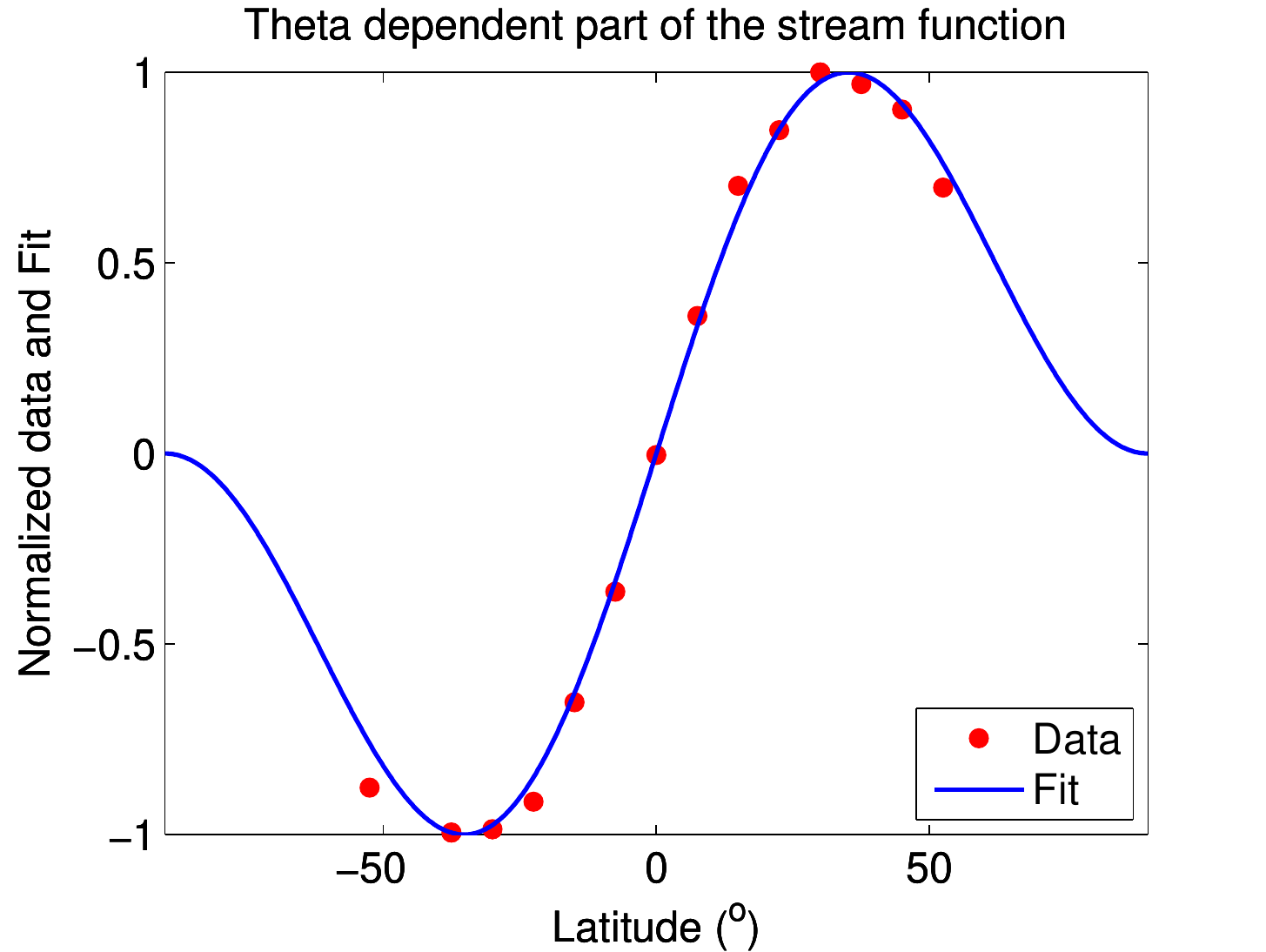} \\
  (c) \\
\end{tabular}
\caption{(a) Measured meridional flow as a function of latitude at different depths (Courtesy Dr.
Irene Gonz\'alez-Hern\'andez), each combination of colors and markers corresponds to a different
depth ranging from $0.97R_\odot$ to $R_\odot$.  (b) Meridional flow after being weighted using solar density.  If we sum all data points at
each latitude we obtain the average velocity.  (c) Normalized average velocity and analytical
fit.}\label{VthFit}
\end{figure}

In order to estimate the parameter $q$ we first take a density weighted average of the
helioseismic data using the values for solar density from the Solar Model S
(Christensen-Dalsgaard et al.\ 1996\nocite{Christensen-Dalsgaardetal96}), such that
\begin{equation}\label{DA}
    \bar{v}_\theta(\theta_j) = \frac{\sum_i v_{\theta}(r_i,\theta_j)\rho(r_i)}{\sum_i \rho(r_i)}.
\end{equation}
In Figure \ref{VthFit}-b we plot the meridional flow at different depths weighted by density, which
we add for each latitude in order to find the average. We then use this average to make a least
squares fit to the analytical expression (eq \ref{Psi_th}), which we can see in Figure \ref{VthFit}-c.
We find that a value of $q = 1$ fits the data best. This therefore constrains the
latitudinal ($\theta$) dependence of the flow profile.


\subsubsection{Constraining the Radial Dependence of the Meridional
Flow}

\begin{figure}[c][h]
\begin{tabular}{c}
  \begin{tabular}{cc}
  \includegraphics[scale=0.5]{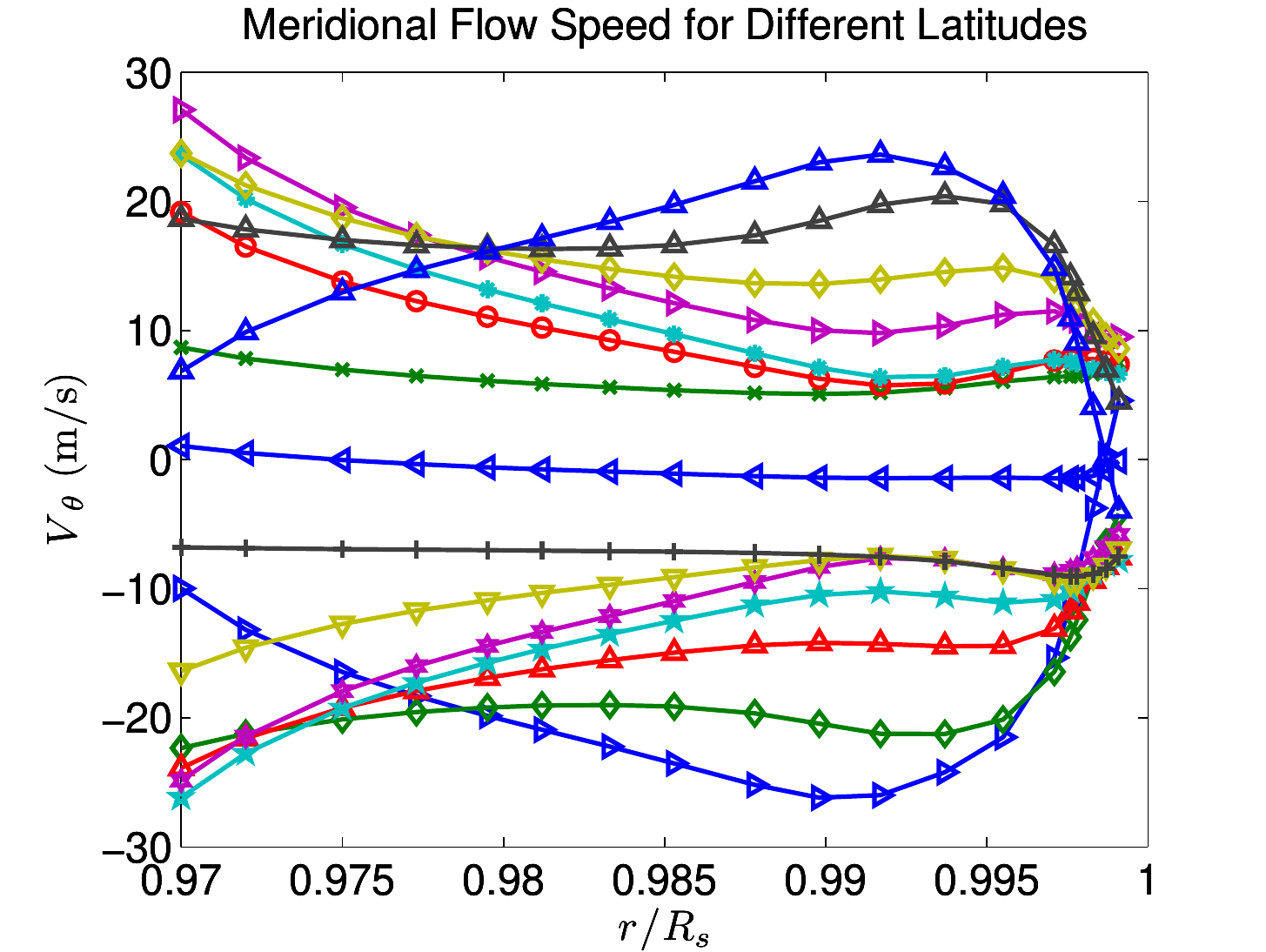} & \includegraphics[scale=0.5]{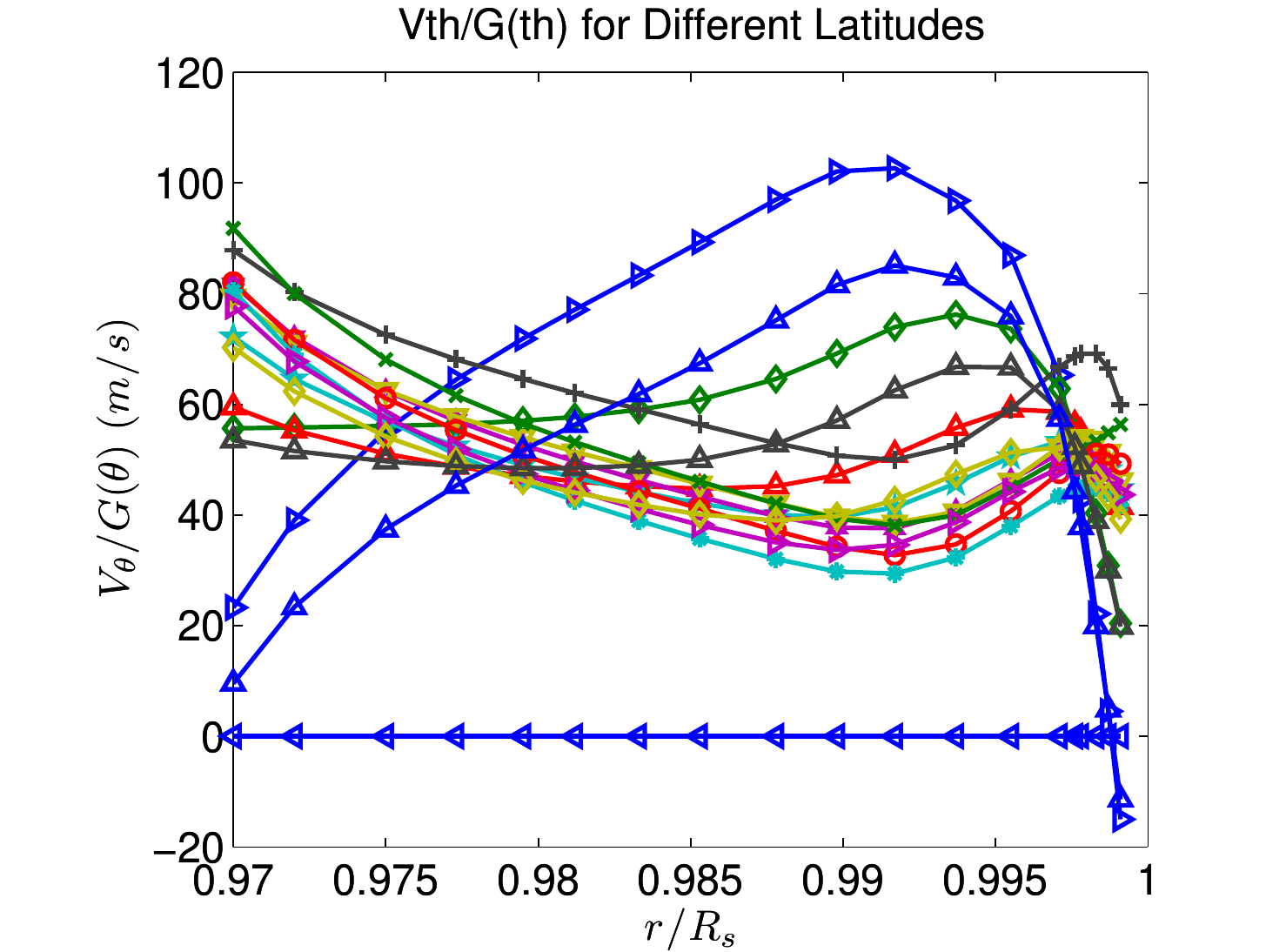}\\
    $(a)$ & $(b)$\\
  \end{tabular} \\
  \includegraphics[scale=0.6]{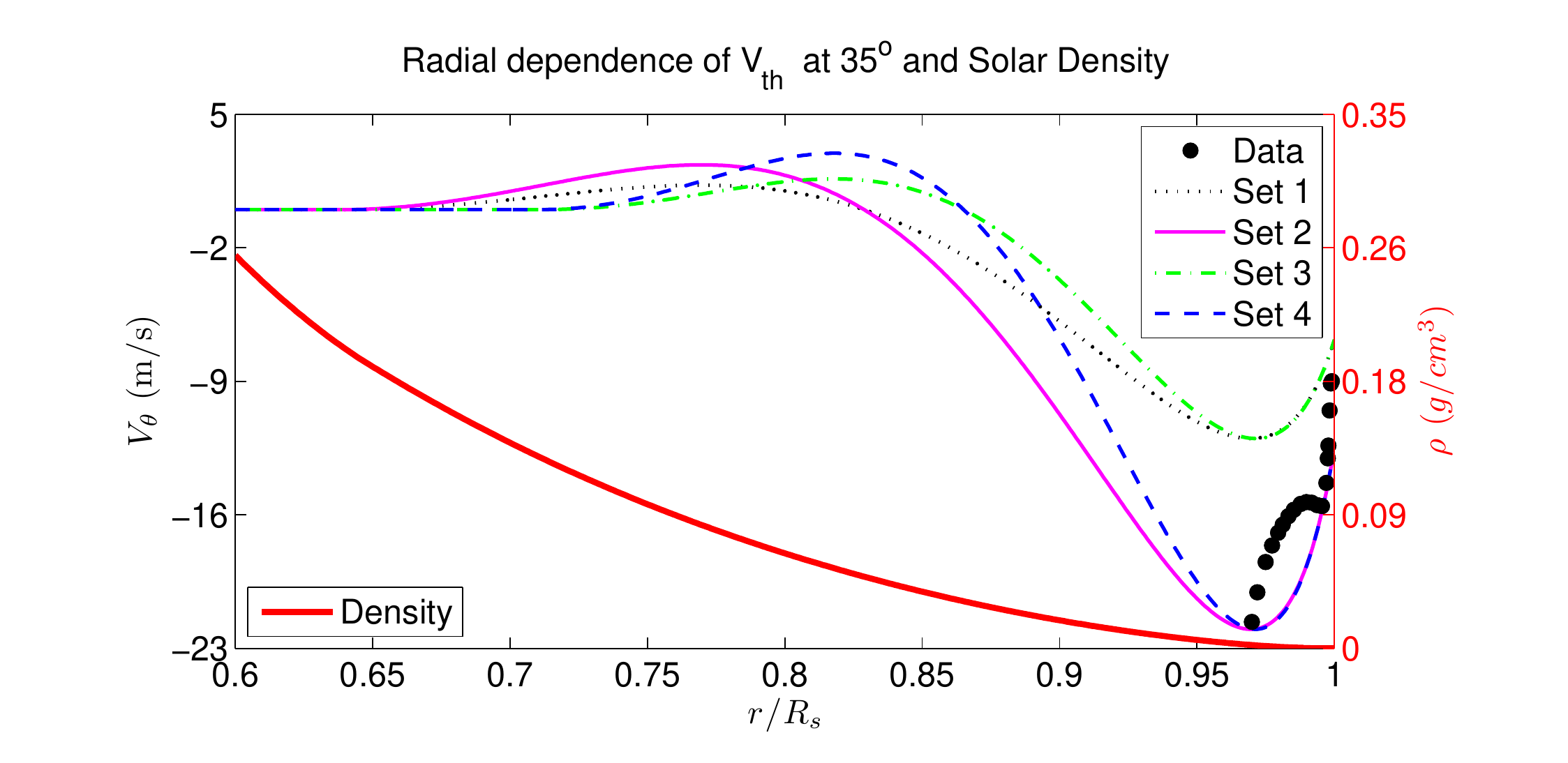} \\
  (c) \\
\end{tabular}
\caption{(a) Measured meridional flow as a function of radius at different latitudes (Courtesy Dr.
Irene Gonz\'alez-Hern\'andez), each combination of colors and markers corresponds to a different
latitude varying from $-52.5$ to $52.5$. (b) Meridional flow after removing the latitudinal dependence, i. e.
$v_\theta/G(\theta)$. The horizontal line with zero latitudinal velocity corresponds to the Equator.
(c) Radial dependence of the latitudinally averaged meridional flow for our
helioseismic data is depicted as large black dots. Other curves correspond to the radial dependence
of the meridional flow profiles used in our simulations and solar density: Set 1 (black dotted)
$R_p = 0.64R_\odot$, $v_o = 12 m/s$; Set 2 (magenta solid) $R_p = 0.64R_\odot$, $v_o = 22 m/s$; Set
3 (green dash-dot) $R_p = 0.71R_\odot$, $v_o = 12 m/s$ and Set 4 (blue dashed line) $R_p =
0.71R_\odot$, $v_o = 22 m/s$. The solar density taken from the solar Model S (Christensen-Dalsgaard
et al.\ 1996) is depicted as a solid red line. The left-vertical axis is in units of velocity and the
right-vertical is in units of density.}\label{VthFitr}
\end{figure}

As opposed to the latitudinal dependence, the radial dependence of the meridional flow is less
constrained since there is no data below $0.97R_\odot$. However, at least some of the parameters
can be constrained: We start with the solar density, for which we perform a least squares fit to
the Solar Model S using the following expression:
\begin{equation}\label{Rho}
    \rho(r) \sim \left( \frac{R_\odot}{r} - \gamma \right)^{m}
\end{equation}
we find that values of $\gamma = 0.9665$ and $m = 1.911$ fit the model best (see Figure
\ref{VthFitr}-c).


In the second step we constrain the radial dependence of the stream function.  We begin with the
function
\begin{equation}\label{Psi_r}
    F(r) = \frac{1}{r}(r - R_p)(r - R_\odot)\sin\left(\pi\frac{r - R_p}{R_1-R_p}\right)^a,
\end{equation}
where $R_\odot$ corresponds to the solar radius, $R_p$ to the maximum penetration depth of the
meridional flow and $a$ and $R_1$ control the location in radius of the poleward peak and the
value of the meridional flow at the surface. In order to constrain them we use the helioseismic
data again, but this time we use the radial dependence (see Figure \ref{VthFitr}-a). We first
remove the latitudinal dependence, which we do by dividing the data of each latitude by
$G(\theta)$ using the value of $q=1$ found in the previous section.  In Figure \ref{VthFitr}-b we
plot the flow data after removing the latitudinal dependence, note that there is no longer any
sign difference between the two hemispheres.  The next step is to generate the latitudinal
average which we can see as black dots in Figure \ref{VthFitr}-c. It is evident from looking at
the radial dependence of the meridional flow that the velocity increases with depth for most
latitudes, and that the point of maximum velocity is not within the depth up to which the data
extends. Since the exact radial dependence of the data is too complex for our functional form to
grasp, the features we concentrate on reproducing are the presence of a maximum inside the
convection zone, as well as the amplitude of the flow at the near surface-layers. The logic here
is to use the fewest possible parameters and a simple, physically transparent profile that does a
reasonable job of matching the data. In view of the lack of better constraints, we assume here
that the peak of the return flow is at $0.97R_\odot$ (which is the depth at which the current
helioseismic data has it's peak).

Following the procedures and steps above we construct profiles to fit the depth-dependence
pointed out in the available helioseismic data; however, this does not constrain how far the
meridional flow can penetrate and therefore we try two different penetration depths, one shallow
$0.71R_\odot$ (i.e., barely beneath the base of the SCZ) and one deep $R_p = 0.64R_\odot$ (into
the radiative interior) -- both of which match the latitudinal and radial constraints as deduced
from the near-surface helioseismic data. Note that observed light element abundance ratios limit
the depth of penetration of the circulation to about $R_p = 0.62R_\odot$ (Charbonneau 2007). Now
we know that the meridional flow speed is highly variable, with fluctuations that can be quite
significant and the measured flow speed can change depending on the phase of the solar cycle
(Hathaway 1996; Gizon \& Rempel 2008\nocite{gizon-rempel08}). The magnetic fields are also
expected to feed back on the flow (Rempel 2006\nocite{rempel06}). Taken together these
considerations point out that the effective meridional flow speed to be used in dynamo
simulations could be less than that implied from the Gonz\'alez-Hern\'andez et al.\ data, a
possibility that is borne out by the work of Braun \& Fan (1998) and Gizon \& Rempel (2008), who
find much lower peak flow speeds in the range $12$--$15$ m/s. Keeping this in mind, we use the
same latitudinal and radial constraint as deduced earlier, but consider in our simulations two
additional profiles with peak flow speeds of $12$ m/s with deep and shallow penetrations.
Therefore, in total we explore four plausible meridional flow profiles in our dynamo simulations
(see Figure \ref{VthFitr}-c and Table \ref{MFlowP} for an overview), the results of which are
presented in the next section.

\begin{table}[h]
  \centering
\begin{tabular}{|c|c|c|c|c|}
  \hline
   & $v_o$& $R_p$ & $a$ & $R_1$\\
  \hline
  \hline
  Set 1 & $12$ m/s & \multirow{2}{*}{$0.64R_\odot$}  & \multirow{2}{*}{$1.795$} & \multirow{2}{*}{$1.027R_\odot$} \\
  \cline{1-2}
  Set 2 & $22$ m/s & & & \\
  \hline
  Set 3 & $12$ m/s & \multirow{2}{*}{$0.71R_\odot$}  & \multirow{2}{*}{$2.03$}  & \multirow{2}{*}{$1.03R_\odot$} \\
  \cline{1-2}
  Set 4 & $22$ m/s & & & \\
  \hline
\end{tabular}
  \caption{Sets of parameters characterizing the different meridional flow profiles used in our dynamo simulations.  $v_o$ corresponds to the meridional flow peak speed, $R_p$ the maximum penetration of the flow, and $a$ and $R_1$ are parameters that control the location of the poleward flow as well as the surface speed.}\label{MFlowP}
\end{table}



\section{Dynamo Simulation: Results and Discussions}

\subsection{Analytic vs. Helioseismic Composite Differential Rotation}

We first compare dynamo solutions found using the helioseismology composite of the differential
rotation (Section \ref{DR_Comp}) and the analytical profile of Charbonneau et al.\
(1999)\nocite{charbonneau-etal99}. In doing so, we perform simulations with a model as described
in Section $2$ (see also numerical methods in the Appendix) and generate field evolution maps for
the toroidal and poloidal fields. From the simulated butterfly diagrams for the evolution of the
toroidal field at the base of the SCZ and the surface-radial field evolution
(Figure~\ref{Comp_An}), we find that large-scale features of the simulated solar cycle are
generally similar across the two different rotation profiles (even with different meridional flow
penetration depths and speeds), specially for the shallowest penetration (sets $3$ and $4$ with
$R_p=0.71R_\odot$).

In order to understand this similarity, it is useful to look at Figures \ref{S1C} and \ref{S4C}
corresponding to simulations using the composite differential rotation, and the meridional flow
sets $1$ ($R_p = 0.64R_\odot$, $v_o = 12 m/s$) and $4$ ($R_p = 0.71R_\odot$, $v_o = 22 m/s$). The
first two columns, from left to right of both figures show the evolution of the shear sources
$(\textbf{B}_r\cdot\nabla_r\Omega)$ and $(\textbf{B}_\theta\cdot\nabla_\theta\Omega)$ -- which
contribute to toroidal field generation by stretching of the poloidal field.  It is evident that
the location and strength of these sources is different -- the radial shear source is mainly
present near the surface whereas the latitudinal shear is spread throughout the convection zone.
It can also be seen that the radial shear source is roughly five times stronger than the
latitudinal. However, if attention is paid to the evolution of the toroidal
field (third column from the right in both figures), it is clear that this radial shear term has
no significant impact on the structure and magnitude of the toroidal field (a similar result was
found by Dikpati et al.\ 2002\nocite{dikpati-etal02}). The reason is that the upper boundary
condition ($B=0$ at $r=R_\odot$), in combination with the high turbulent diffusivity (and thus
short diffusive time scale) there, imposes itself very quickly on the toroidal field generated by
the radial shear -- washing it out. This greatly reduces the relative role of the surface shear as a
source of toroidal magnetic field, effectively making the surface dynamics very similar across
simulations using the analytical rotation profile (without any surface shear) or the composite
helioseismic profile.


Once the surface layers are ruled out as important sources of toroidal field generation we are
faced with the fact that the strongest source of toroidal field is the latitudinal shear inside the
convection zone (and not the tachocline radial shear), as is evident in Figures \ref{S1C} and
\ref{S4C}. This goes against the commonly held perception that the tachocline is where most of the
toroidal field is produced.  However, the importance of the latitudinal shear term in the SCZ is
clearly demonstrated here where we have plotted the shear source terms, which is not normally done
by dynamo modelers.

The establishment of the SCZ as an importance source region of toroidal field is of relevance
when regarding the similarity of the solutions obtained using the composite data and the
analytical profile.  This is because the region where the shear of the analytical profile and the
helioseismic composite data differ most (both for radial and latitudinal shear) is in the
tachocline. This is evident in Figure~\ref{DRDH_Res} where we plot the residual of subtracting
the radial and latitudinal shear of the analytical profile from the shear of the composite data.
The similarity between solutions is specially important for shallow meridional flow profiles with
low penetration -- which does not transport any poloidal field into the deeper tachocline,
thereby further diminishing the role of the tachocline shear.



\subsection{Shallow versus Deep Penetration of the Meridional Flow}

\begin{figure}[c]
  \begin{tabular}{ccc}
   \includegraphics[scale=0.3]{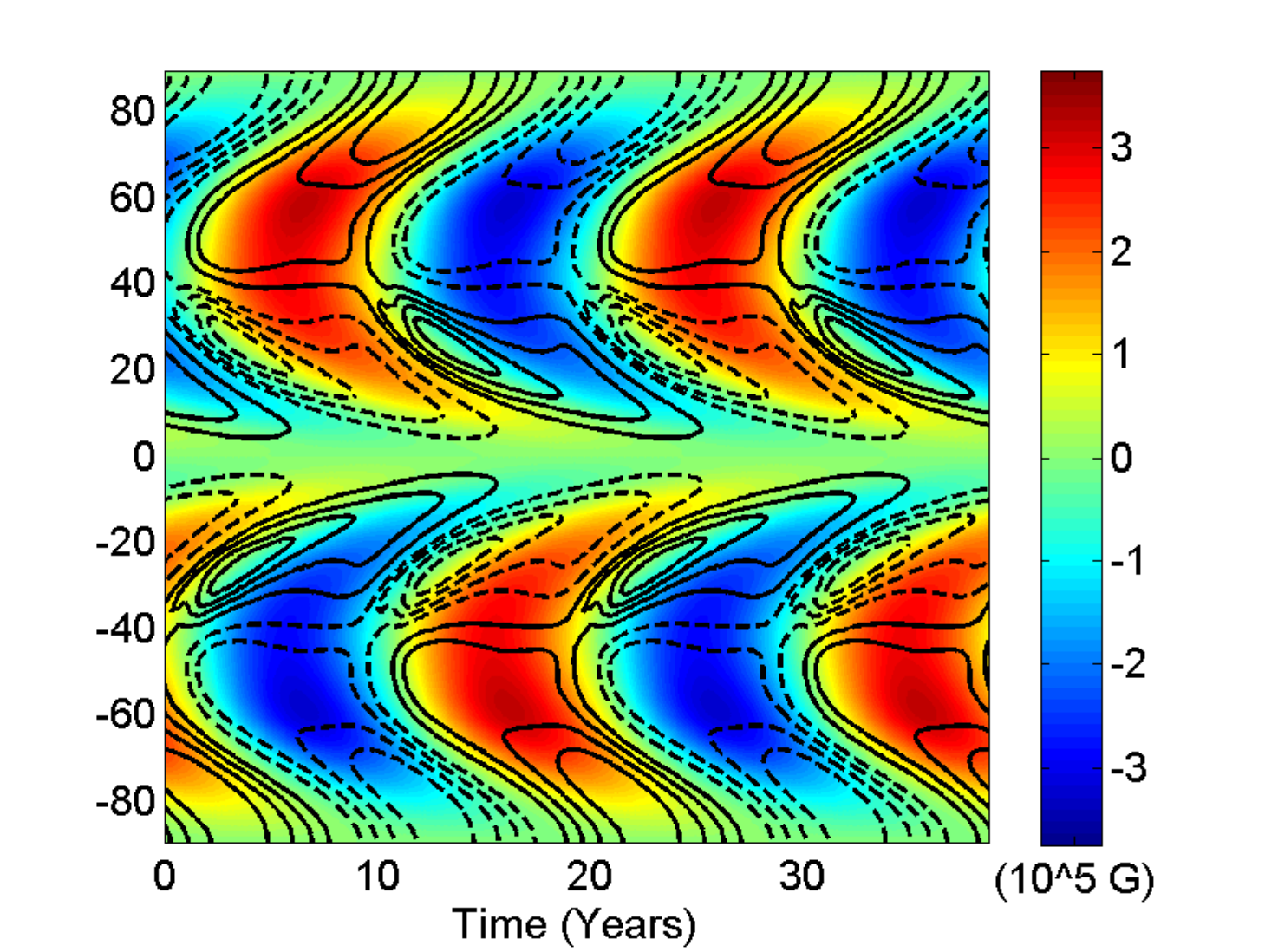} & $R_p = 0.64R_\odot$& \includegraphics[scale=0.3]{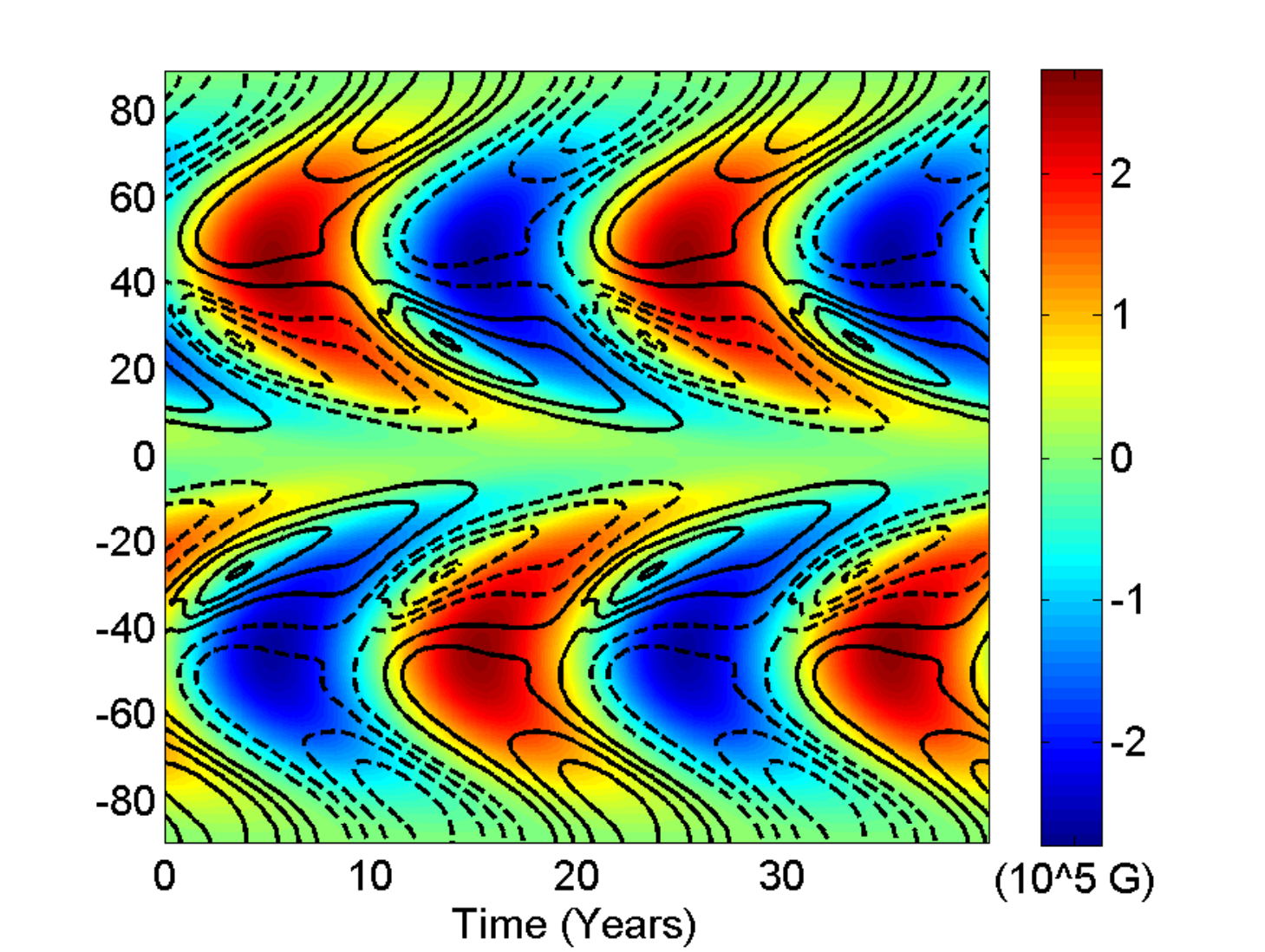}\\
   & $v_o = 12$m/s& \\
   \includegraphics[scale=0.3]{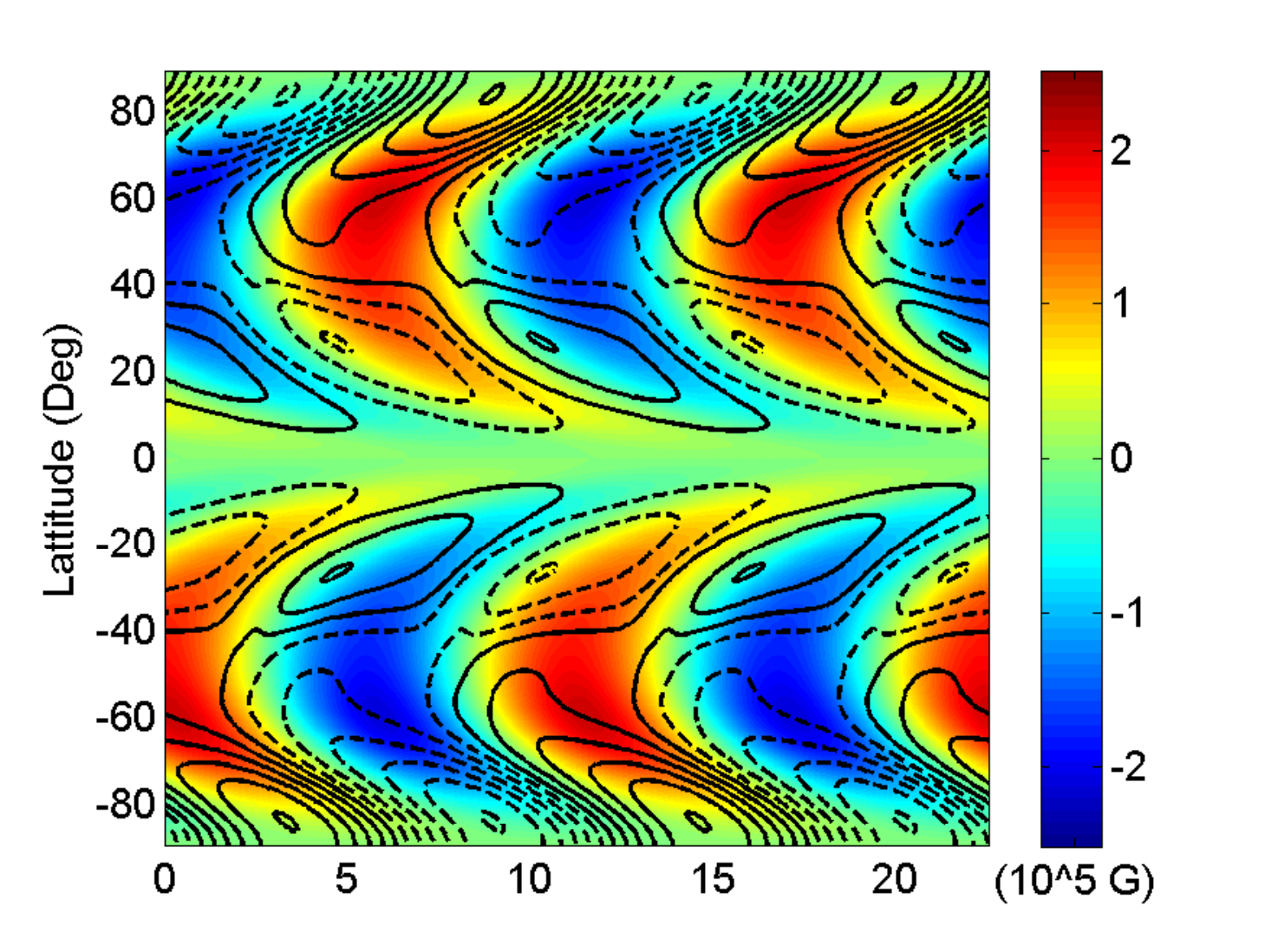} &$R_p = 0.64R_\odot$ & \includegraphics[scale=0.3]{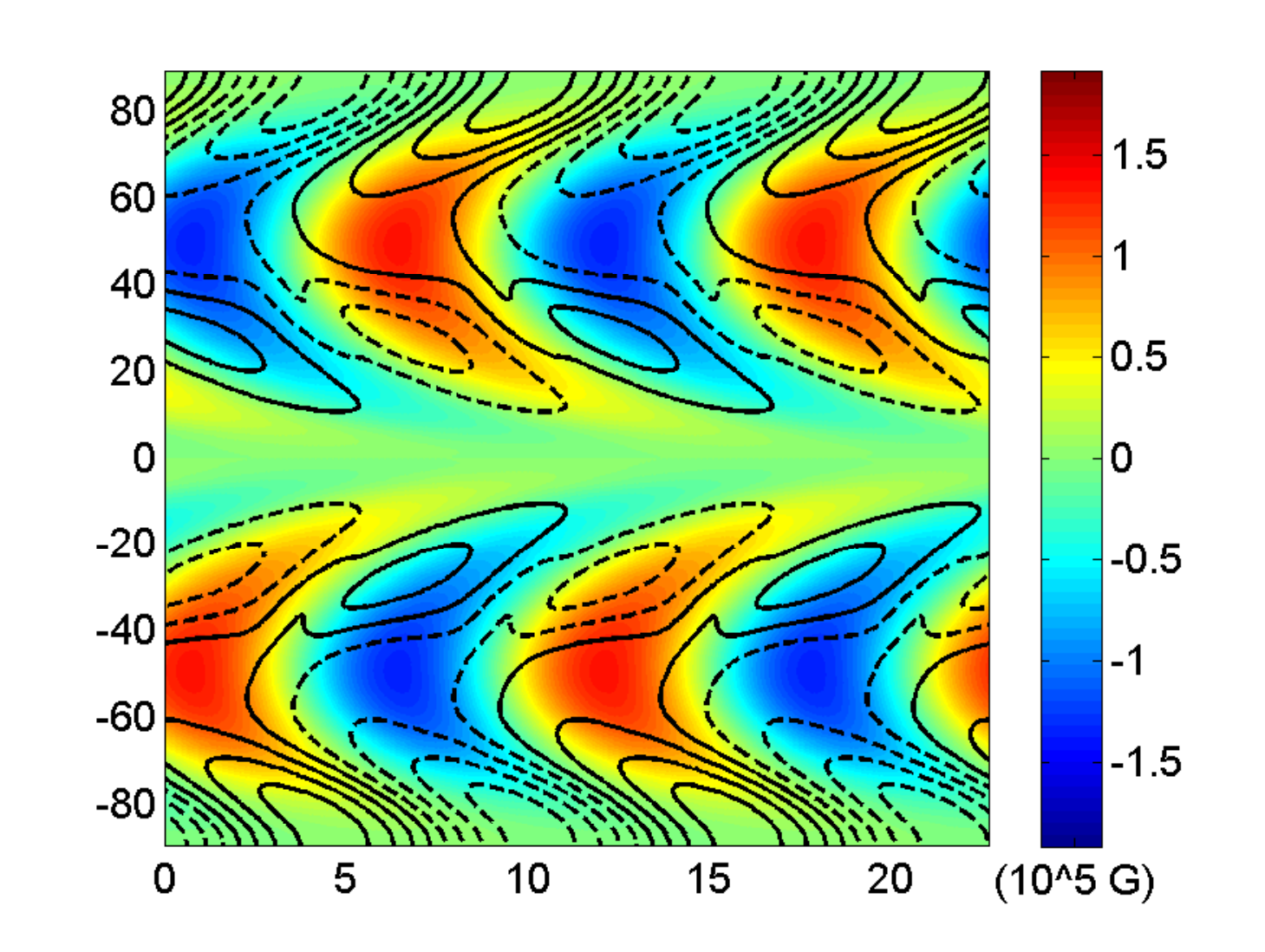}\\
   & $v_o = 22$m/s& \\
   \includegraphics[scale=0.3]{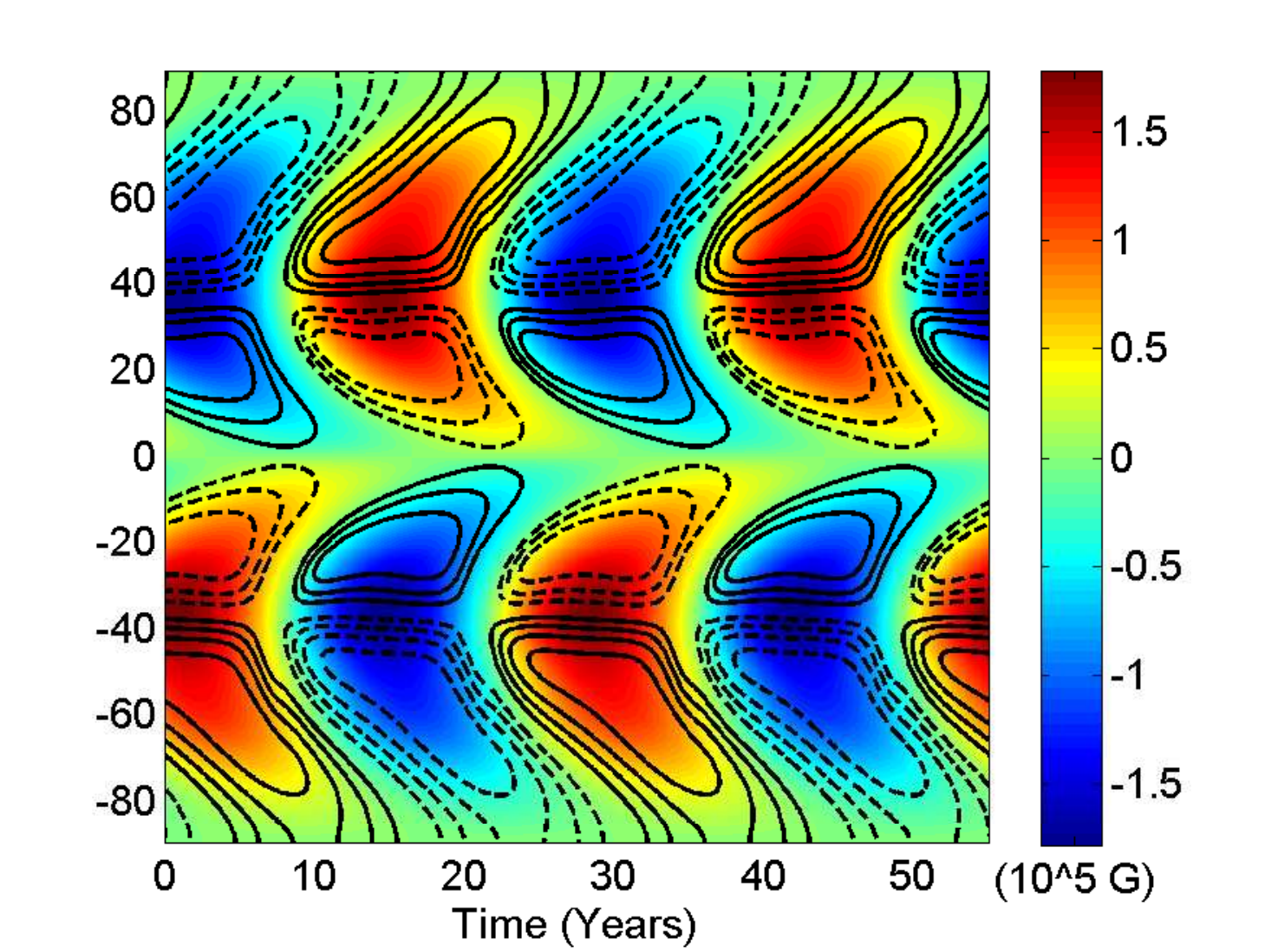} & $R_p = 0.71R_\odot$& \includegraphics[scale=0.3]{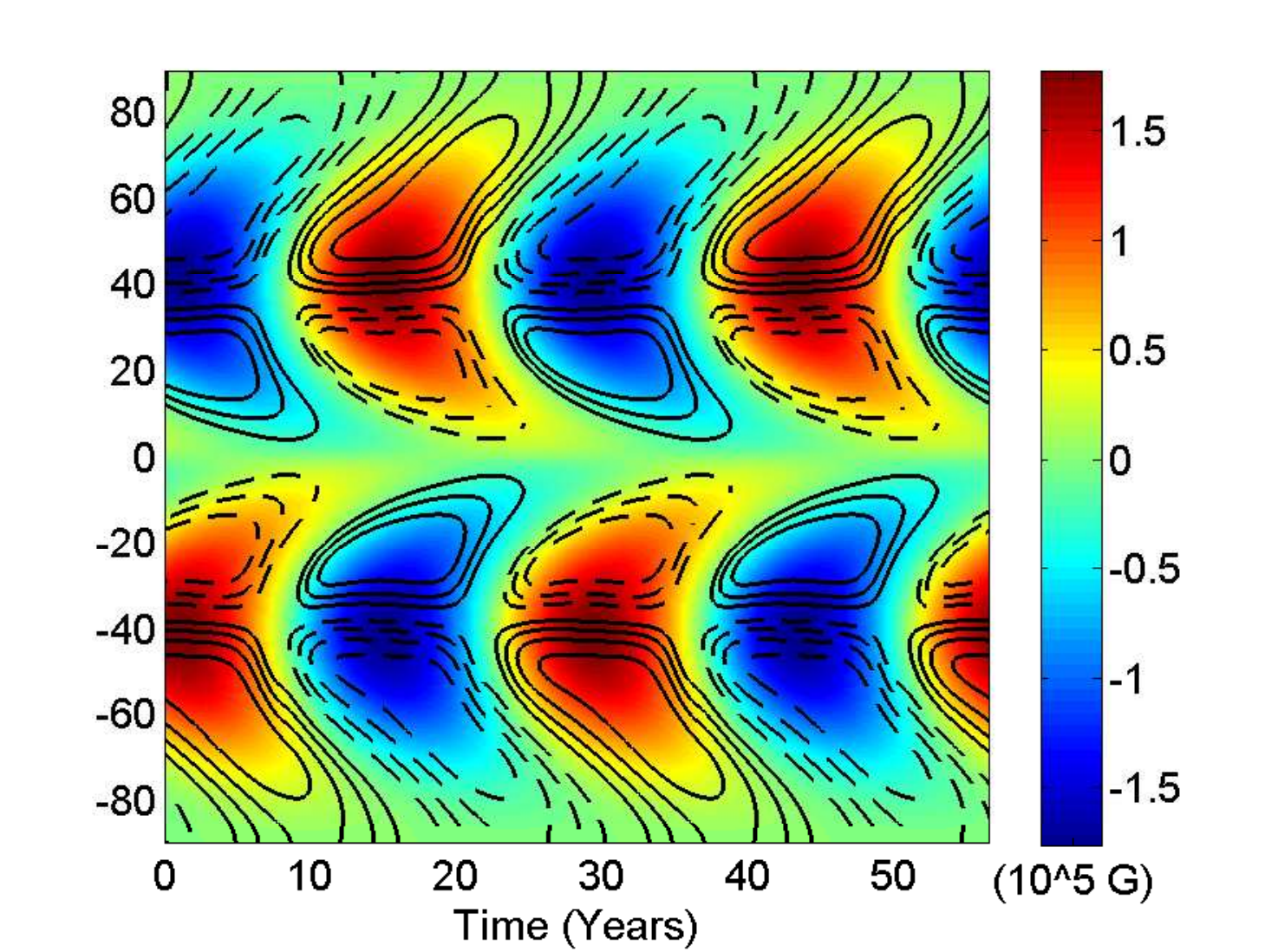}\\
   & $v_o = 12$m/s& \\
   \includegraphics[scale=0.3]{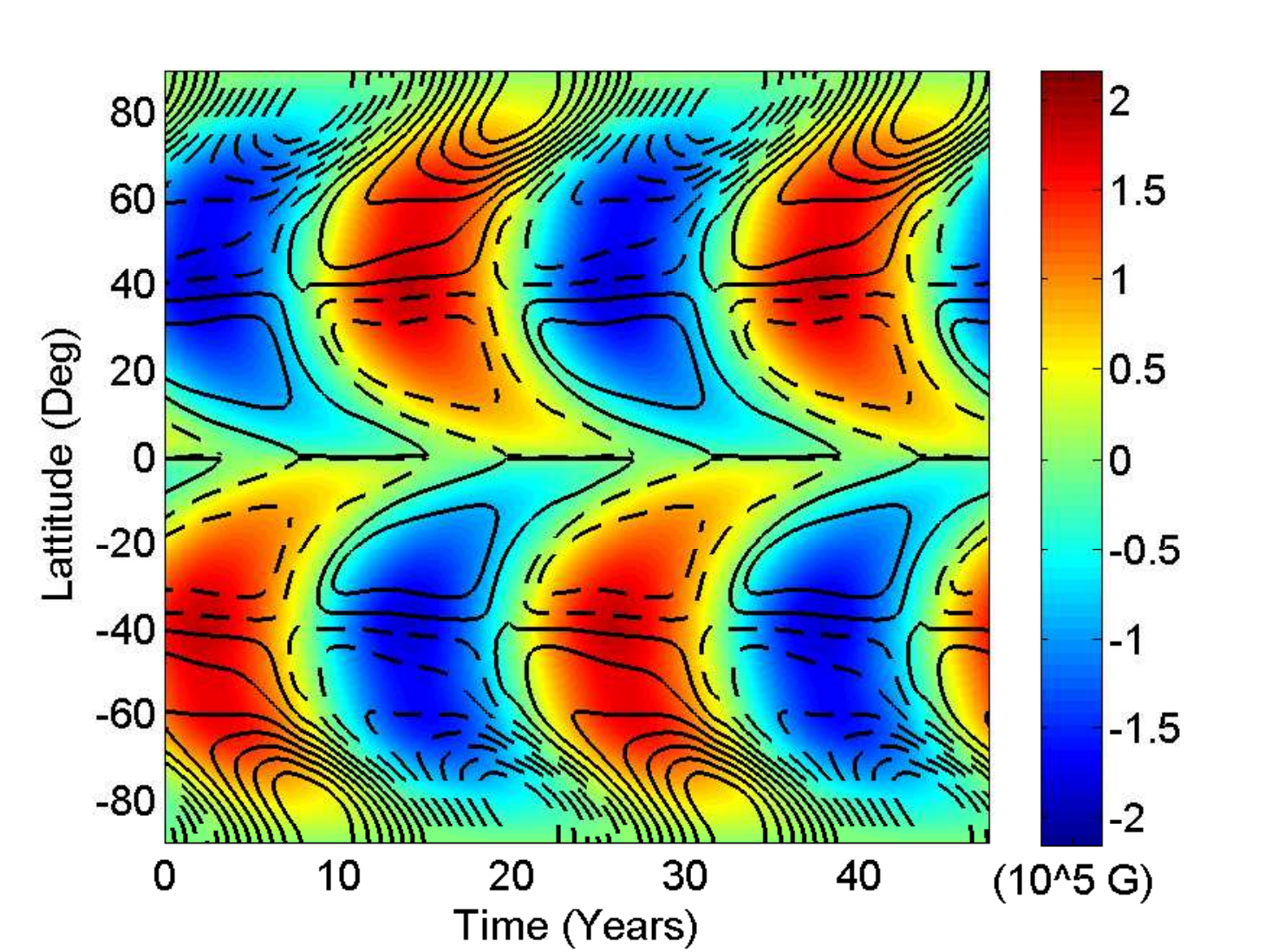} &$R_p = 0.71R_\odot$ & \includegraphics[scale=0.3]{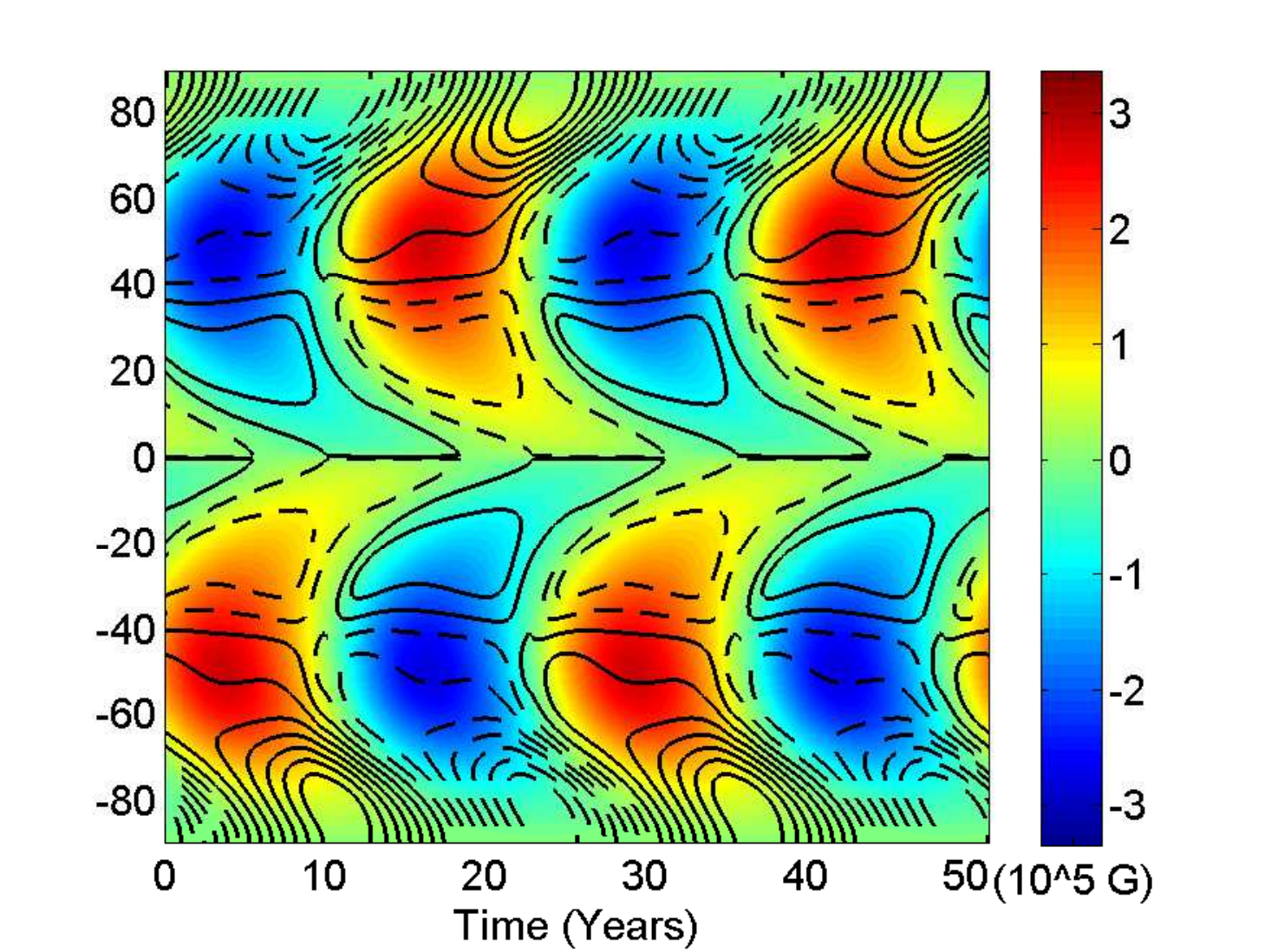}\\
   Composite DR & $v_o = 22$m/s& Analytical DR
  \end{tabular}
\caption{Butterfly diagram of the toroidal field at the bottom of the convection zone (color) with
radial field at the surface (contours) superimposed on it. Each row corresponds to one of the
different meridional circulation sets. The left column corresponds to runs using the
helioseismic composite and the right one to runs using the analytical profile.}\label{Comp_An}
\end{figure}

\begin{figure}[c][h]
  \begin{tabular}{cc}
  \includegraphics[scale=0.5]{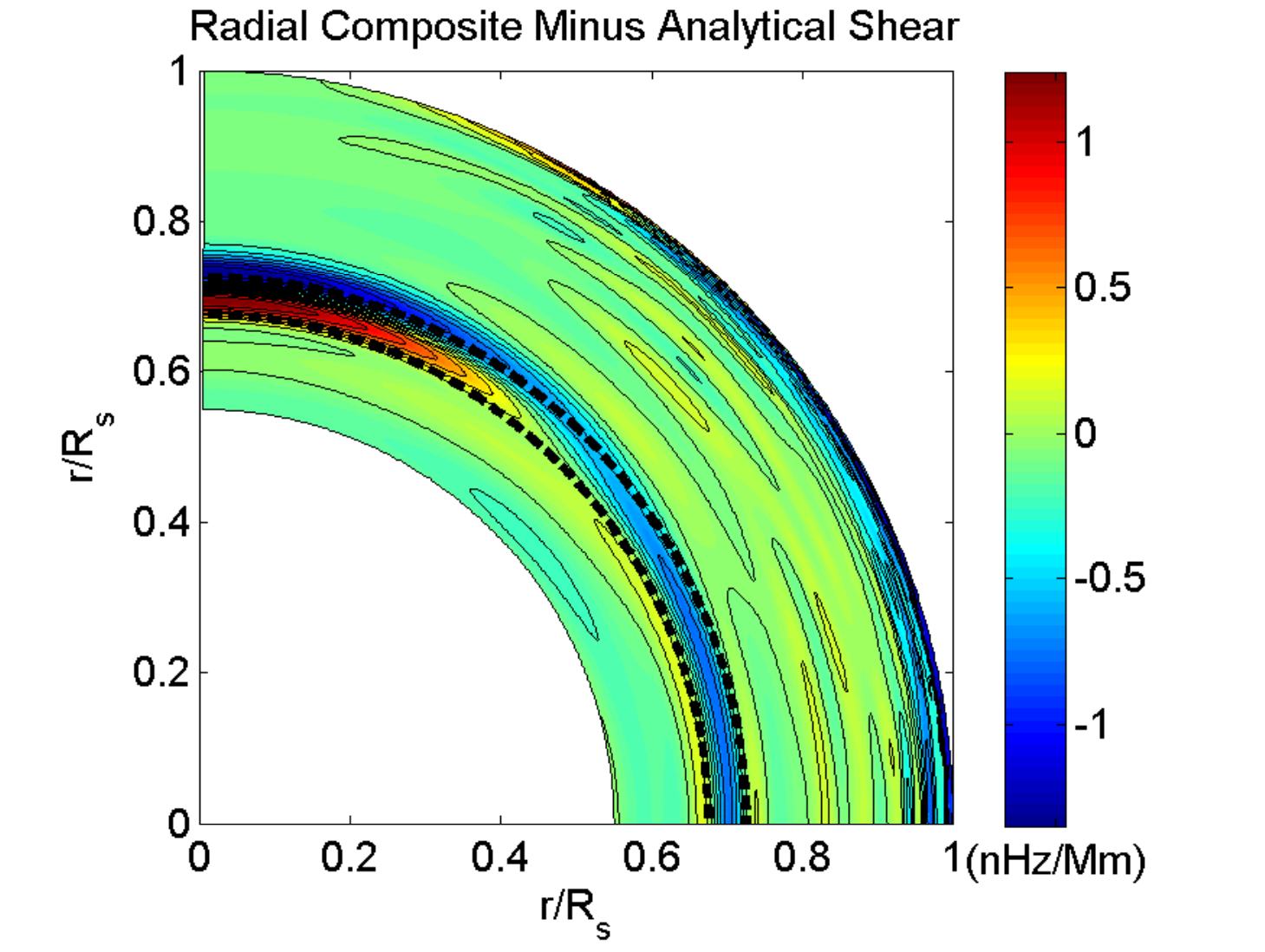} & \includegraphics[scale=0.5]{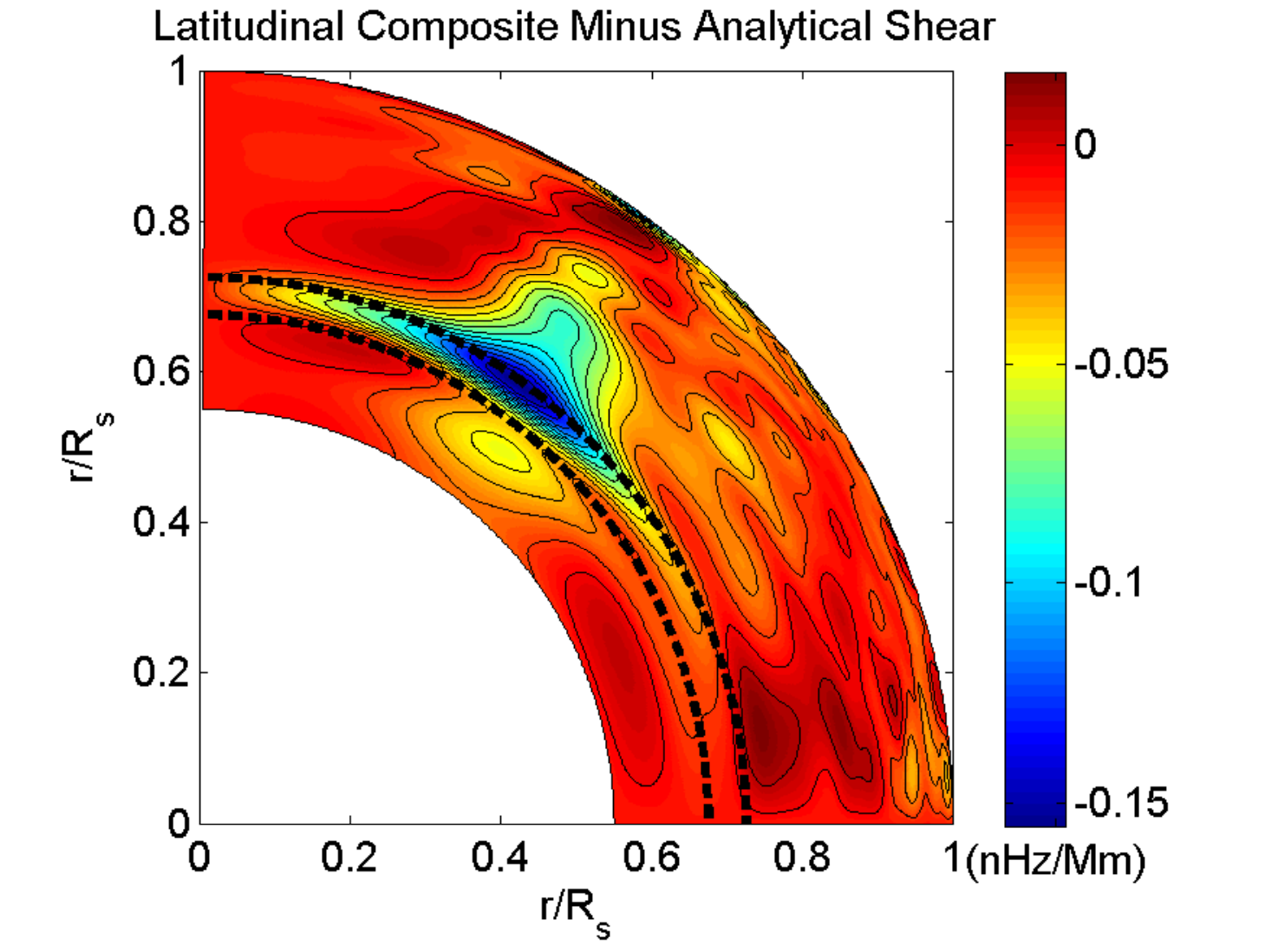}\\
    $(a)$ & $(b)$
  \end{tabular}
\caption{(a) Residual after subtracting the radial shear of the analytical profile commonly used by
the community from the radial shear of our composite data. (b) Residual of subtracting the latitudinal shear of
the analytical profile commonly used by the community from the latitudinal shear of our composite
data.}\label{DRDH_Res}
\end{figure}

\begin{figure}[c]
  \begin{tabular}{cccc}
  $(\textbf{B}_r\cdot\nabla_r\Omega)$            & $(\textbf{B}_\theta\cdot\nabla_\theta\Omega)$       & Toroidal Field                         & Poloidal Field \\
  \includegraphics[scale=0.25]{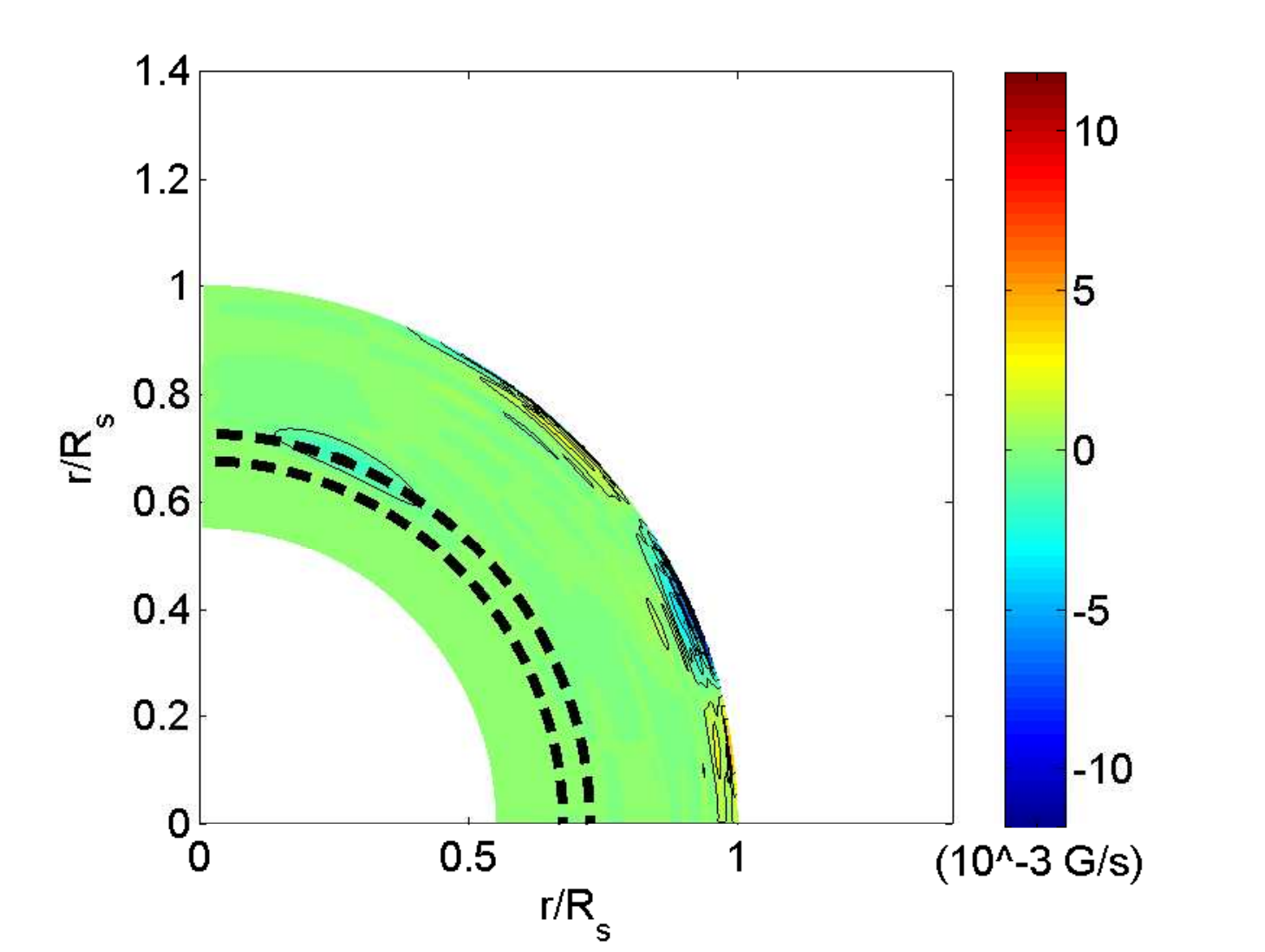} & \includegraphics[scale=0.25]{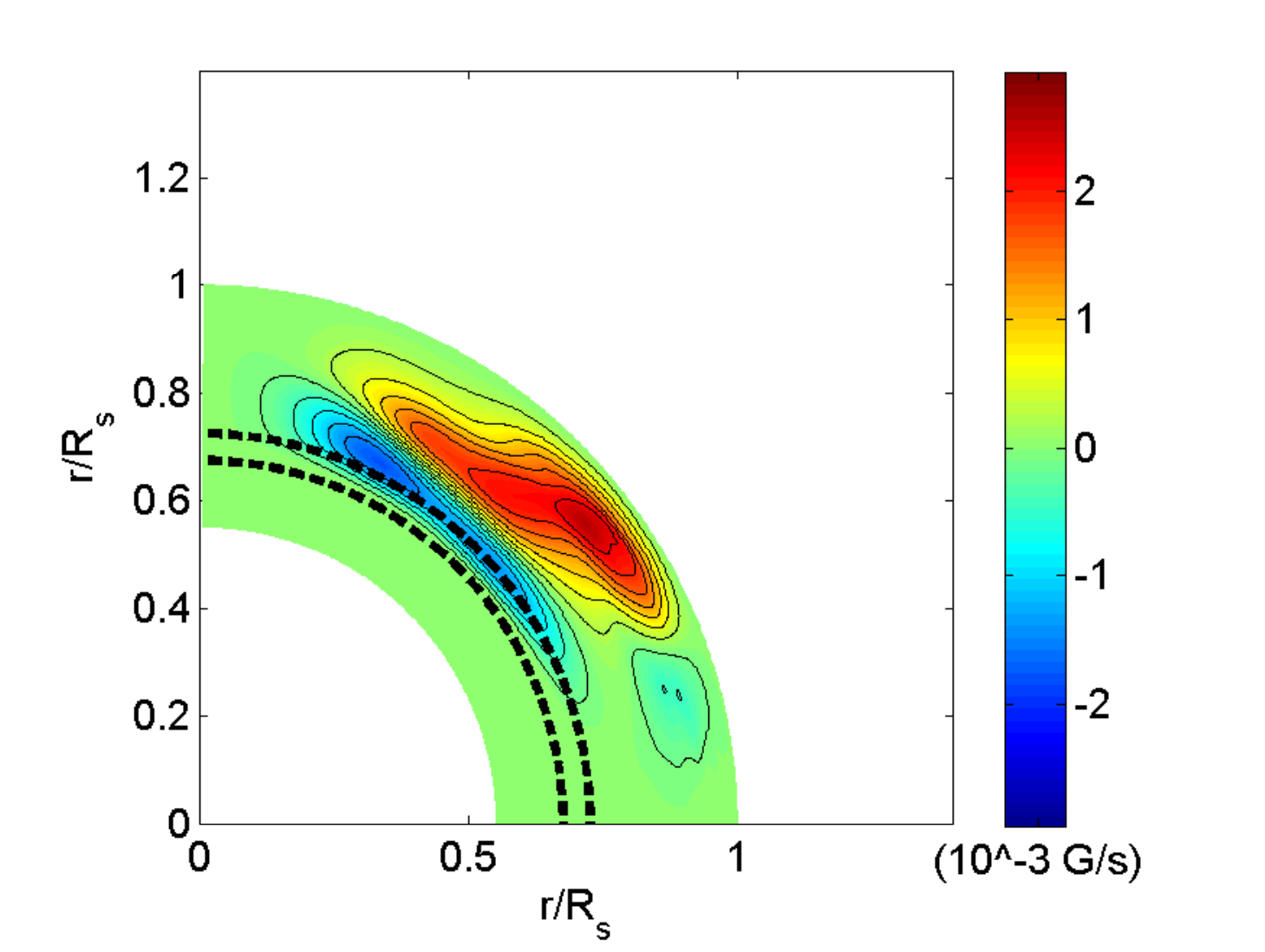} & \includegraphics[scale=0.25]{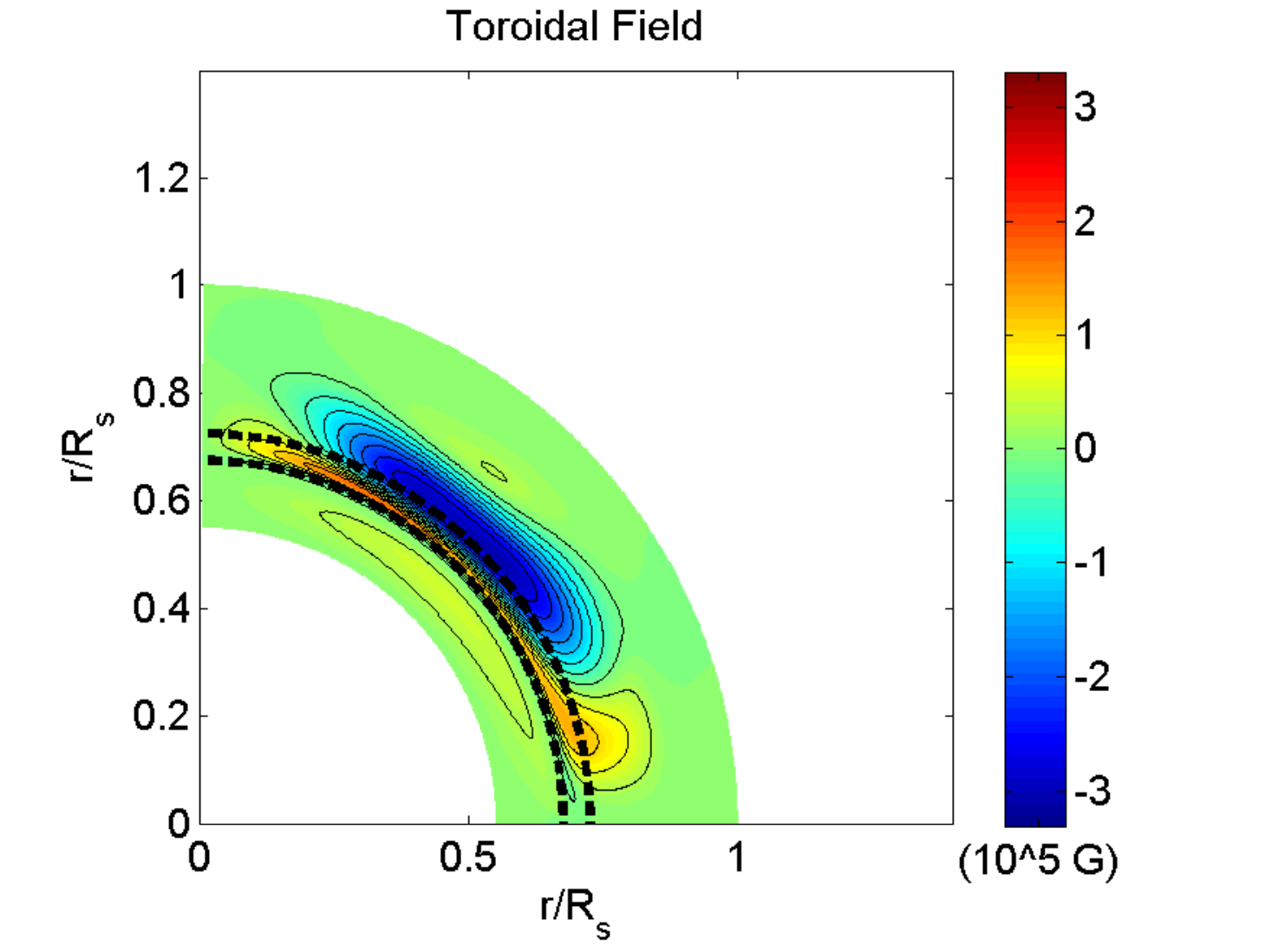} & \includegraphics[scale=0.25]{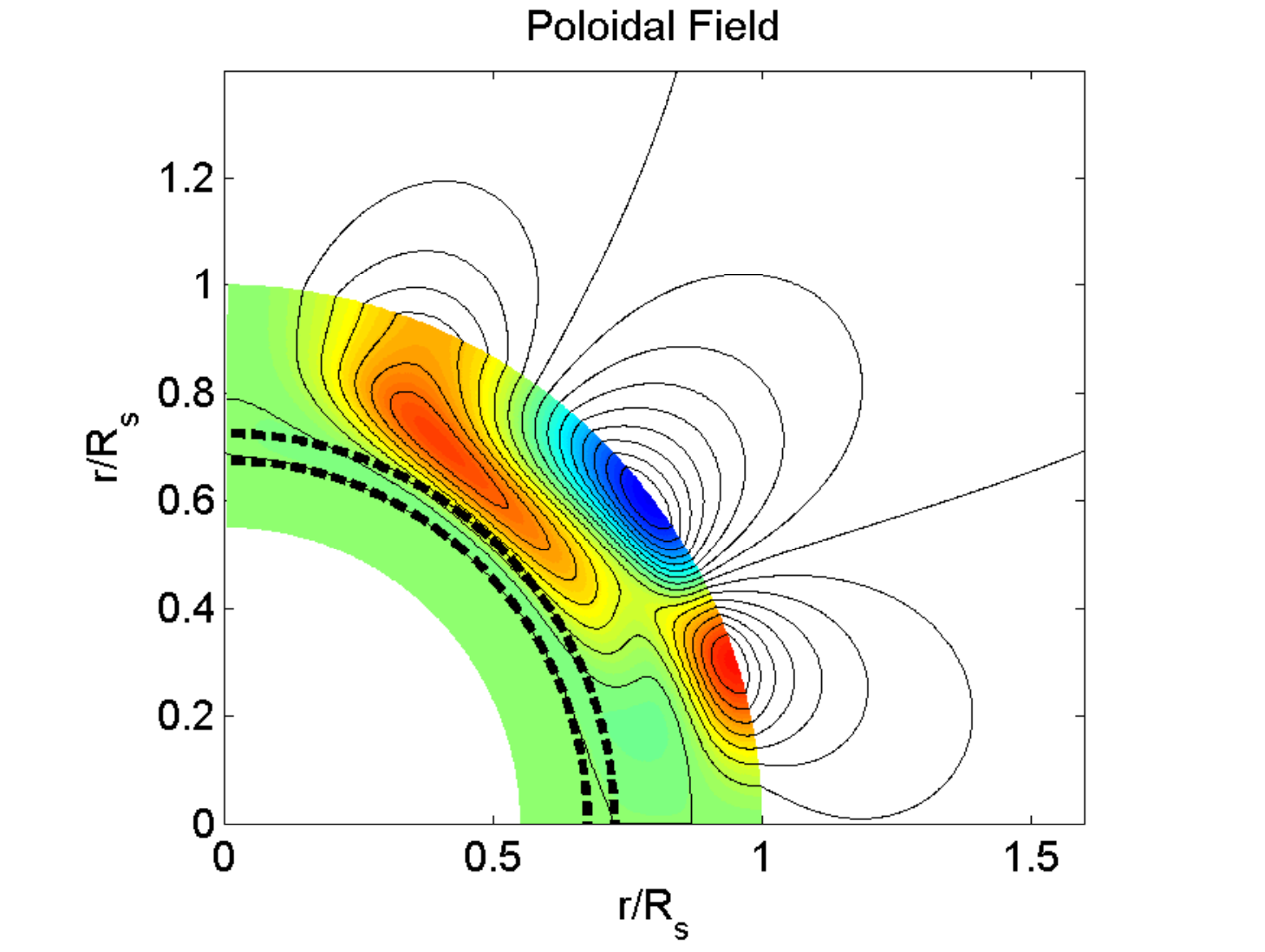}\\
  \includegraphics[scale=0.25]{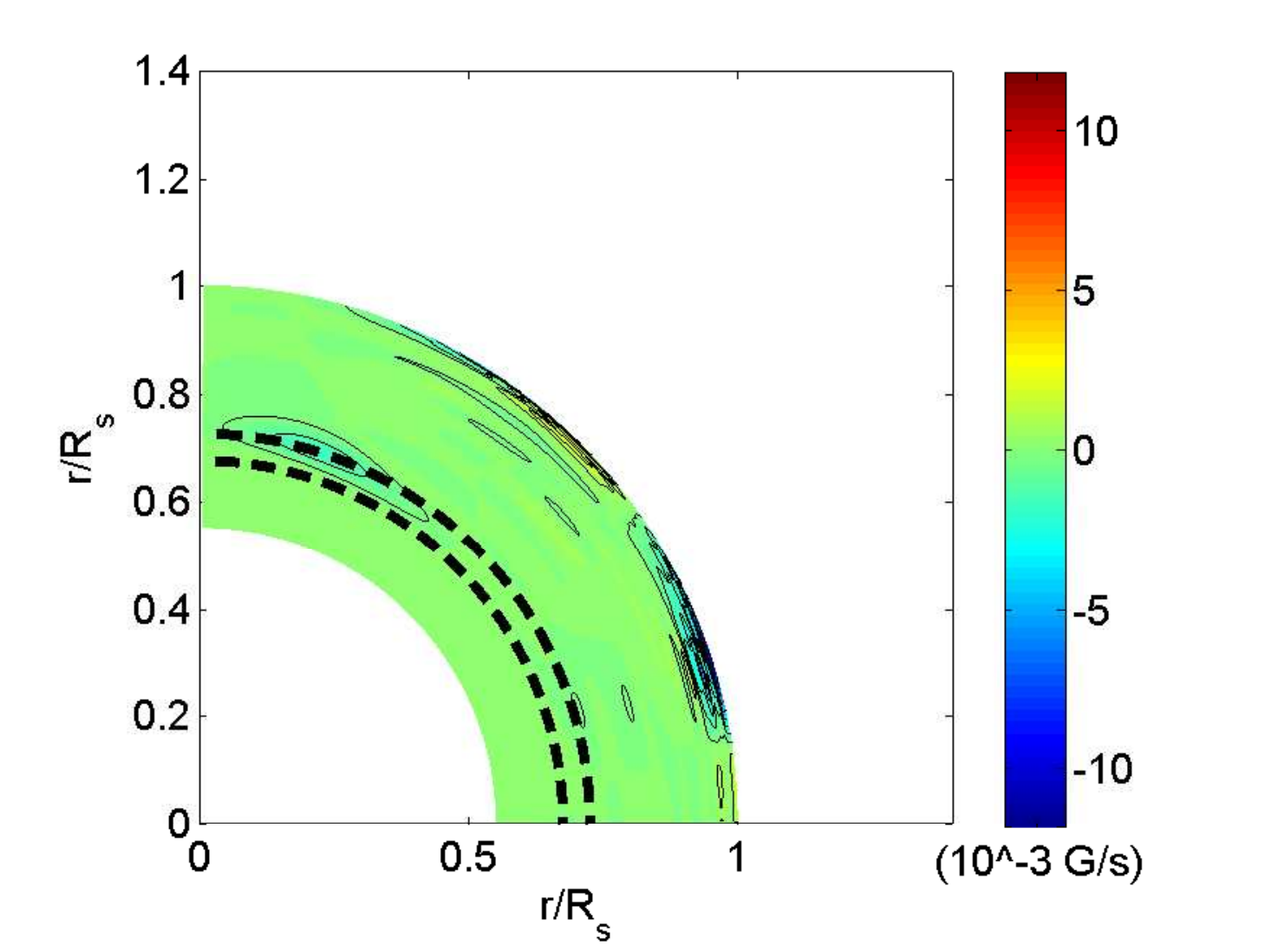} & \includegraphics[scale=0.25]{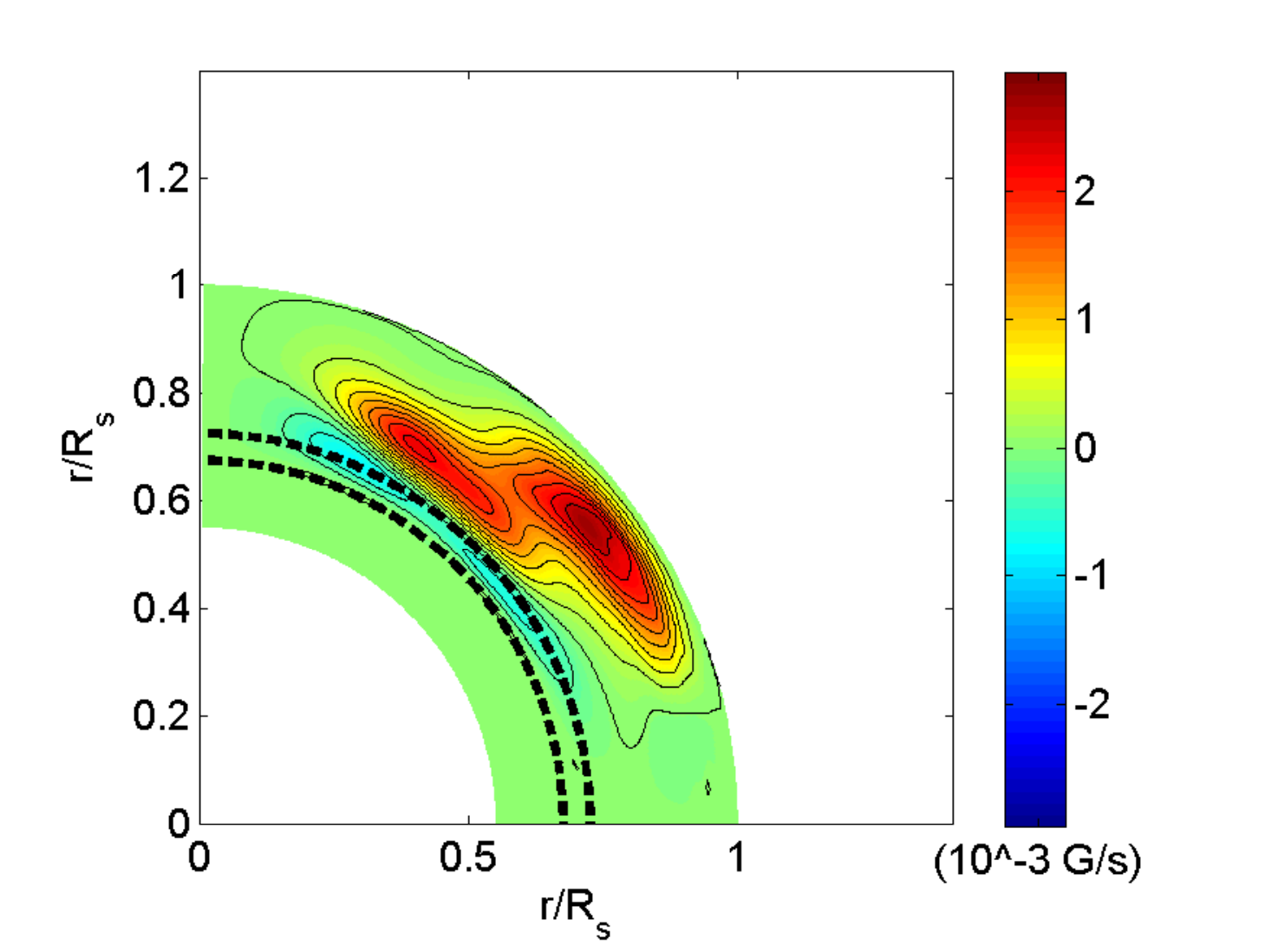} & \includegraphics[scale=0.25]{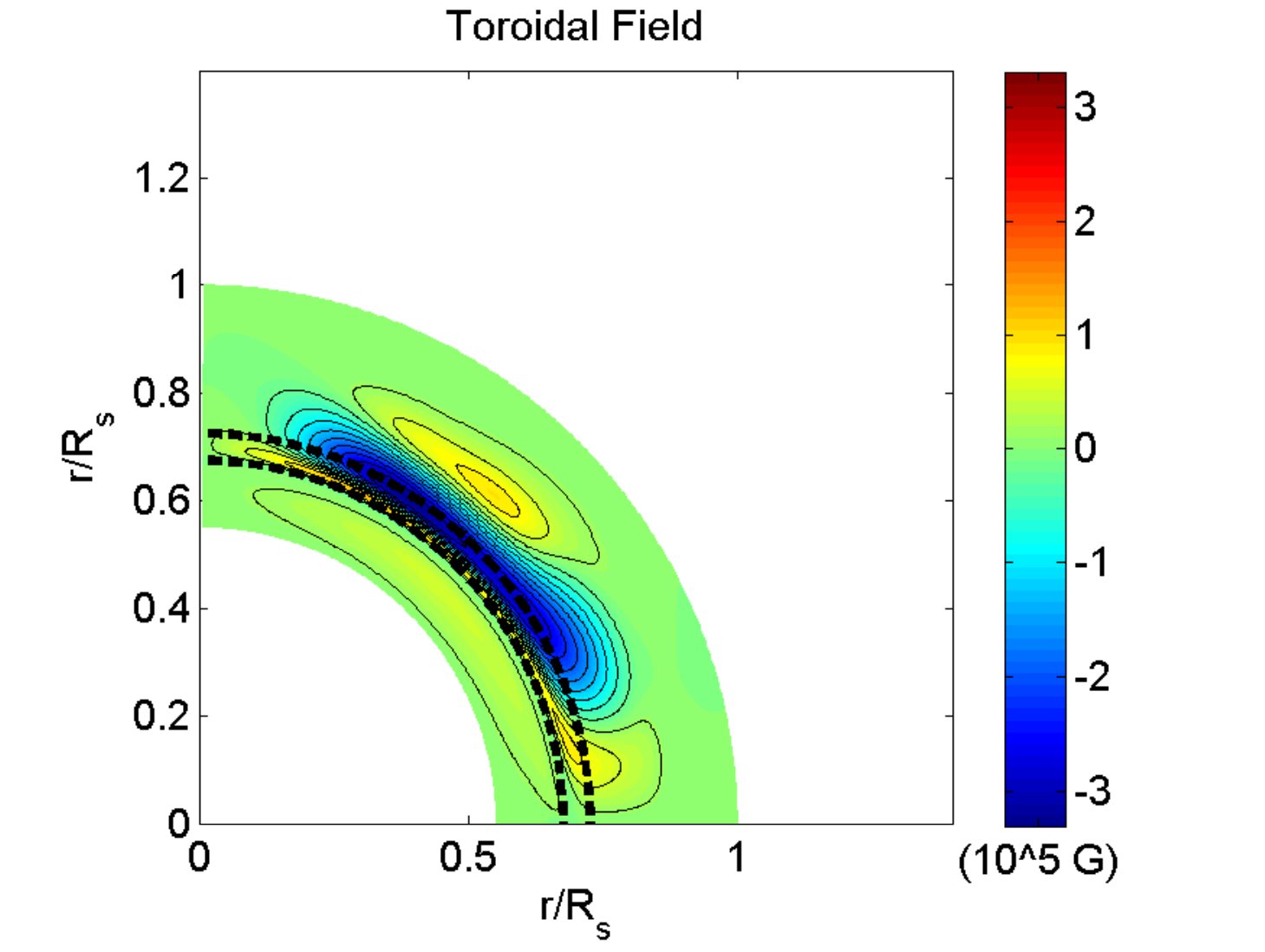} & \includegraphics[scale=0.25]{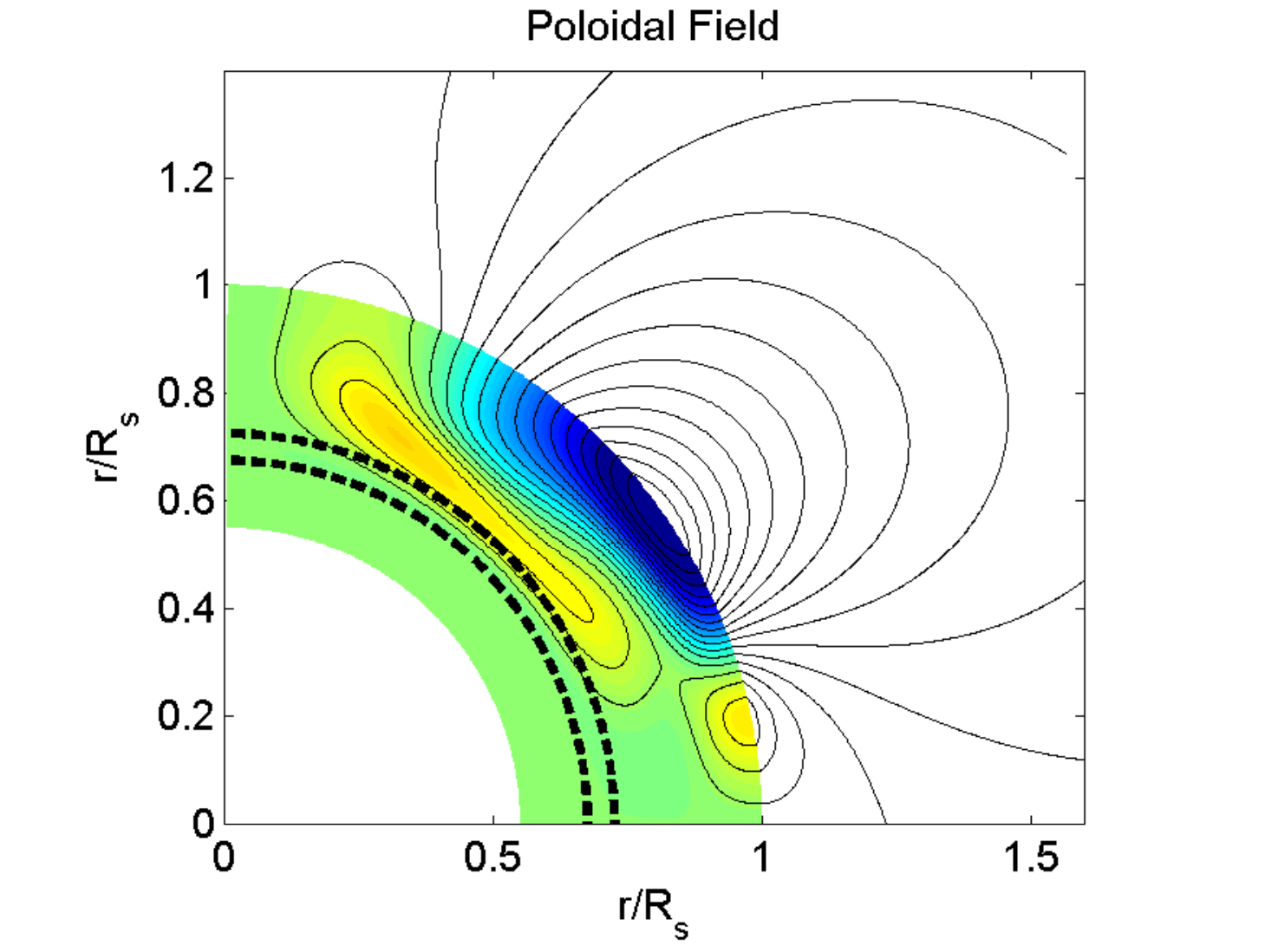}\\
  \includegraphics[scale=0.25]{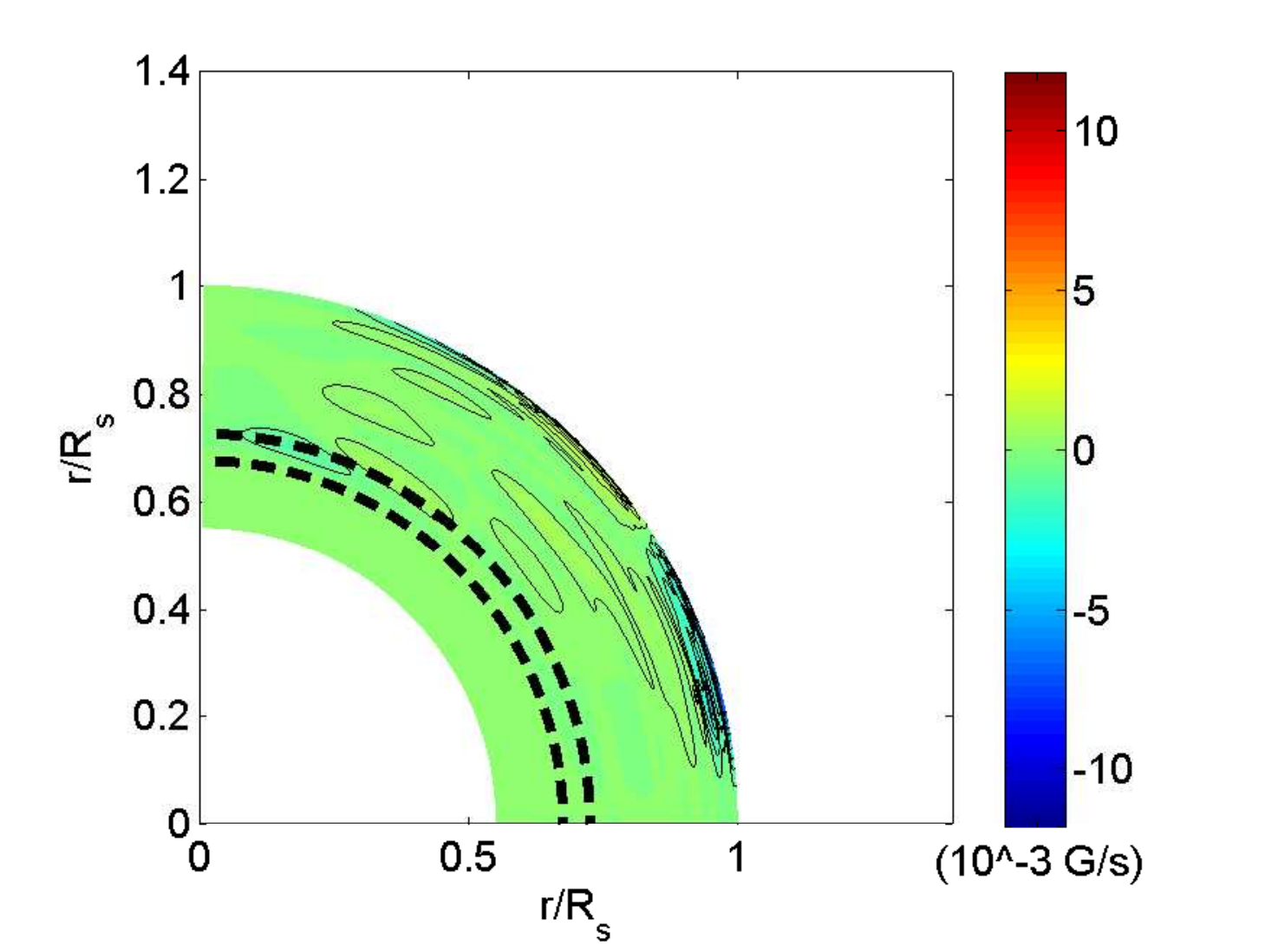} & \includegraphics[scale=0.25]{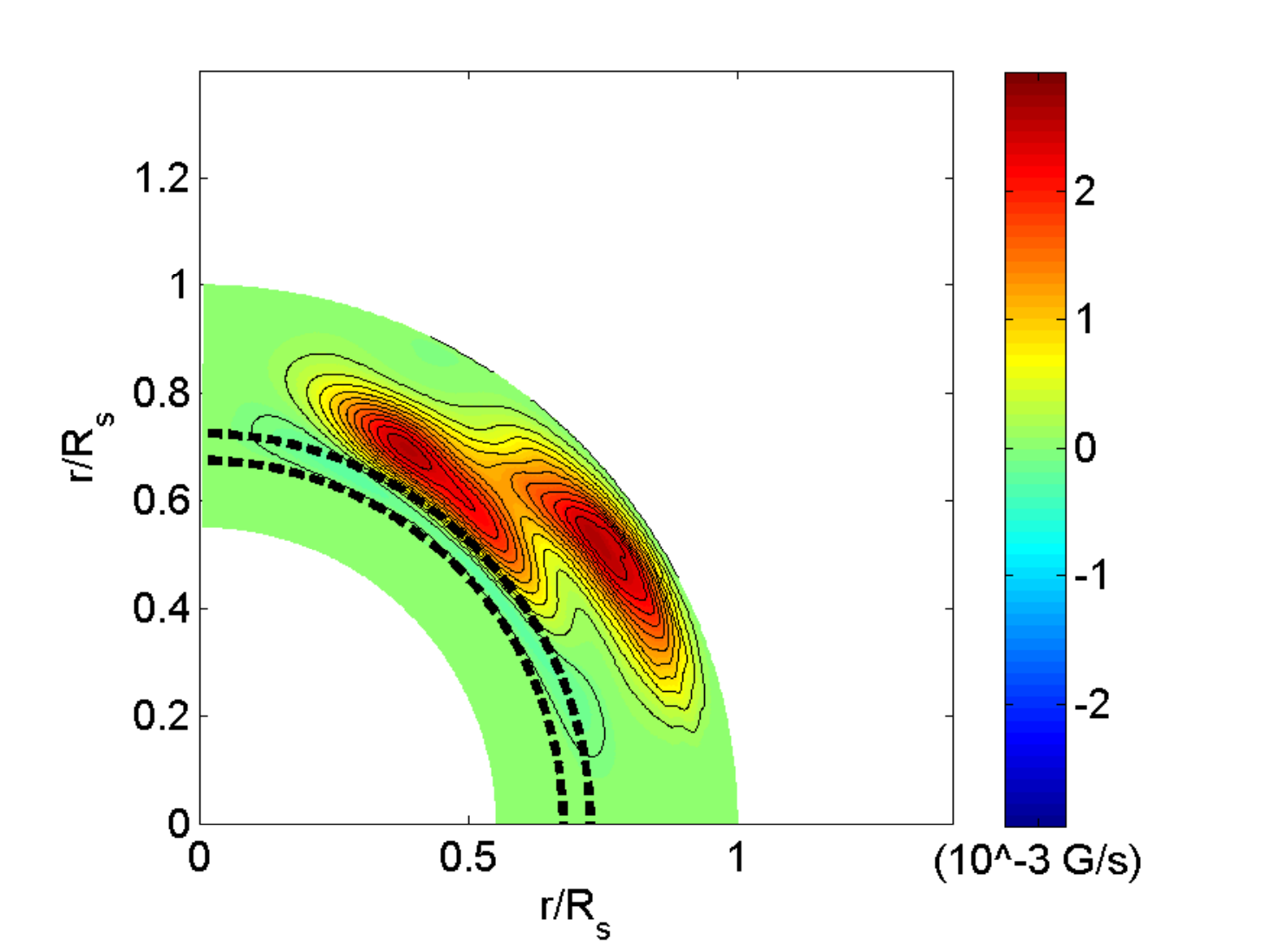} & \includegraphics[scale=0.25]{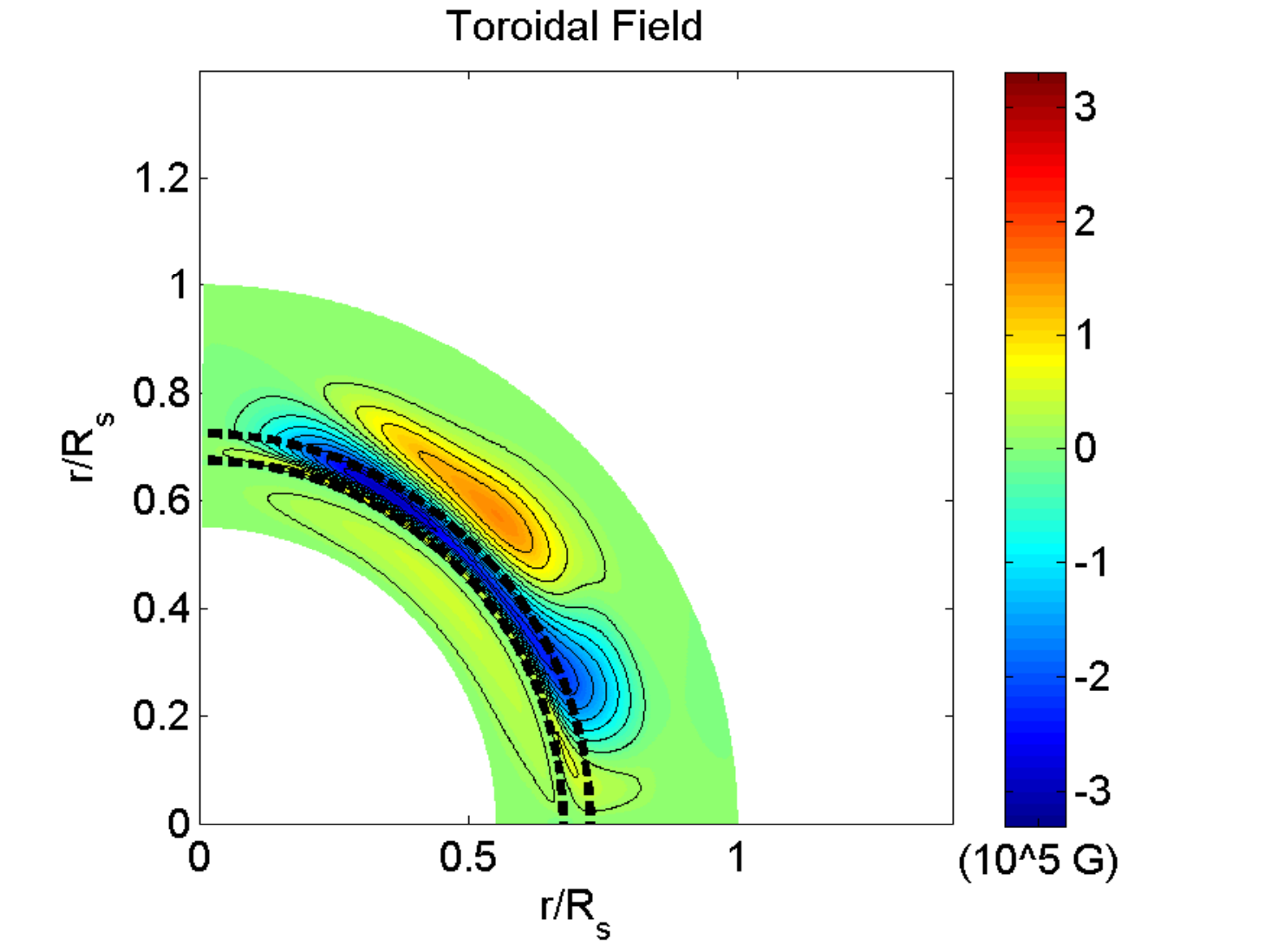} & \includegraphics[scale=0.25]{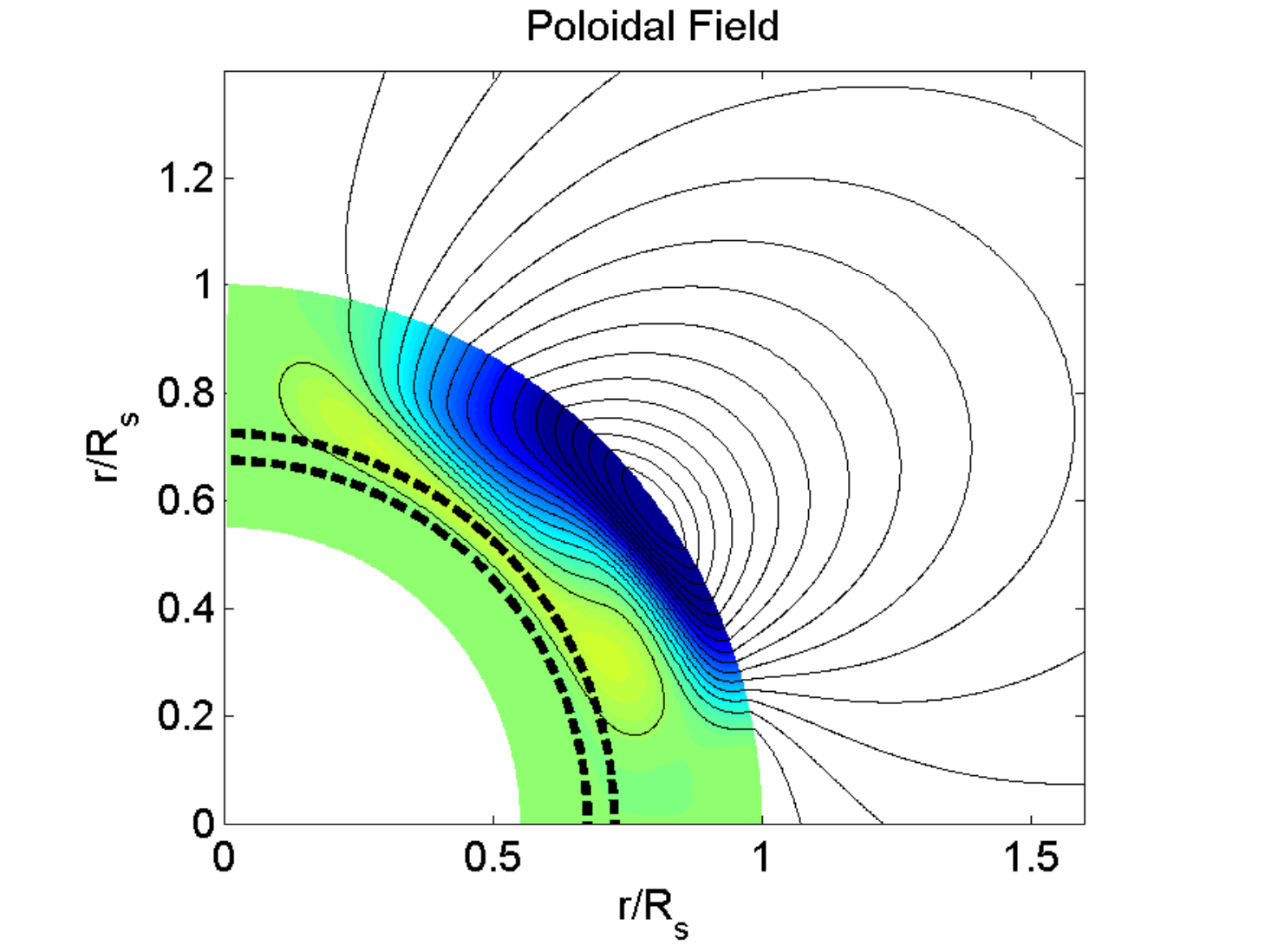}\\
  \includegraphics[scale=0.25]{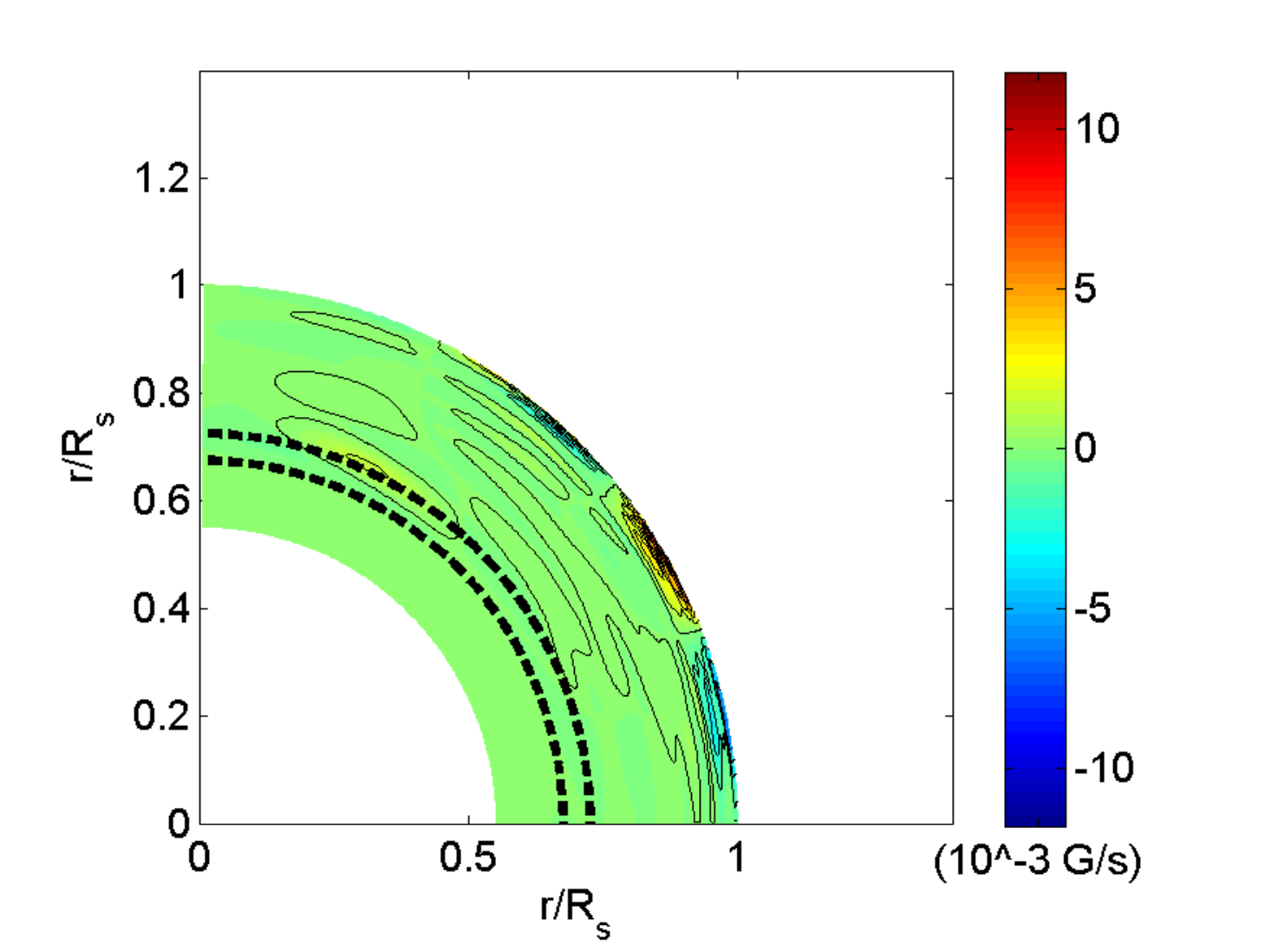} & \includegraphics[scale=0.25]{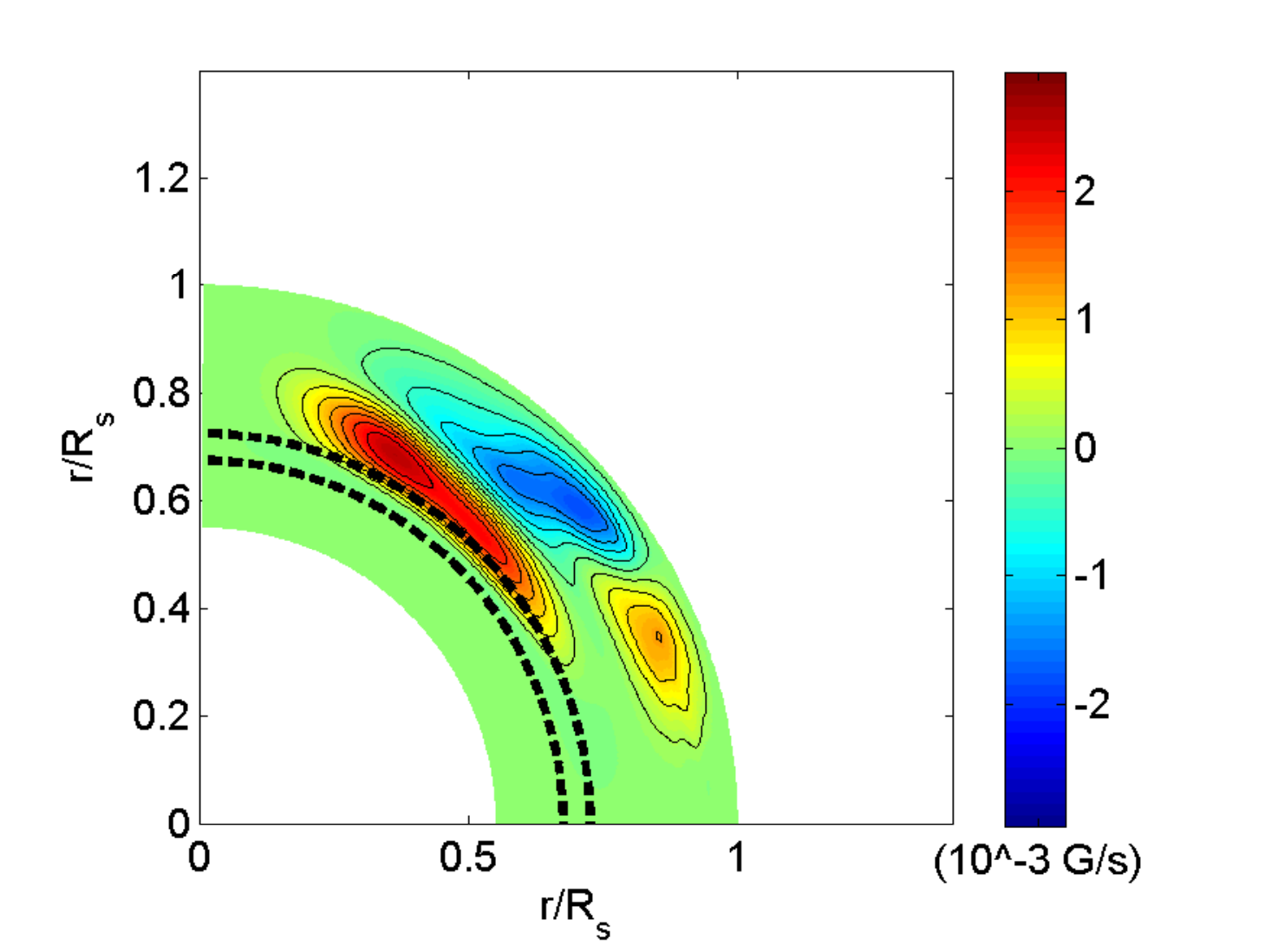} & \includegraphics[scale=0.25]{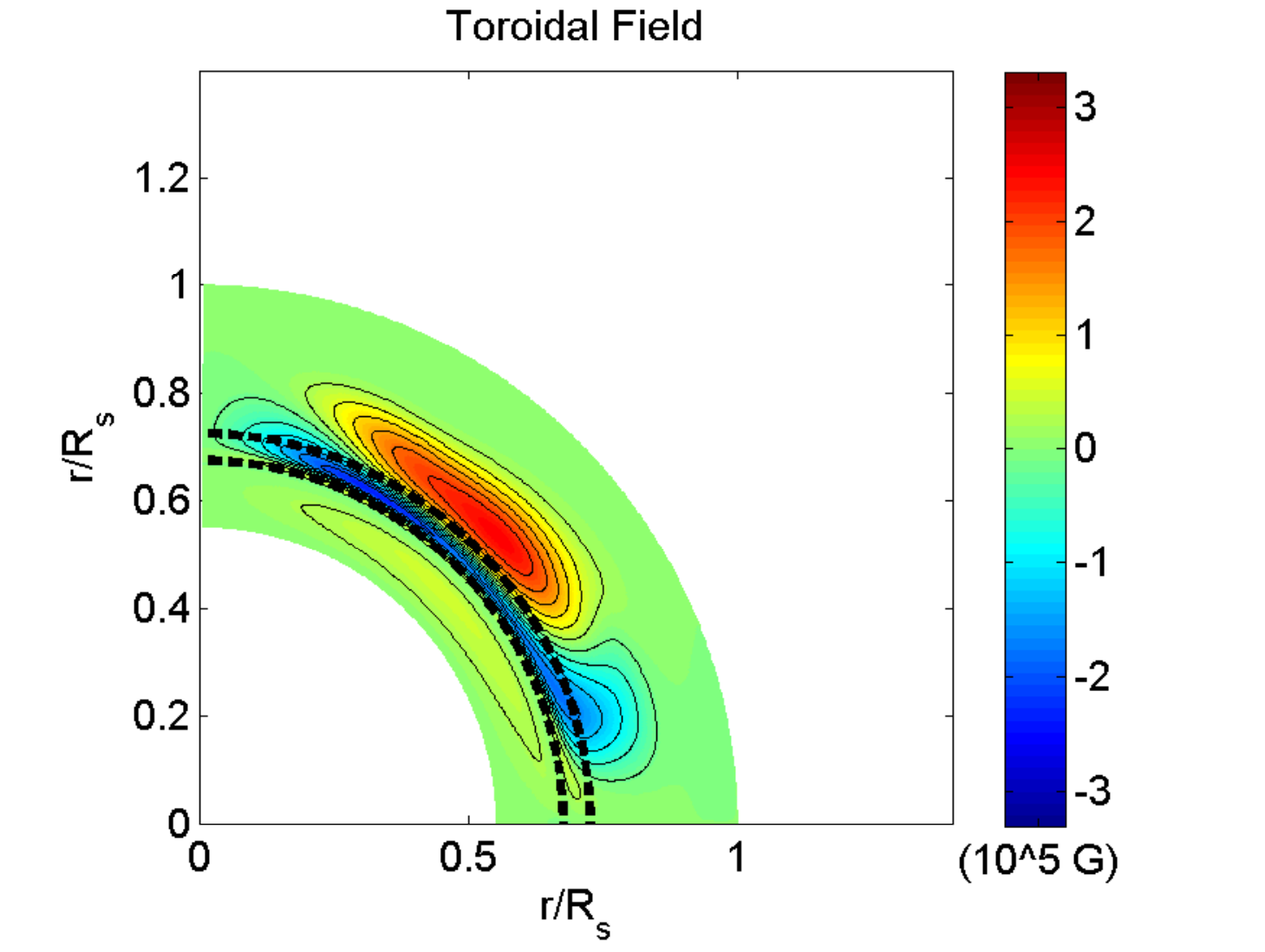} & \includegraphics[scale=0.25]{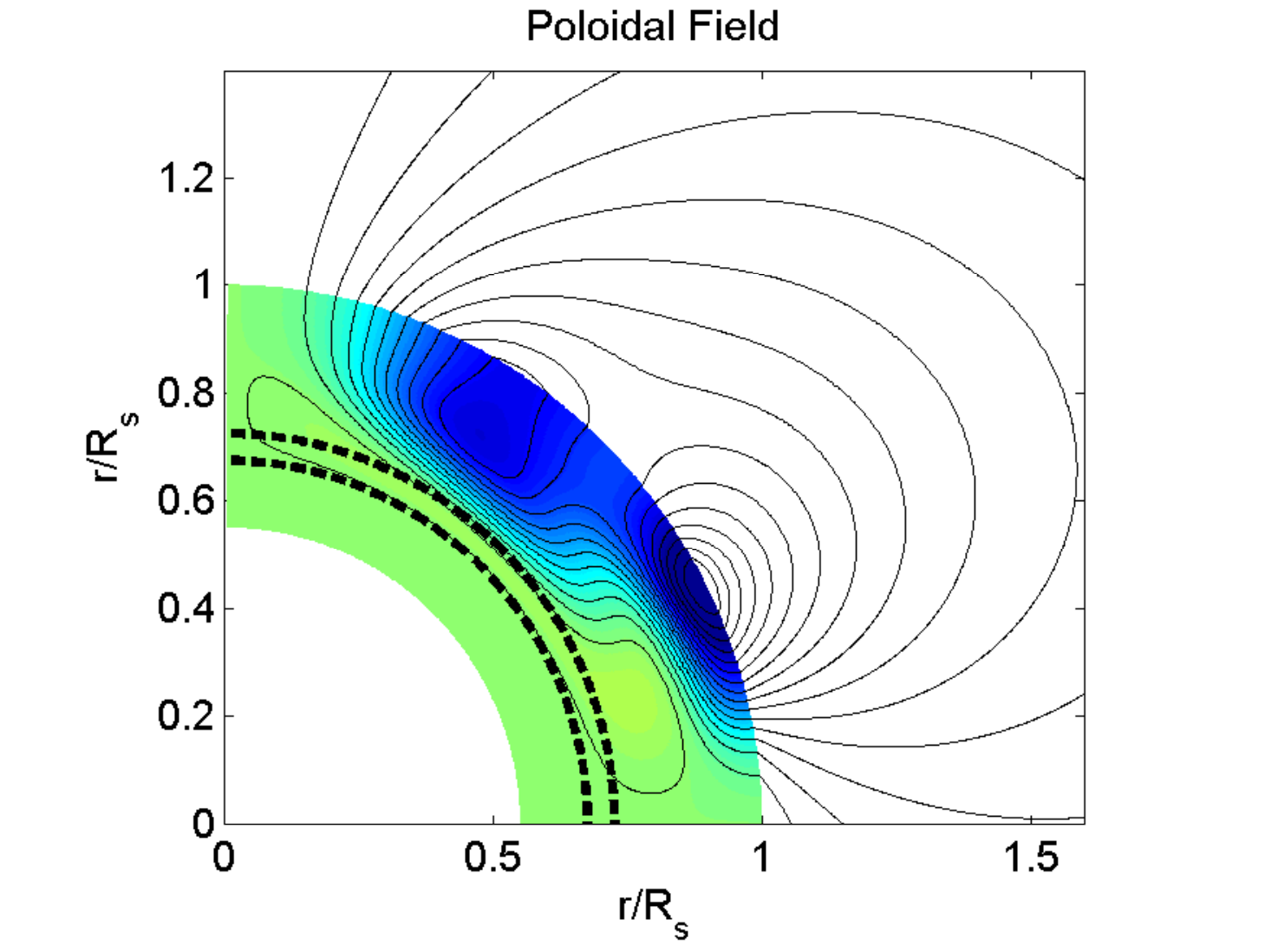}\\
  \end{tabular}
\caption{Snapshots of the shear source terms and the magnetic field over half a dynamo cycle (a
sunspot cycle). Each row is advanced by an eight of the dynamo cycle (a quarter of the sunspot
cycle) i.e., from top to bottom $t = 0, \tau/8, \tau/4$ and $3\tau/8$. The solution corresponds to
the composite differential rotation and meridional flow Set 1 (deepest penetration with a peak flow
of 12 m/s)}\label{S1C}
\end{figure}

\begin{figure}[c]
  \begin{tabular}{cccc}
  $(\textbf{B}_r\cdot\nabla_r\Omega)$            & $(\textbf{B}_\theta\cdot\nabla_\theta\Omega)$       & Toroidal Field                         & Poloidal Field \\
  \includegraphics[scale=0.25]{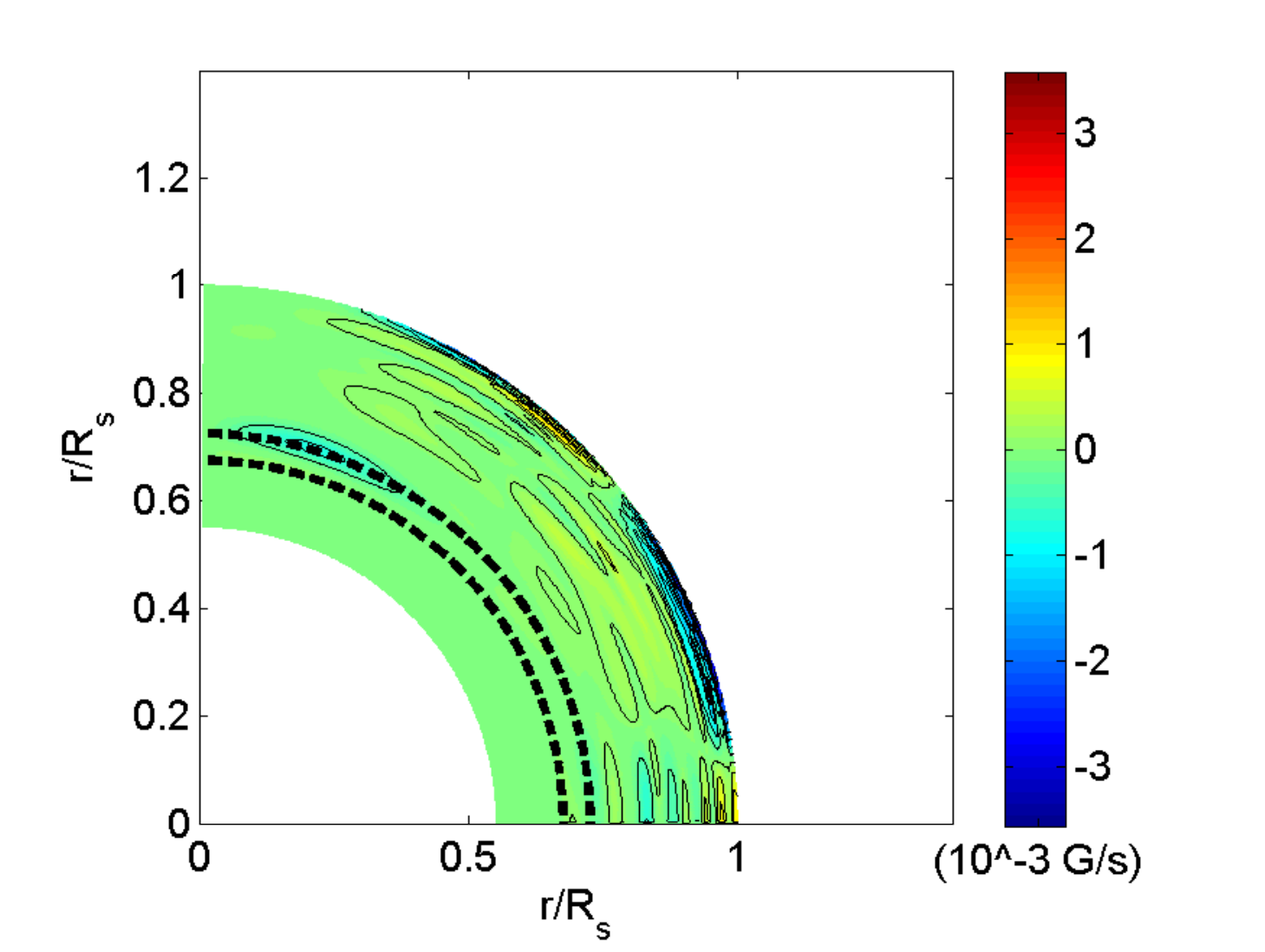} & \includegraphics[scale=0.25]{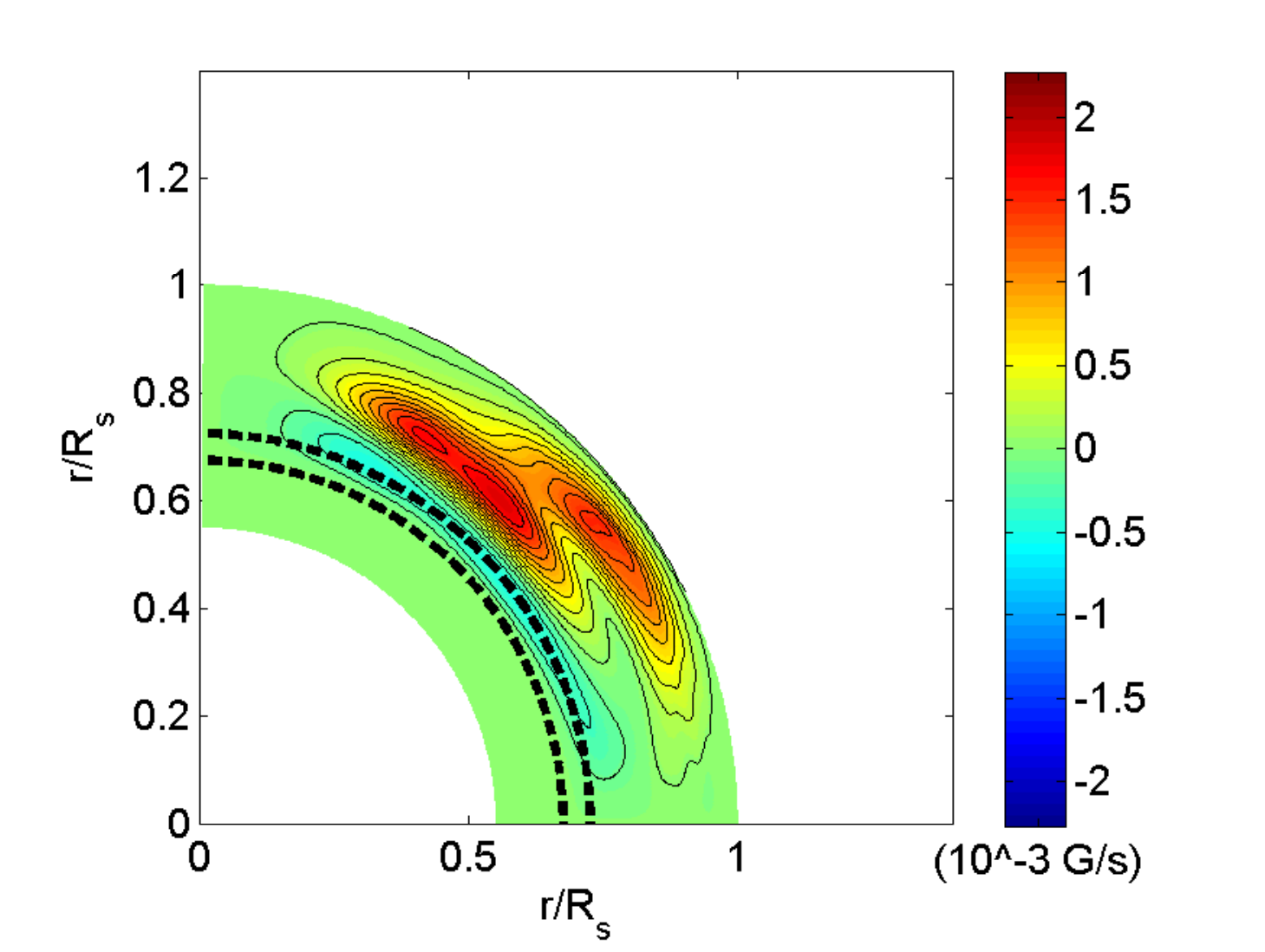} & \includegraphics[scale=0.25]{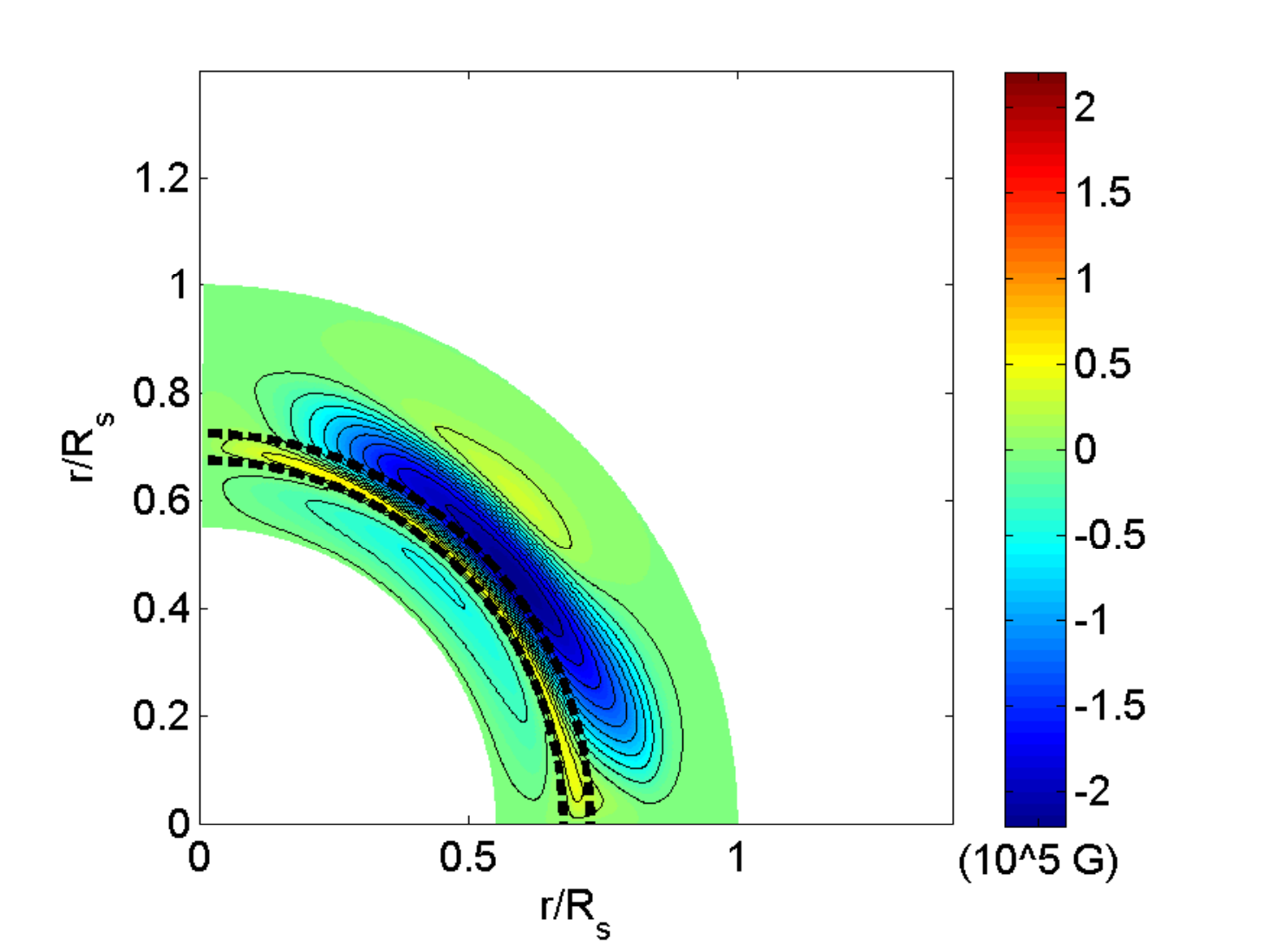} & \includegraphics[scale=0.25]{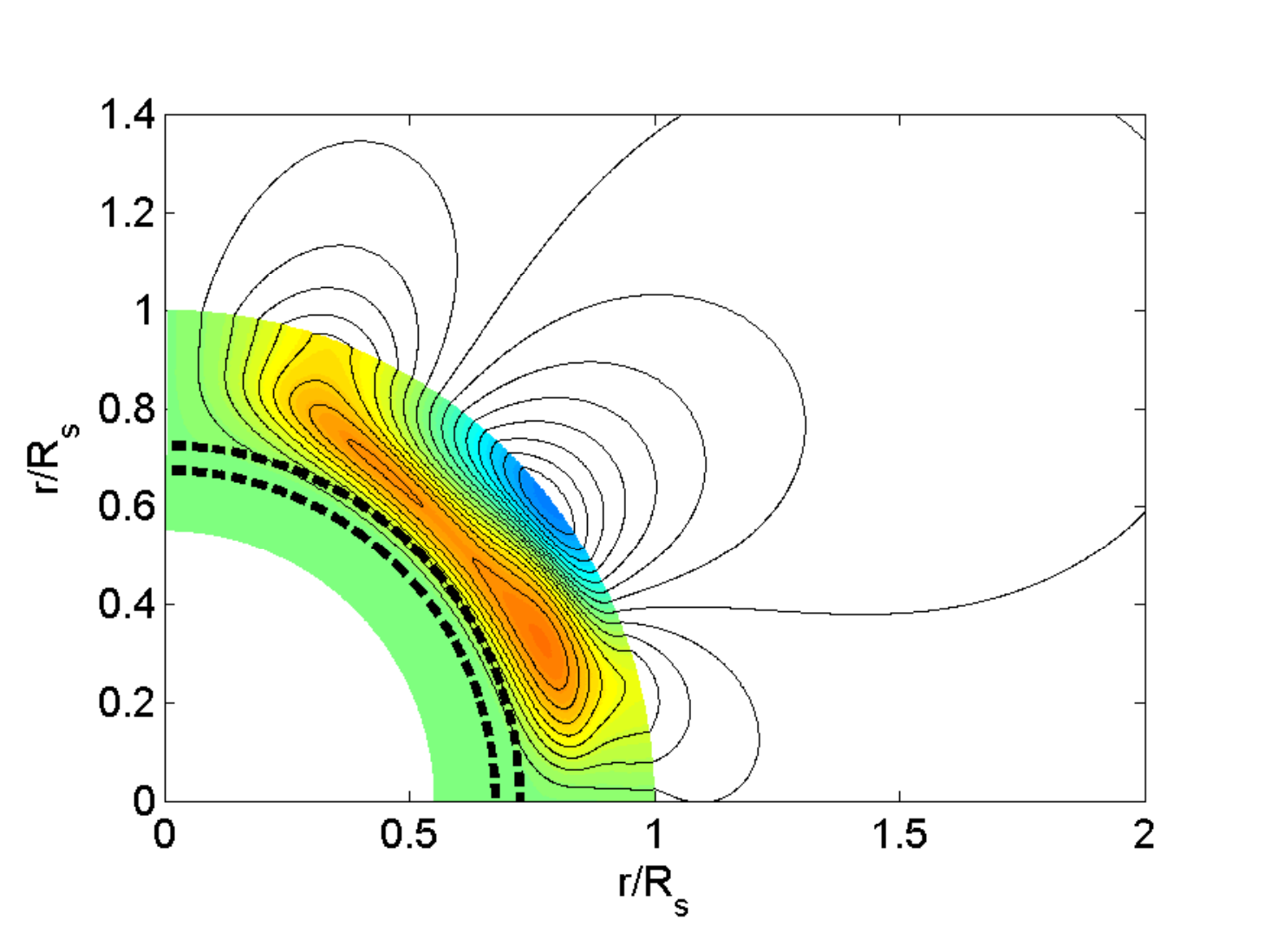}\\
  \includegraphics[scale=0.25]{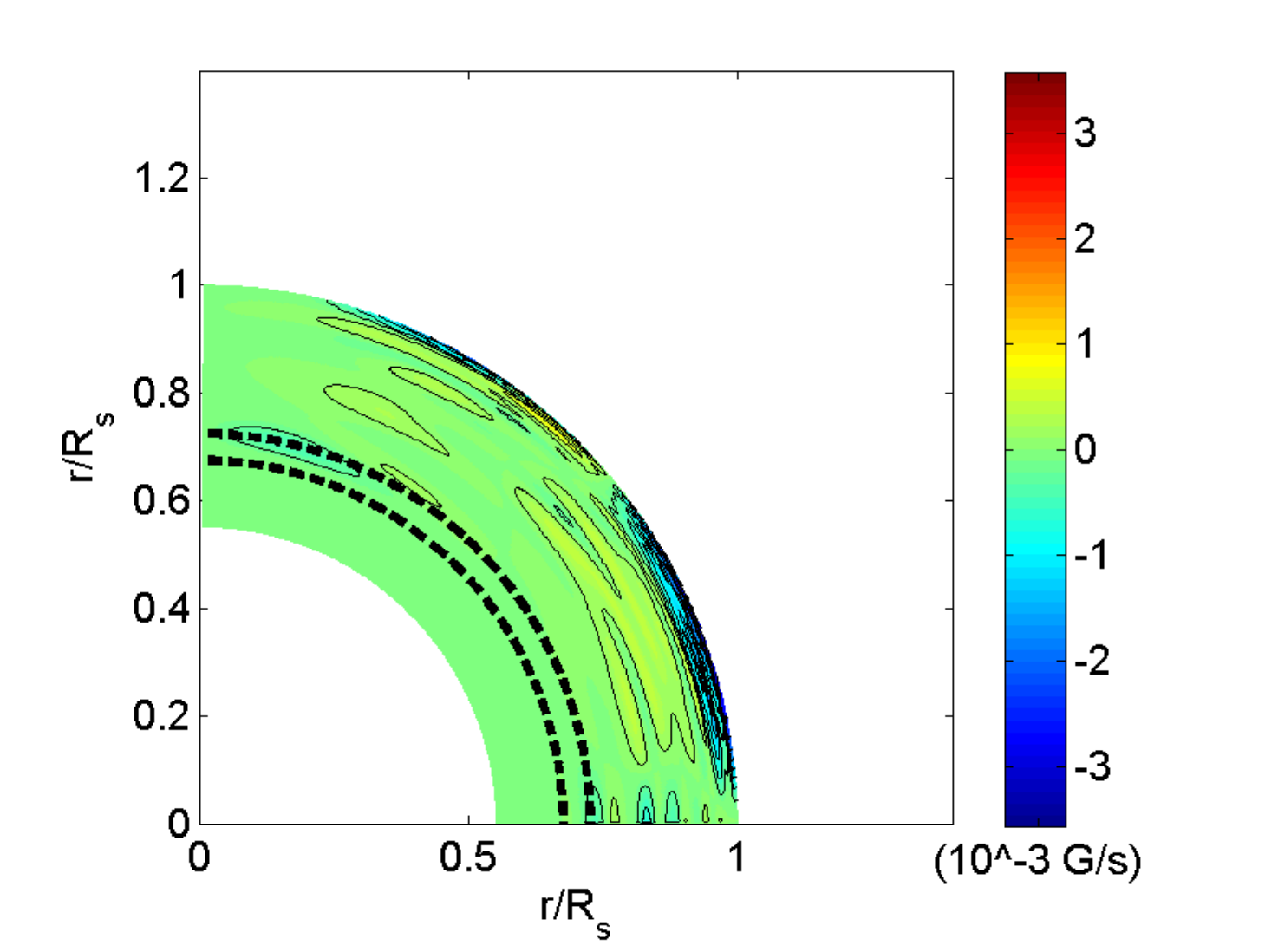} & \includegraphics[scale=0.25]{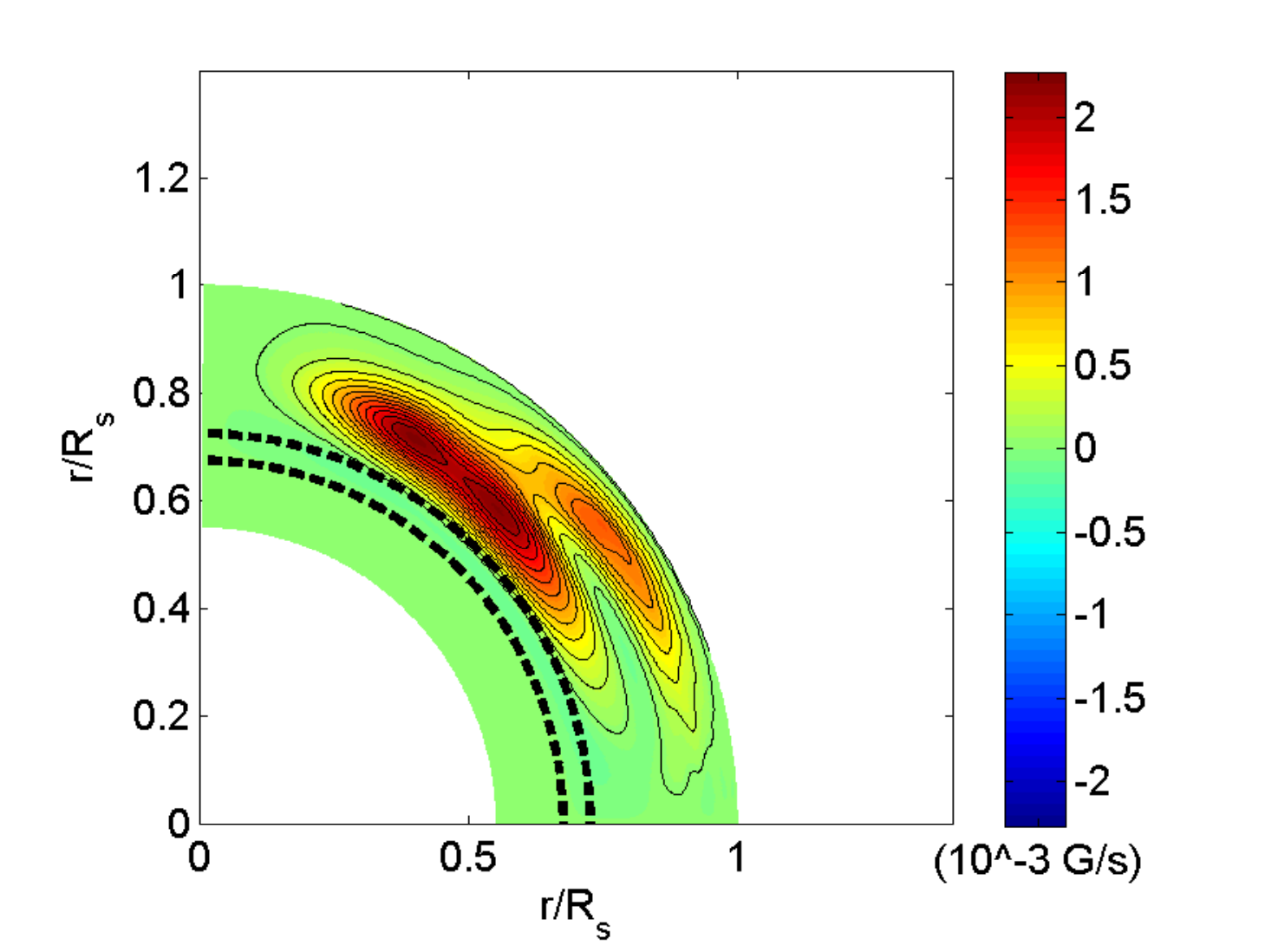} & \includegraphics[scale=0.25]{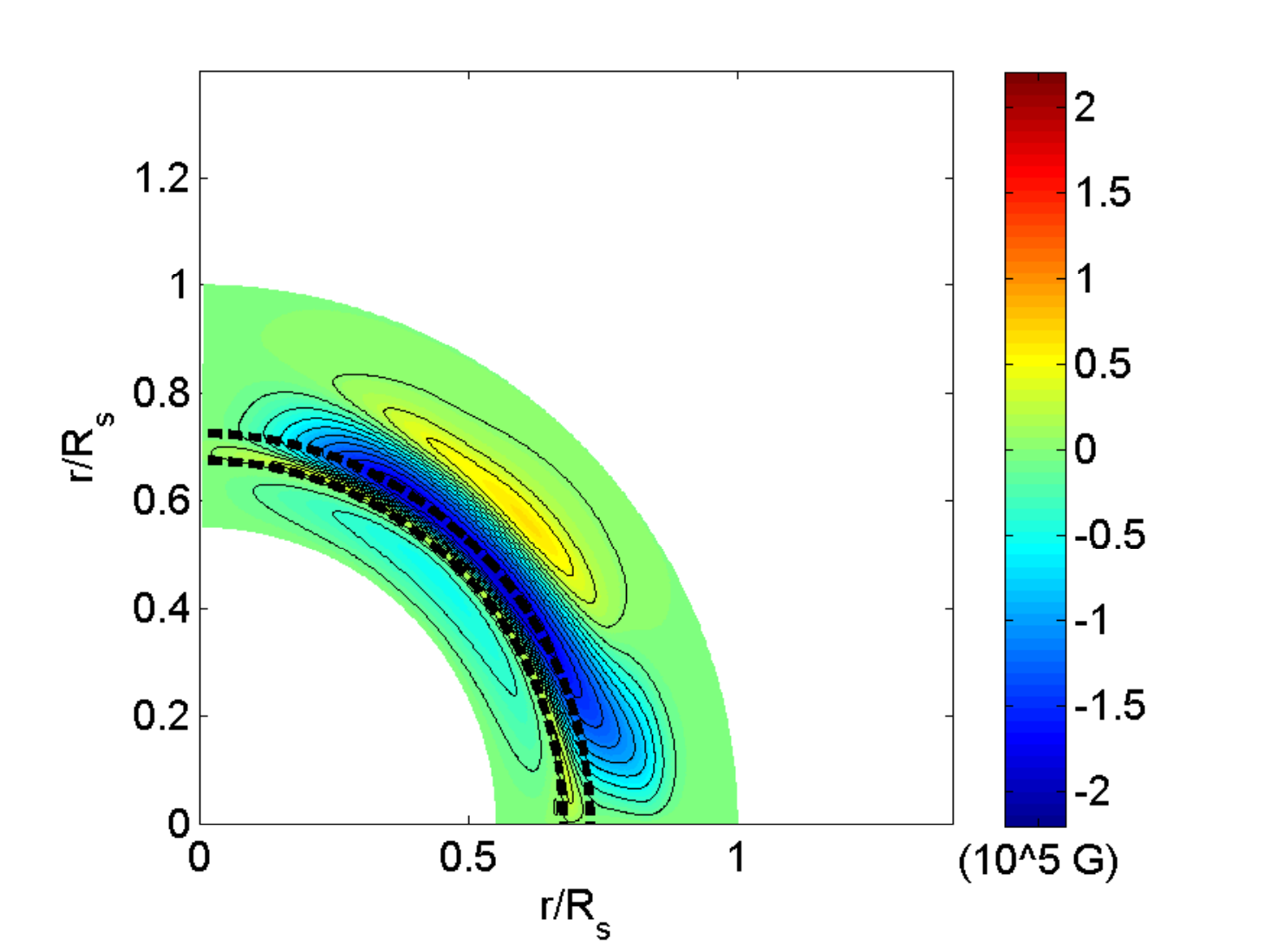} & \includegraphics[scale=0.25]{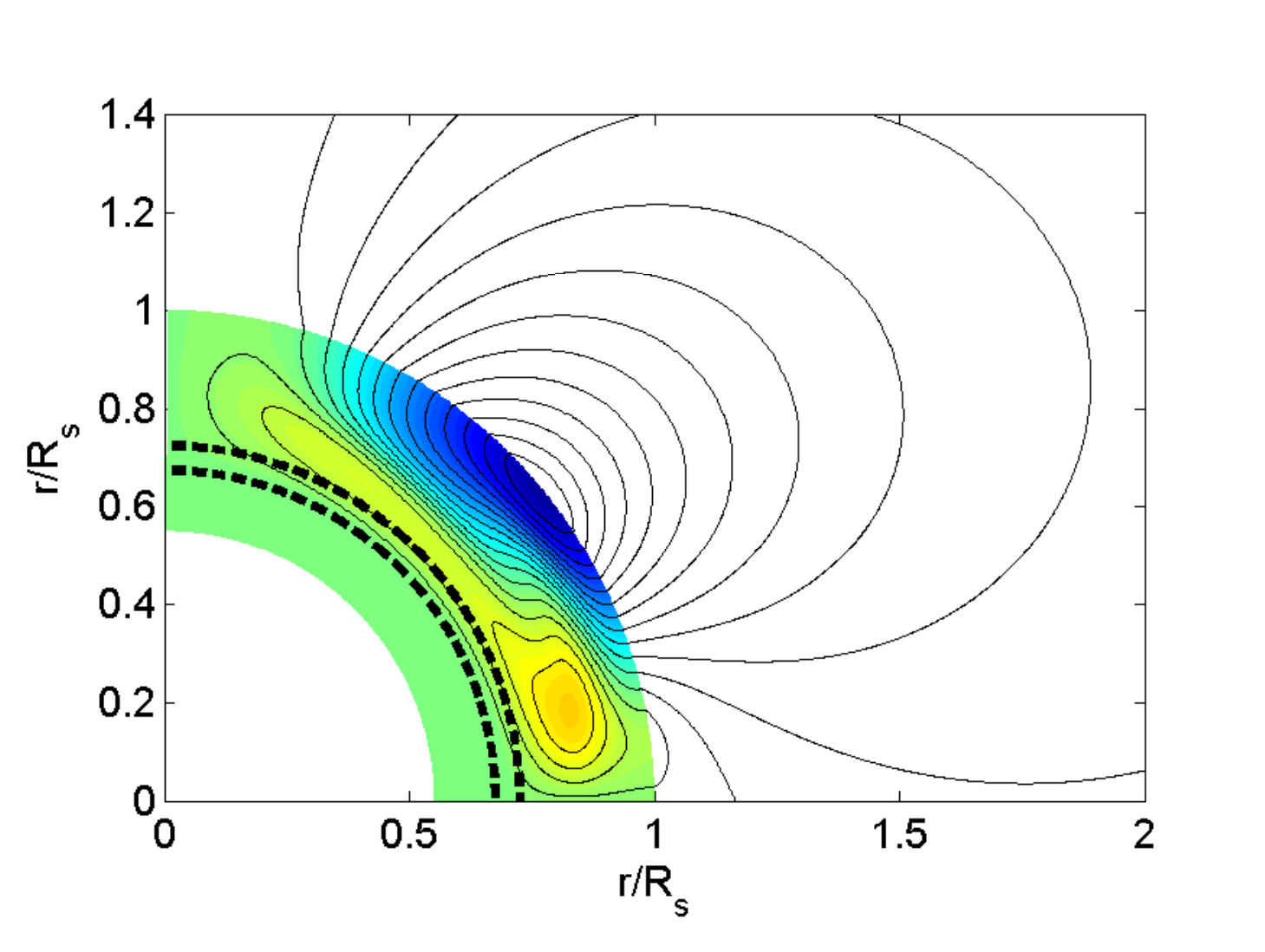}\\
  \includegraphics[scale=0.25]{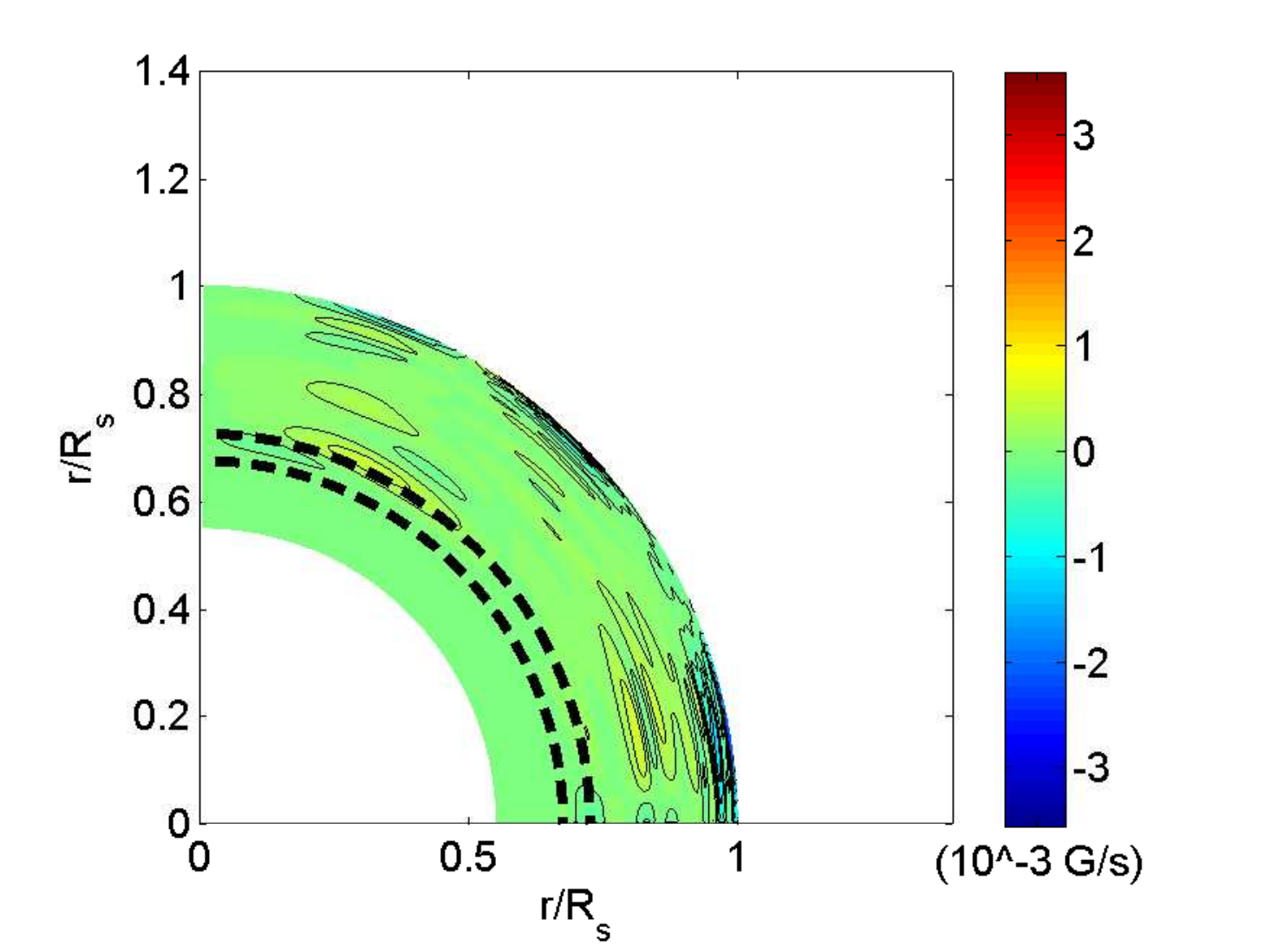} & \includegraphics[scale=0.25]{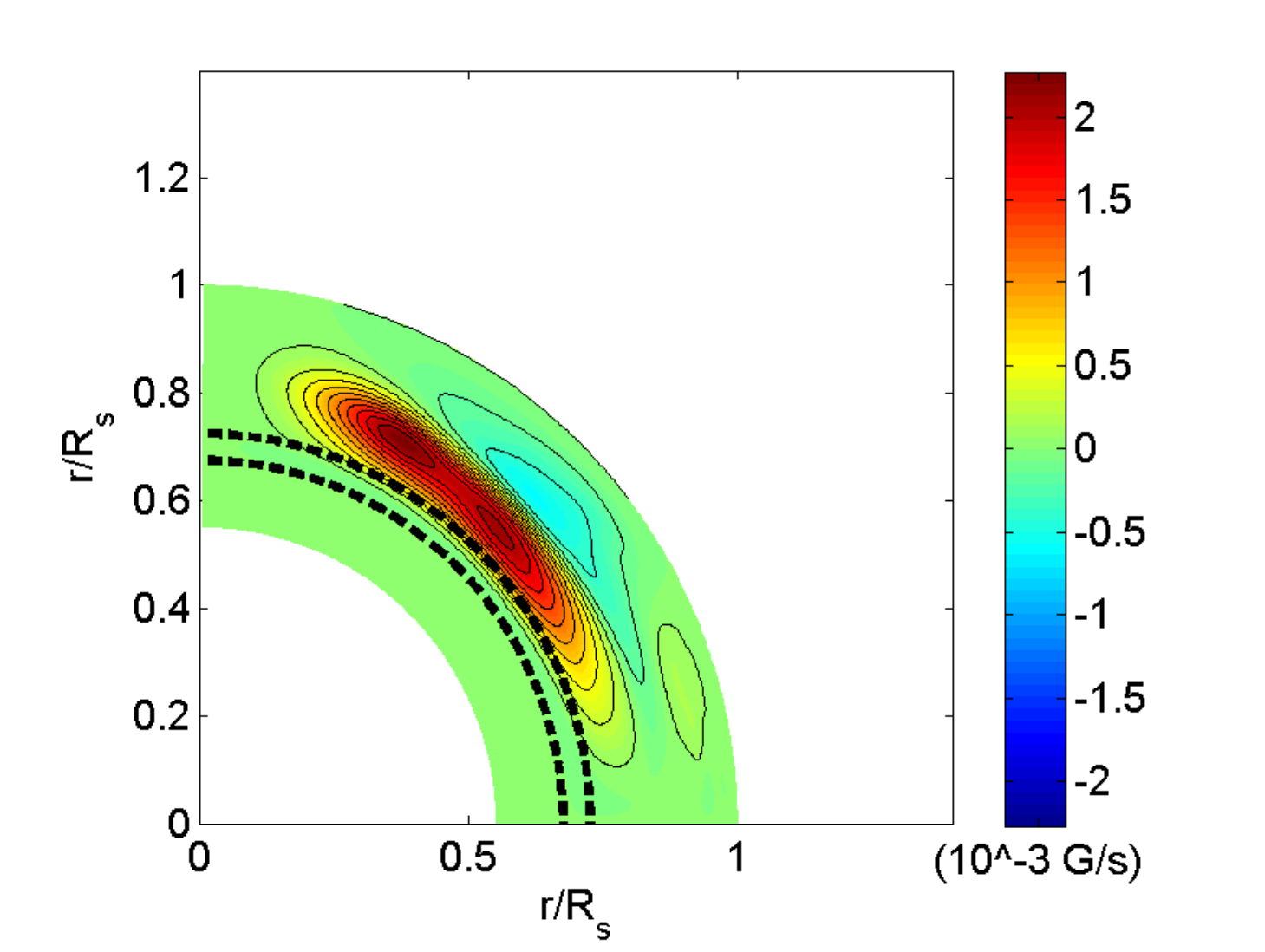} & \includegraphics[scale=0.25]{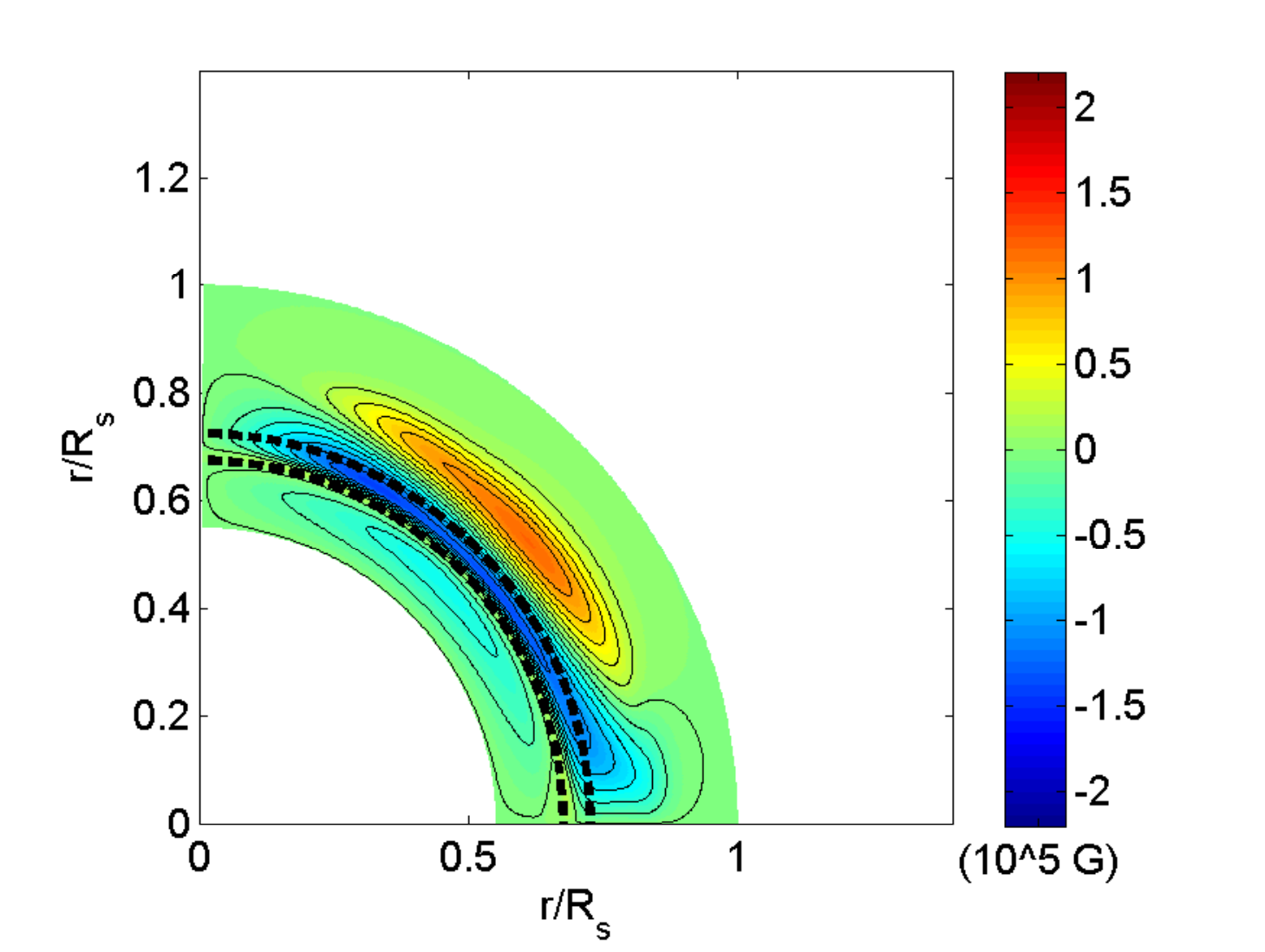} & \includegraphics[scale=0.25]{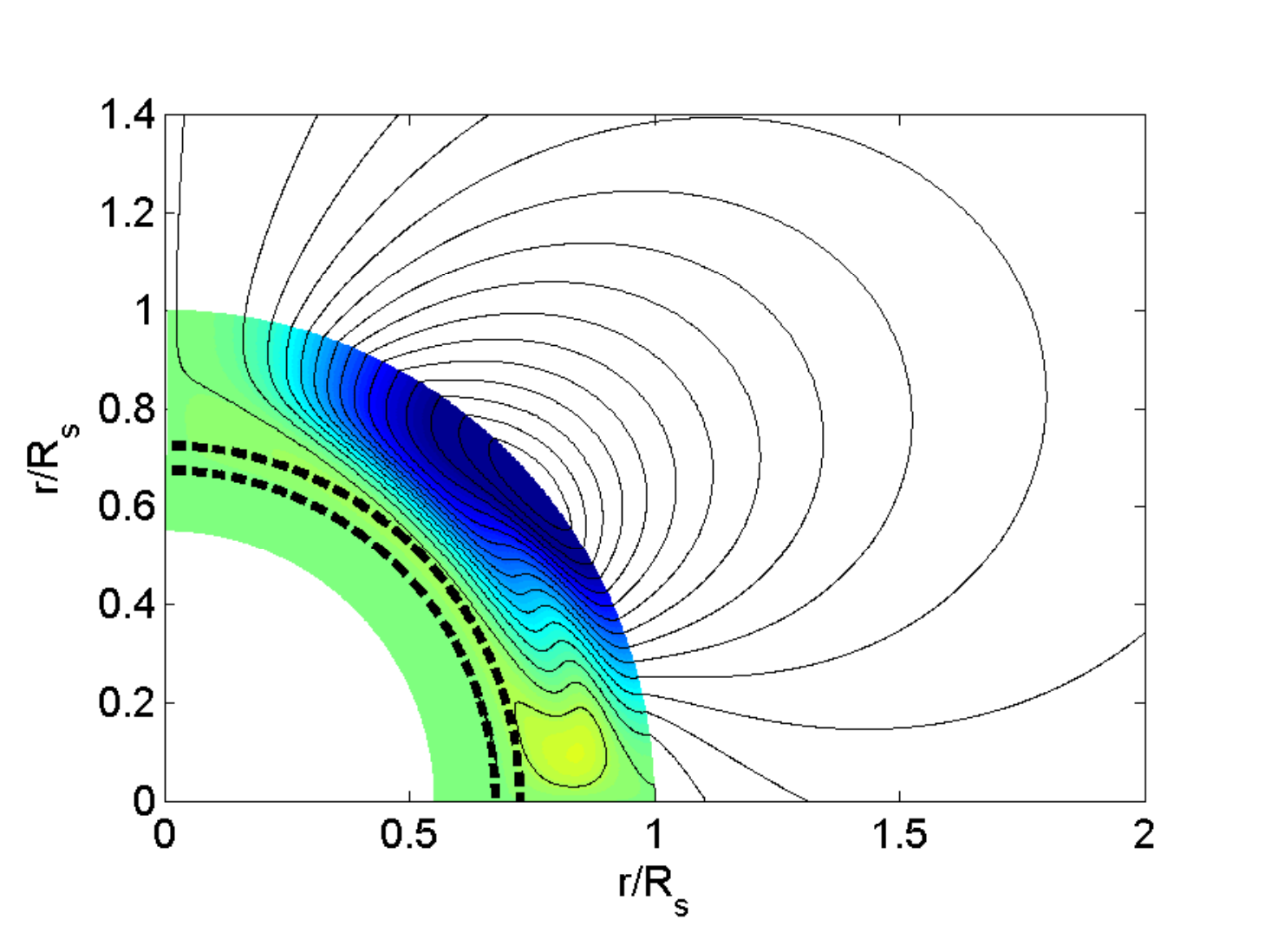}\\
  \includegraphics[scale=0.25]{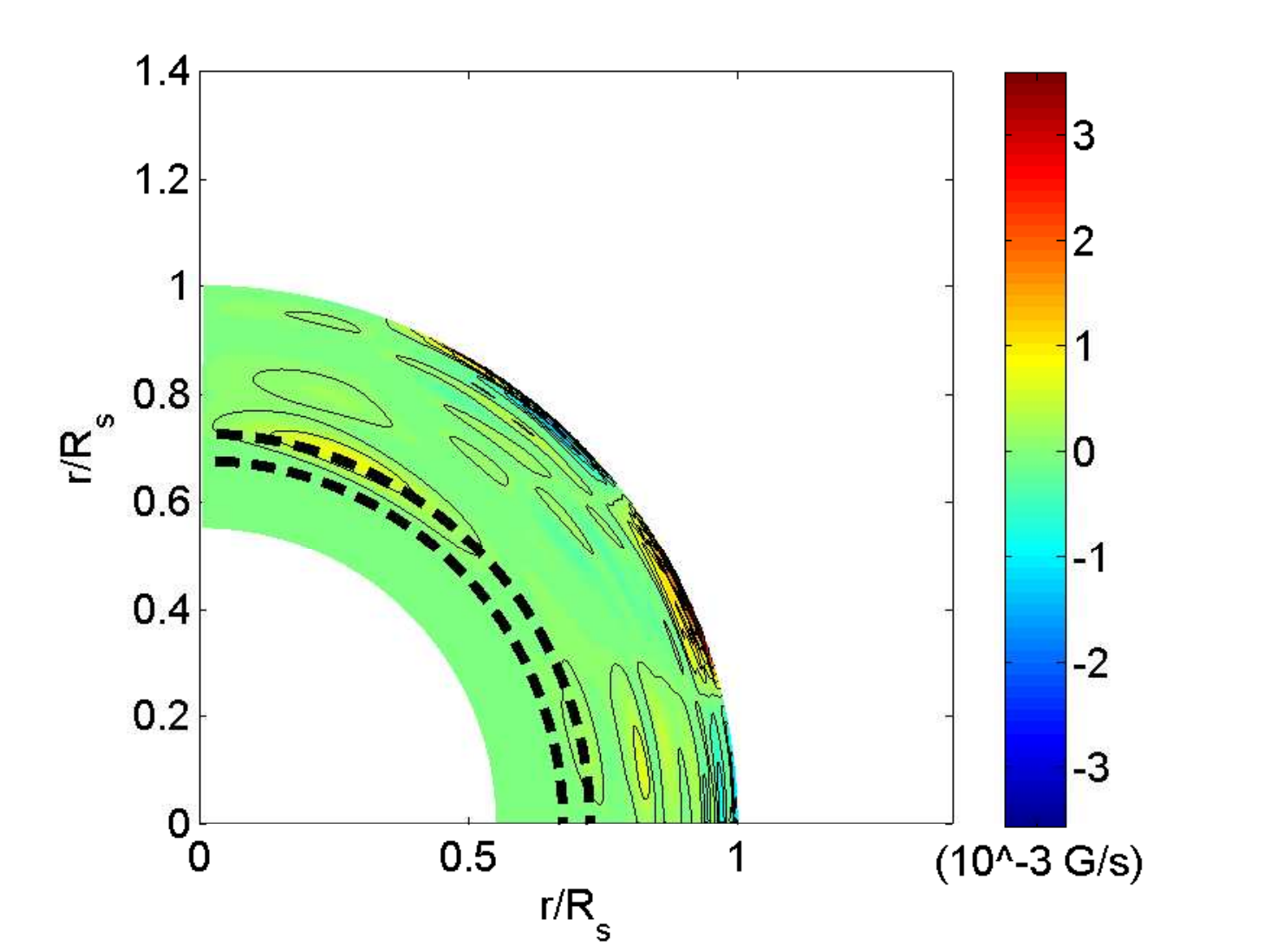} & \includegraphics[scale=0.25]{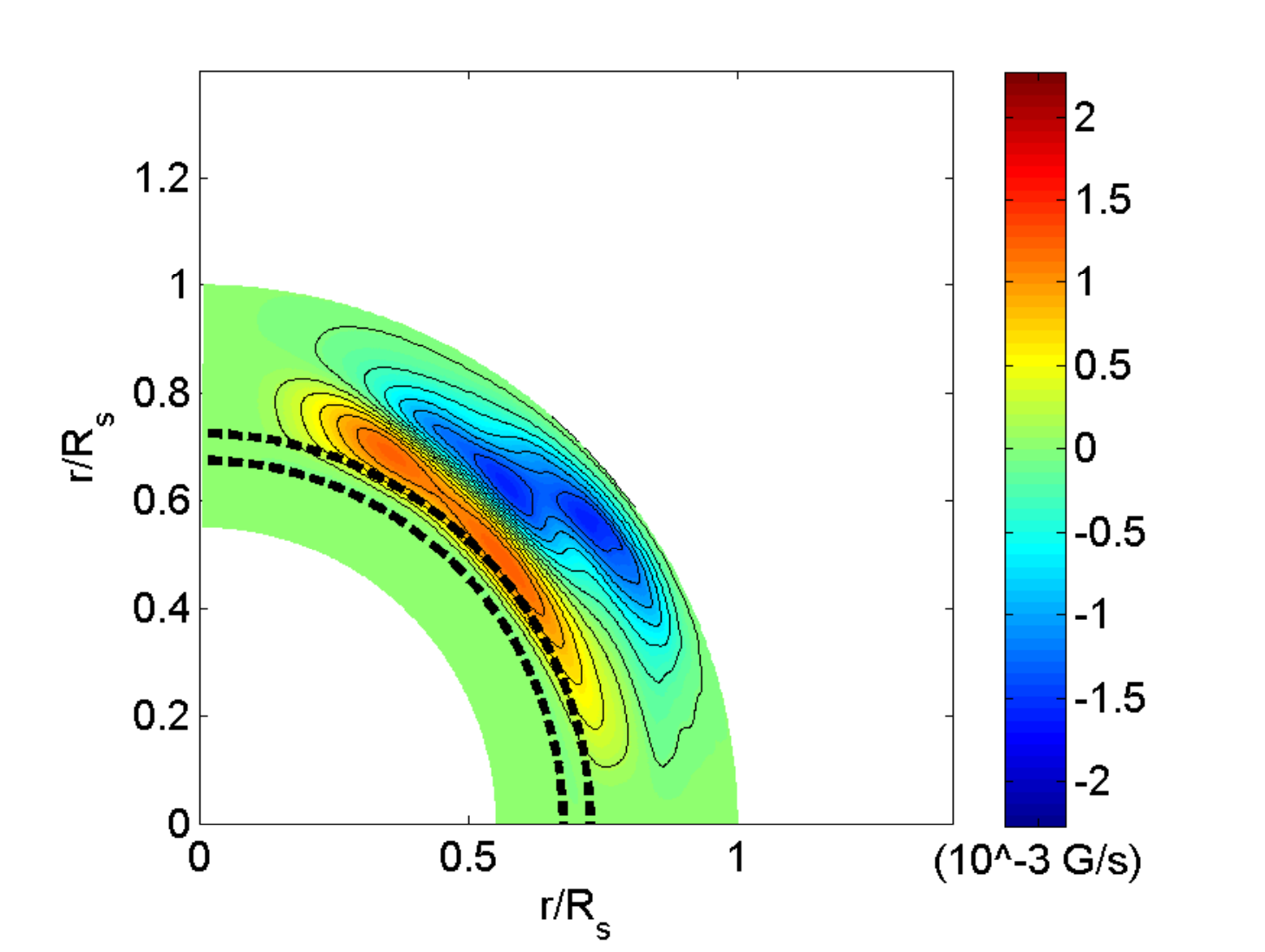} & \includegraphics[scale=0.25]{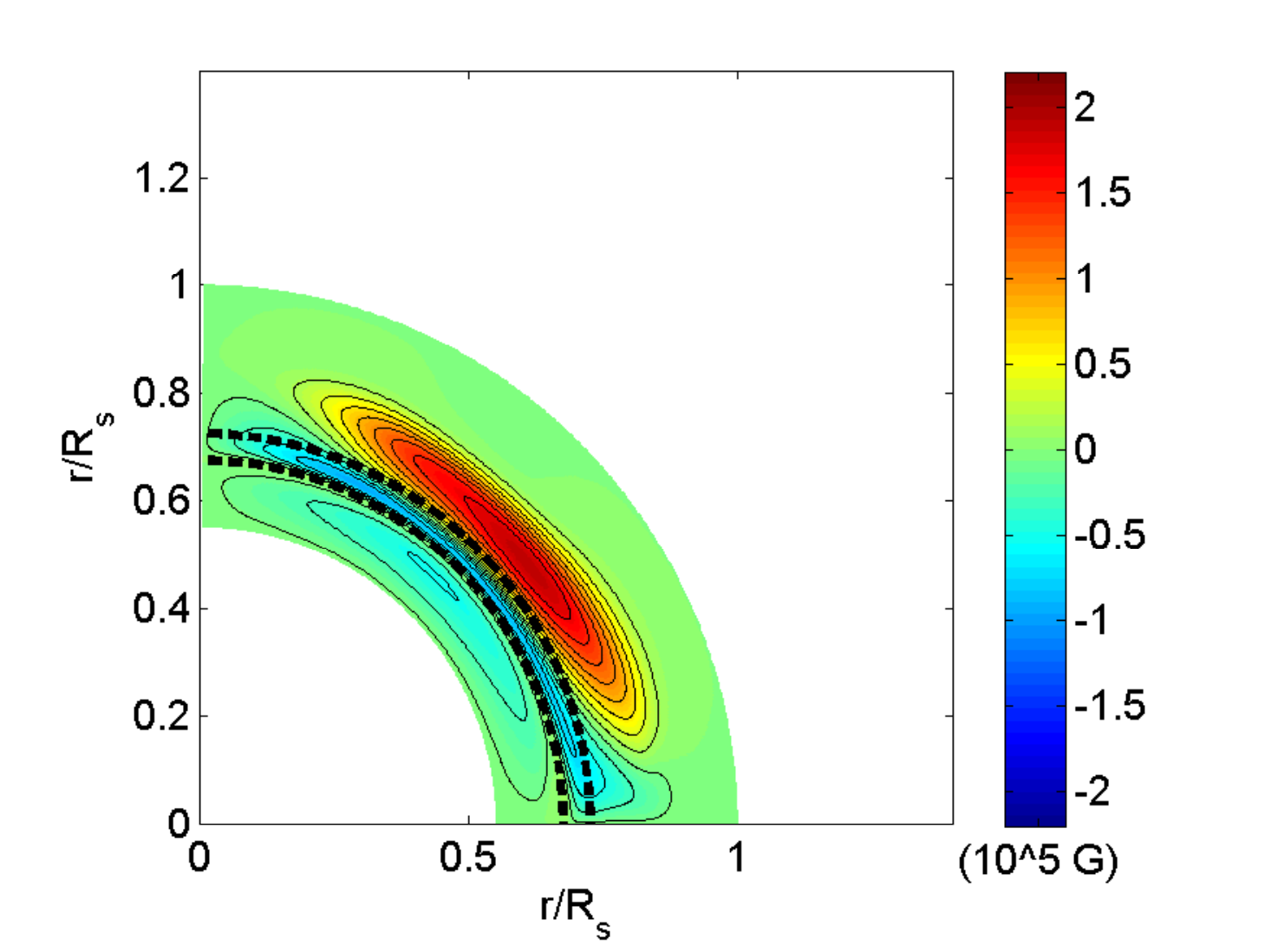} & \includegraphics[scale=0.25]{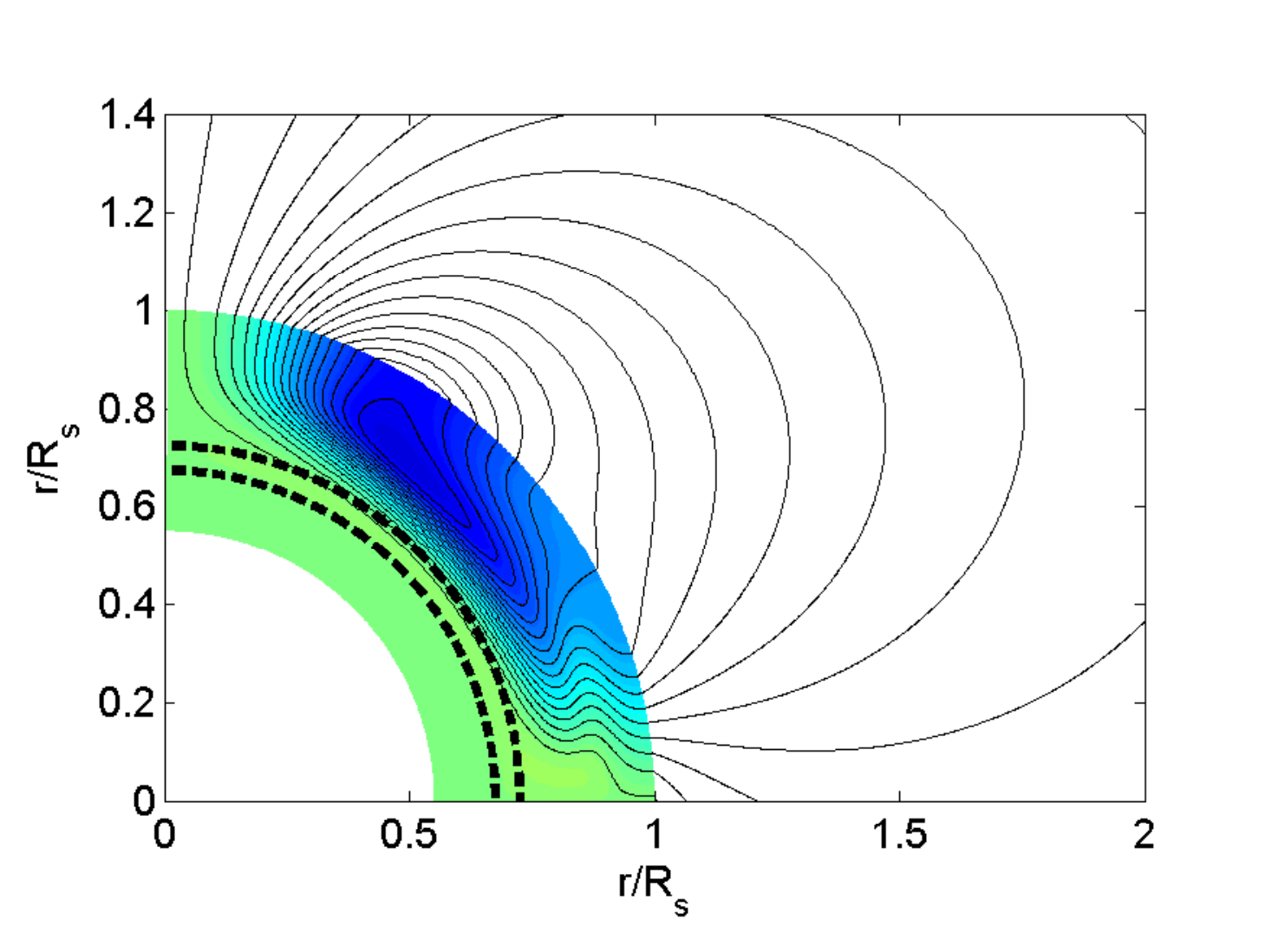}\\
  \end{tabular}
\caption{Snapshots of the shear source terms and the magnetic field over half a dynamo cycle (a
sunspot cycle). Each row is advanced by an eight of the dynamo cycle (a quarter of the sunspot
cycle) i.e., from top to bottom $t = 0, \tau/8, \tau/4$ and $3\tau/8$.  The solution corresponds to
the composite differential rotation and meridional flow Set 4 (shallowest penetration with a peak
flow of 22 m/s)}\label{S4C}
\end{figure}

In the second part of our work we compare dynamo solutions obtained for each of the four
different meridional flow profiles with two different penetration depth and with two different
peak flow speeds. First, from Figure~\ref{Comp_An} it is evident that the shape of the solutions
changes with varying penetration depth; this is caused by the increasing role of the tachocline
shear in generation and storage of the toroidal field as the penetration depth of the meridional
flow increases. This is apparent when one compares Figure~\ref{S1C} (deep penetration) to
Figure~\ref{S4C} (shallow penetration). If we look at the inductive shear sources it is evident
that for the shallowest penetration no field is being generated inside the tachocline. In the
poloidal field plots (right column) we see that no poloidal field is advected into the tachocline
region for the shallowest penetration, but some is advected for the deepest.

Second we compare the periods of our solutions in Table~\ref{MFT}: We find that most solutions
have a sunspot cycle (i.e., half-dynamo cycle) period that is comparable to that of the Sun, with
the exception of the fast flow with deep penetration and the slow flow with shallow penetration,
which have respectively a comparatively smaller and larger period. As the meridional flow is
buried deeper, one expects the length of the advective circuit to increase, thereby resulting in
larger dynamo periods. However, it is evident that as we increase the penetration, the period
decreases -- even if the length of the flow loop that supposedly transports magnetic flux
increases; this is counterintuitive but has a simple explanation. Our simulations and exhaustive
analysis points out that it is not how deep the flow penetrates that governs the cycle period,
but it is the magnitude of the meridional counterflow right at the base of the SCZ ($R_p =
0.713R_\odot$) that is most relevant. This is because most of the poloidal field creation at
near-surface layers is coupled to buoyant eruptions of toroidal field from this layer of
equatorward migrating toroidal field belt at the base of the SCZ (see third columns in Figures
\ref{S1C} and \ref{S4C}) and it is the flow speed at this region that governs the dynamo period.
In Figure~\ref{VthFitr} it is clear that the speed of the counterflow in this convection
zone--radiative interior interface increases as the flow becomes more penetrating (a consequence
of the constraints set by mass conservation and the fits to the near-surface helioseismic data),
thereby reducing the dynamo period.

Overall, an evaluation of the butterfly diagram (Figure~\ref{Comp_An}), points out that the
toroidal field belt extends to lower latitudes (where sunspots are observed) for deeper penetrating
meridional flow, although there is a polar branch as well. For the shallow flow, we find that the
toroidal field belt is concentrated around mid-latitudes with almost symmetrical polar and
equatorial branches -- a signature of the convection zone latitudinal shear producing most of the
toroidal field (as in the interface dynamo models, see, e.g., Parker 1993 and Charbonneau \& MacGregor 1997).


\begin{table}[h]
  \centering
\begin{tabular}{|c|c|c|c|c|}
  \hline
   & $R_p$ & $v_o$ & $\tau$ - HS data (yrs) & $\tau$ - Analytical (yrs)\\
  \hline
  \hline
  Set 1 & \multirow{2}{*}{$0.64R_\odot$} & $12$ m/s  & $9.67$ & $10.00$\\
  \cline{1-1}
  \cline{3-5}
  Set 2 & & $22$ m/s & $5.63$ & $5.67$\\
  \hline
  Set 3 & \multirow{2}{*}{$0.71R_\odot$} & $12$ m/s & $14.67$ & $14.02$\\
  \cline{1-1}
  \cline{3-5}
  Set 4 & & $22$ m/s & $11.85$& $12.85$\\
  \hline
\end{tabular}
  \caption{Simulated sunspot cycle period for the different sets of meridional flow parameters. $R_p$ corresponds to the maximum penetration depth of the meridional flow, $v_o$ to the peak speed in the poleward flow and $\tau$ is the period of the solutions in units of years.  For the rest of the parameters in each set please refer to Table \ref{MFlowP}}\label{MFT}
\end{table}

\subsection{Dependence of the solutions on changes in the turbulent diffusivity profile}

Although a detailed exploration of the turbulent diffusivity parameter space is outside the scope of this work, we study two special cases in which we vary a single parameter while leaving the rest fixed.

\begin{figure}[c]
  \begin{tabular}{ccc}
   \includegraphics[scale=0.3]{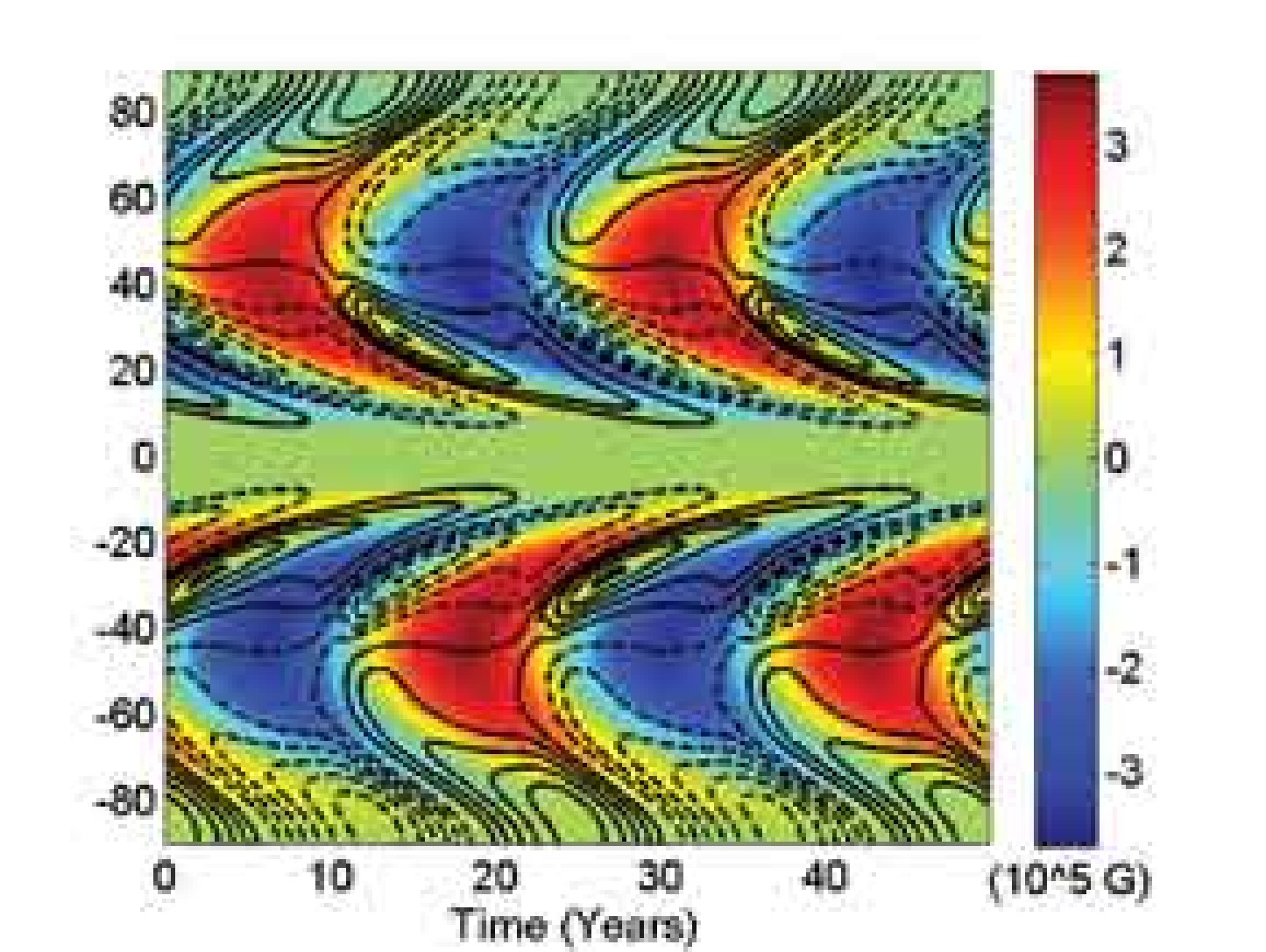} & $R_p = 0.64R_\odot$& \includegraphics[scale=0.3]{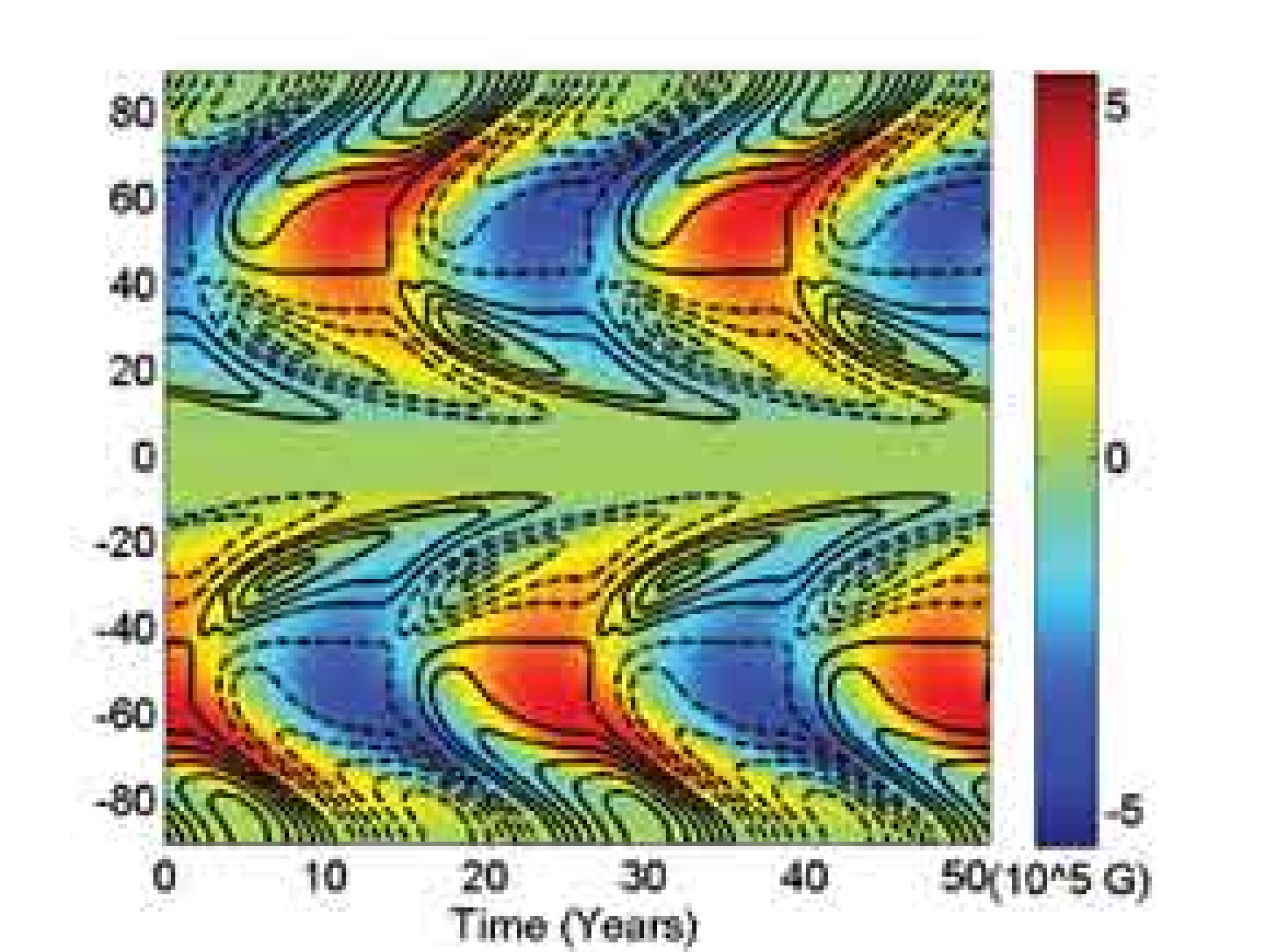}\\
   & $v_o = 12$m/s& \\
   \includegraphics[scale=0.3]{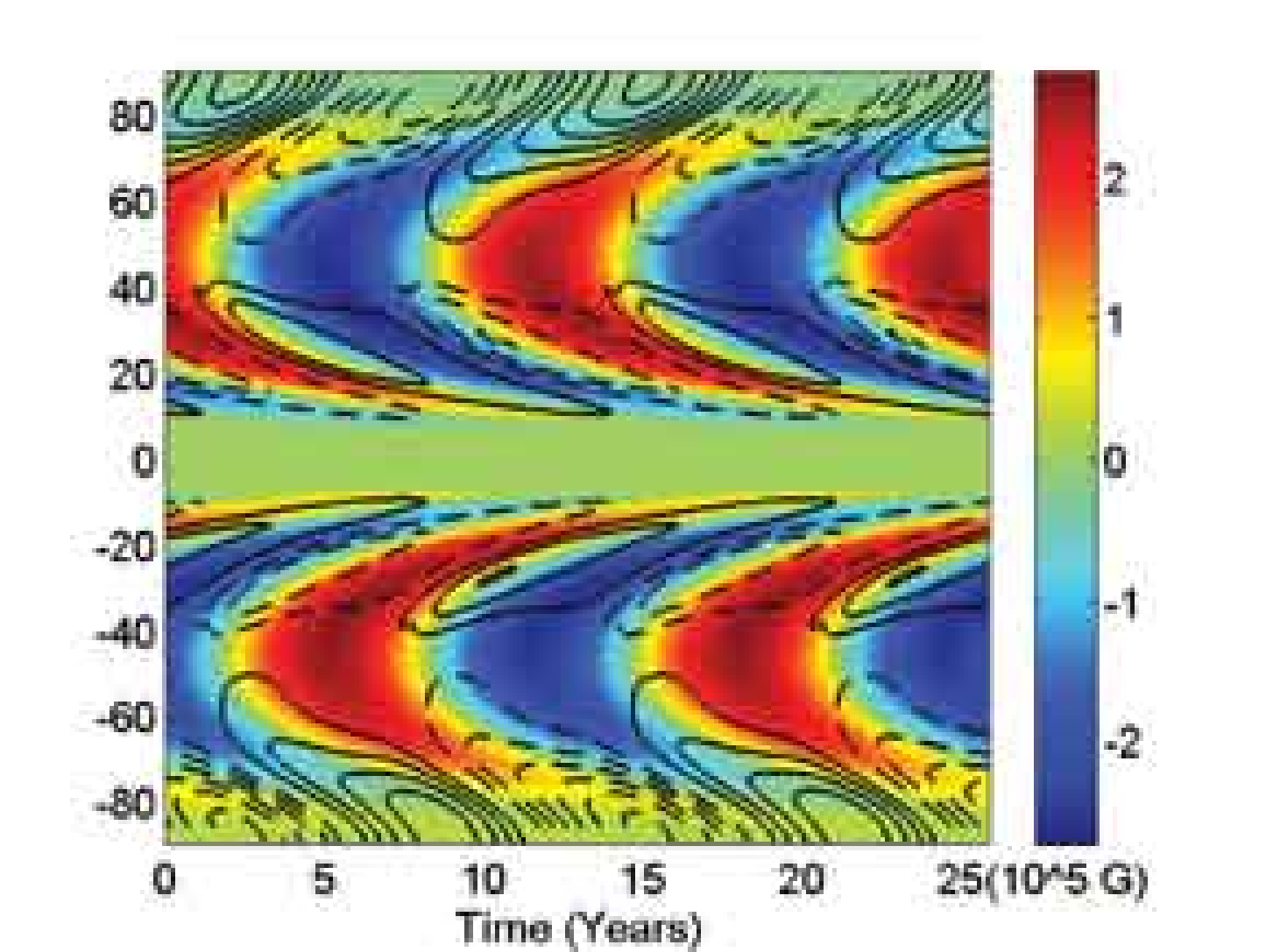} &$R_p = 0.64R_\odot$ & \includegraphics[scale=0.3]{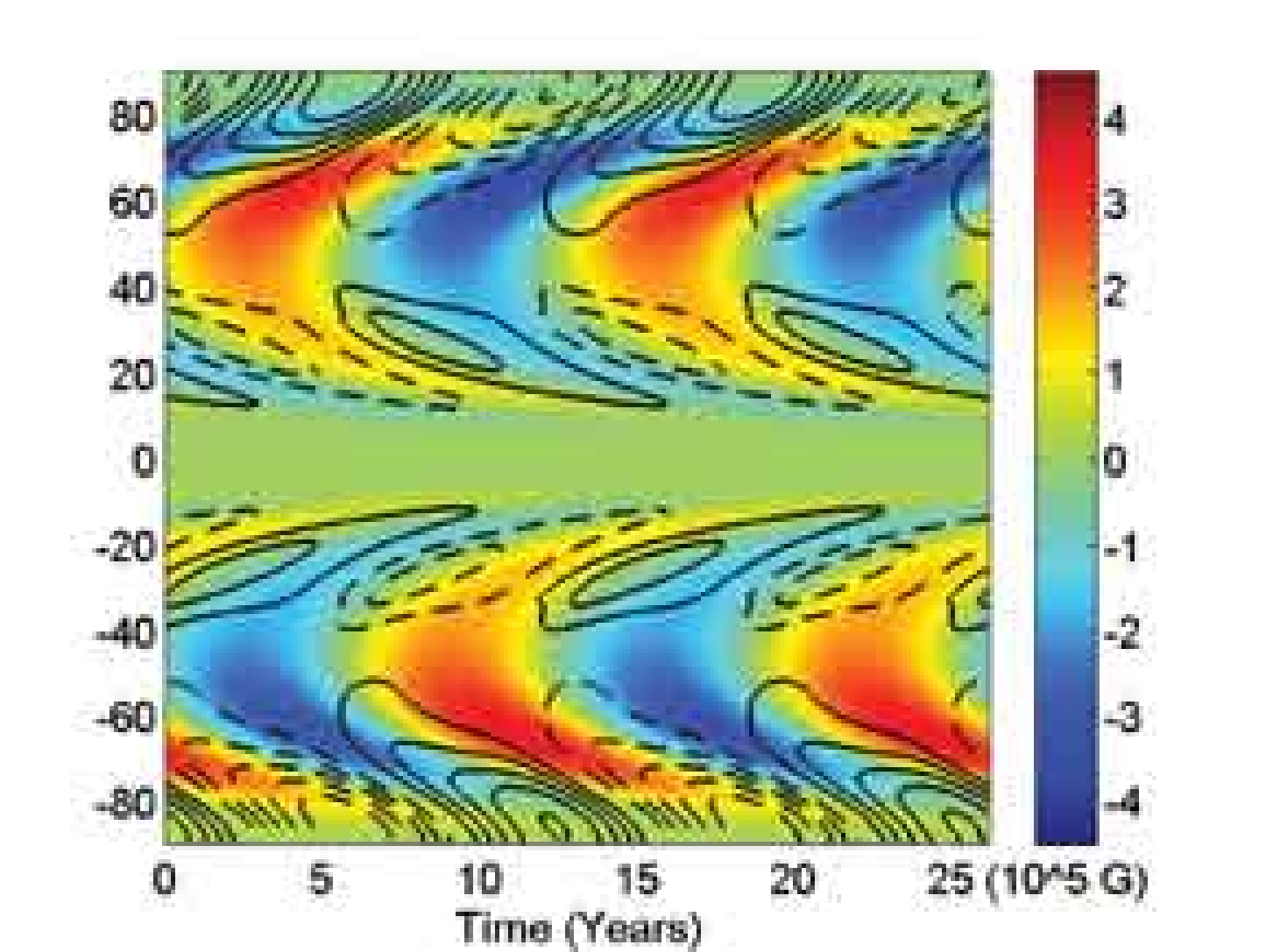}\\
   & $v_o = 22$m/s& \\
   \includegraphics[scale=0.3]{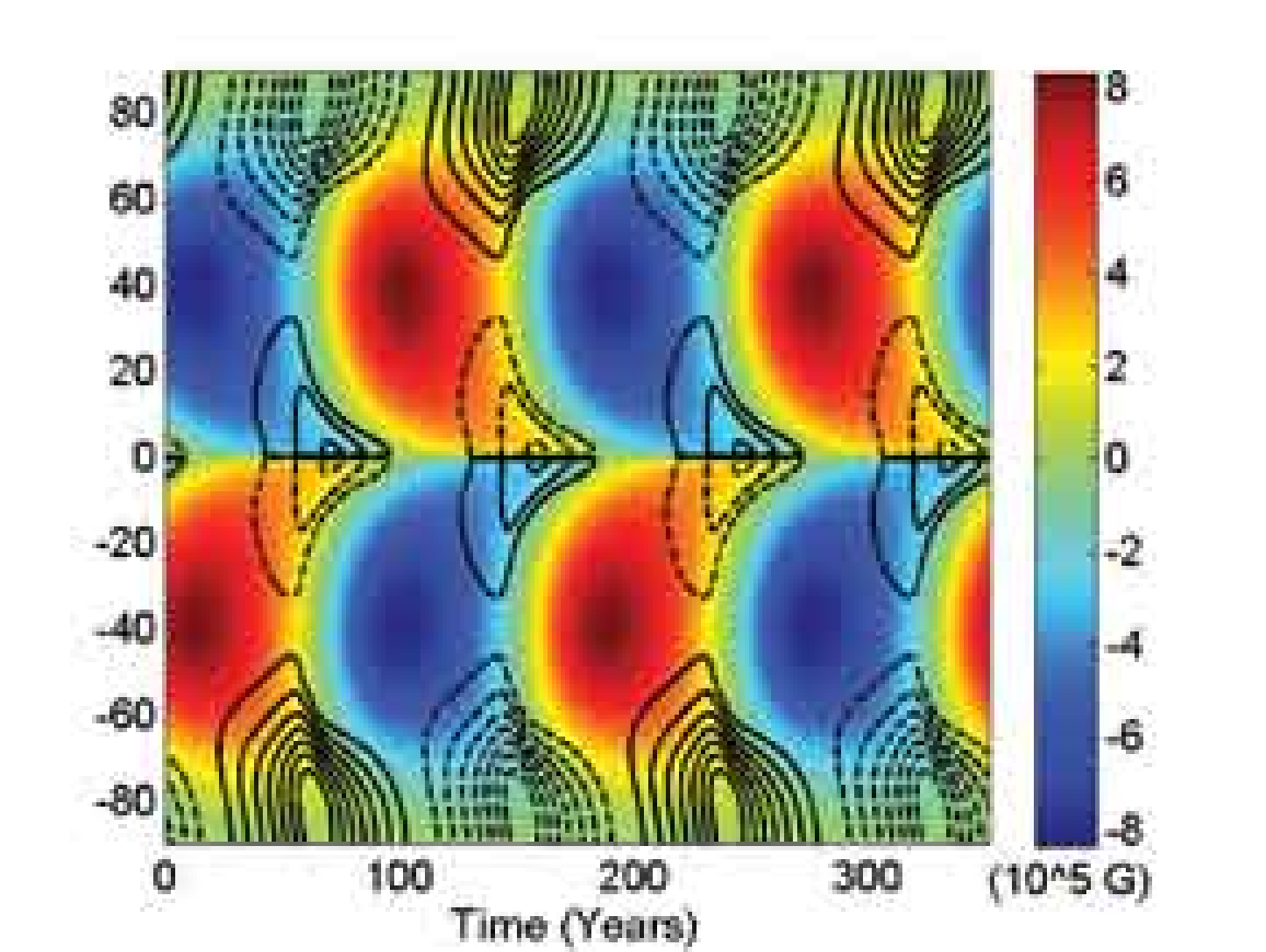} & $R_p = 0.71R_\odot$& \includegraphics[scale=0.3]{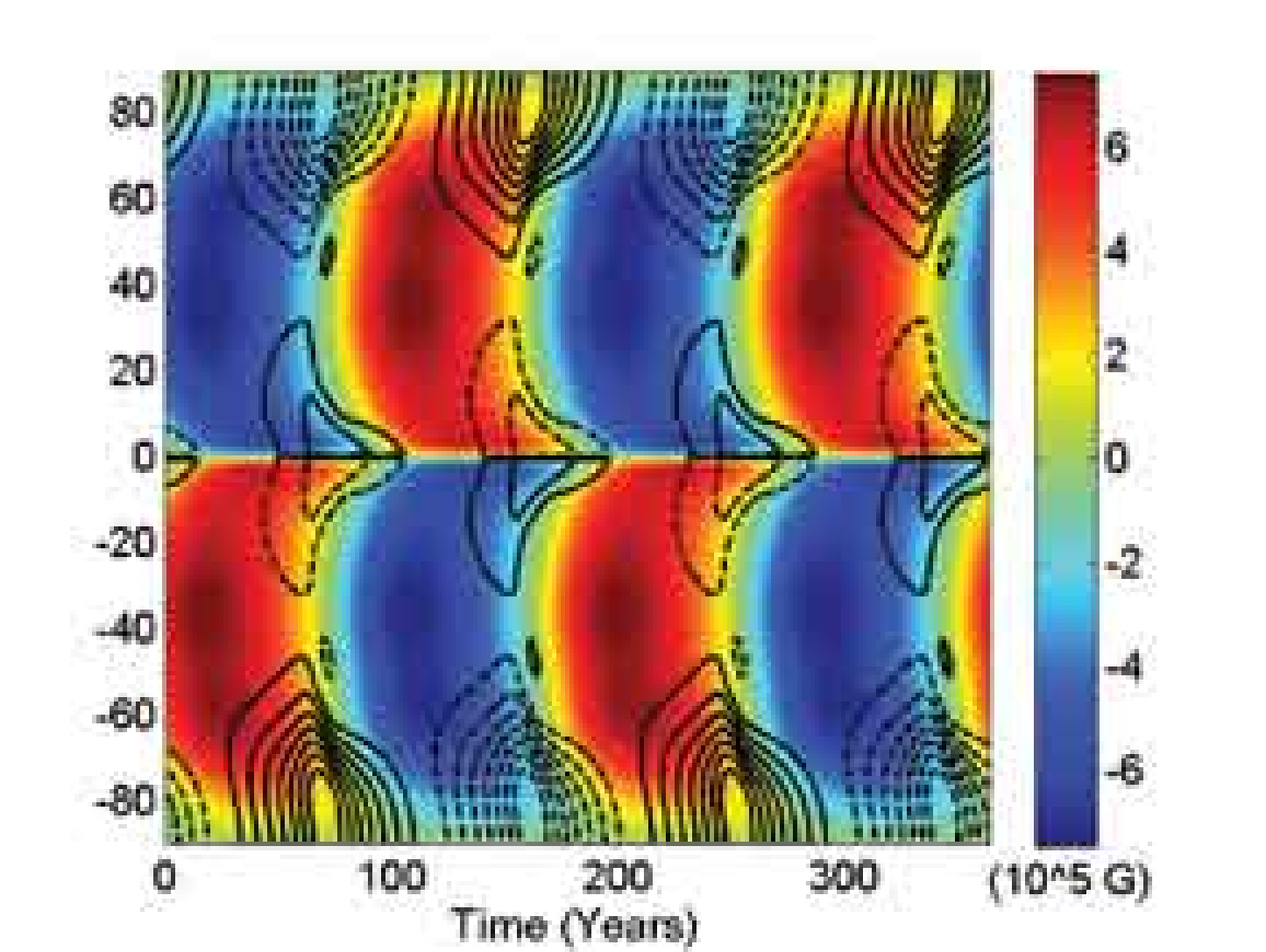}\\
   & $v_o = 12$m/s& \\
   \includegraphics[scale=0.3]{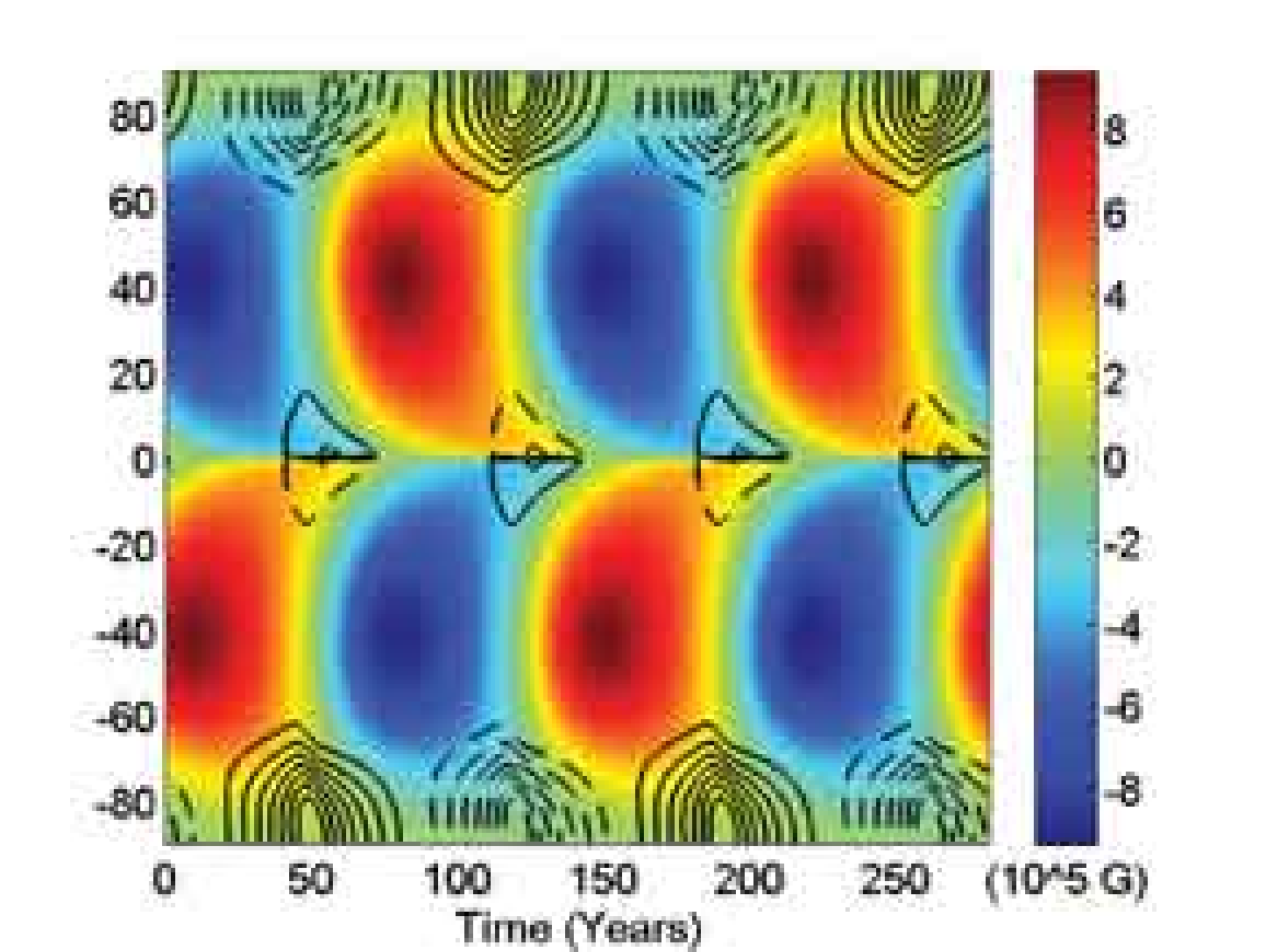} &$R_p = 0.71R_\odot$ & \includegraphics[scale=0.3]{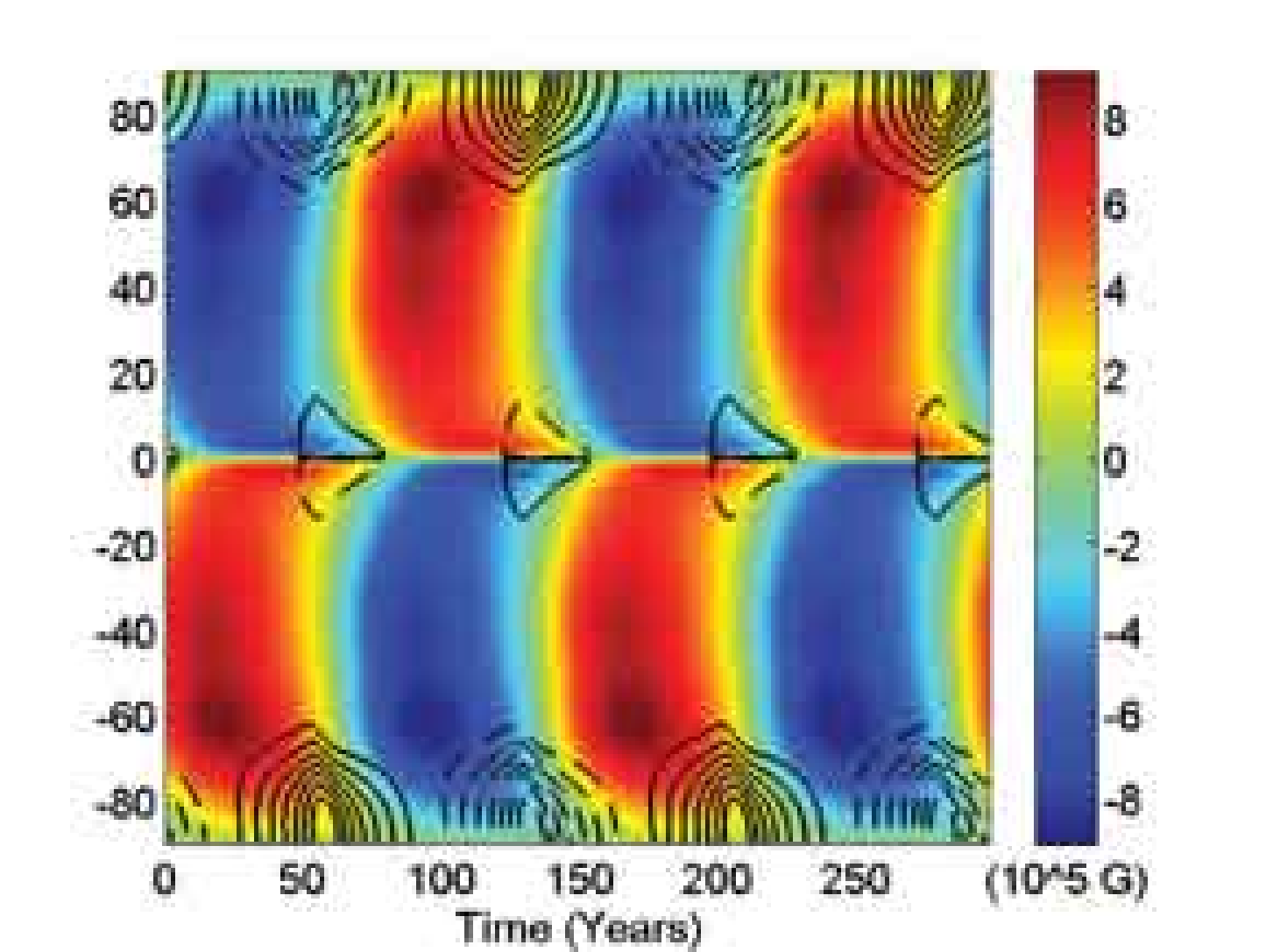}\\
   Composite DR & $v_o = 22$m/s& Analytical DR
  \end{tabular}
\caption{Butterfly diagram of the torodial field at the bottom of the convection zone (Color) with
radial field at the surface (Contours) superimposed on it when using a low diffusivity in the convection zone ($\eta_{cz}=10^{10} cm^2/s$). Each row corresponds to one of the
different meridional circulation sets. The left column corresponds to runs using the
helioseismology composite and the right one to runs using the analytical profile.}\label{Comp_An_ld}
\end{figure}

For the first case we lower the diffusivity in the convection zone, $\eta_{cz}$, from $10^{11} cm^2/s$ to $10^{10} cm^2/s$.  As can be seen in Figure~\ref{Comp_An_ld}, this introduces two important changes in the dynamo solutions:  The first one is an overall increase in magnetic field magnitude due to the reduction in diffusive decay while keeping the strength of the field sources constant. The second is a drastic increase in the dynamo period (which can be seen tabulated in Table~\ref{MFT_LD}) for the solutions that use a meridional flow with a penetration of $R_p = 0.71R_\odot$. The reason behind such a change resides in the nature of the transport processes at the bottom of the convection zone, which are a combination of both advection and diffusion.  In the case of flow profiles with deep penetration the velocity at the bottom is high enough for downward advection to transport flux into the tachocline.  On the other hand, in the case of low penetration, the last bit of downward transport into the tachocline is done by diffusive transport and thus dominated by diffusive timescales.  Because of this, by decreasing turbulent diffusivity by an order of magnitude, we drastically increase the period of the solutions.



\begin{table}[h]
  \centering
\begin{tabular}{|c|c|c|c|c|}
  \hline
   & $R_p$ & $v_o$ & $\tau$ - HS data (yrs) & $\tau$ - Analytical (yrs)\\
  \hline
  \hline
  Set 1 & \multirow{2}{*}{$0.64R_\odot$} & $12$ m/s  & $12.46$ & $12.86$\\
  \cline{1-1}
  \cline{3-5}
  Set 2 & & $22$ m/s & $6.47$ & $6.55$\\
  \hline
  Set 3 & \multirow{2}{*}{$0.71R_\odot$} & $12$ m/s & $87.78$ & $90.92$\\
  \cline{1-1}
  \cline{3-5}
  Set 4 & & $22$ m/s & $70.48$& $74.41$\\
  \hline
\end{tabular}
  \caption{Simulated sunspot cycle period for the different sets of meridional flow parameters when using a low diffusivity in the convection zone ($\eta_{cz}=10^{10} cm^2/s$). $R_p$ corresponds to the maximum penetration depth of the meridional flow, $v_o$ to the peak speed in the poleward flow and $\tau$ is the period of the solutions in units of years.}\label{MFT_LD}
\end{table}

\begin{figure}[c]
  \begin{tabular}{cc|cc}
  \multicolumn{2}{c}{Analitical DR} & \multicolumn{2}{c}{Composite DR}\\
  Toroidal Field  & Poloidal Field      & Toroidal Field                         & Poloidal Field \\
  \includegraphics[scale=0.25]{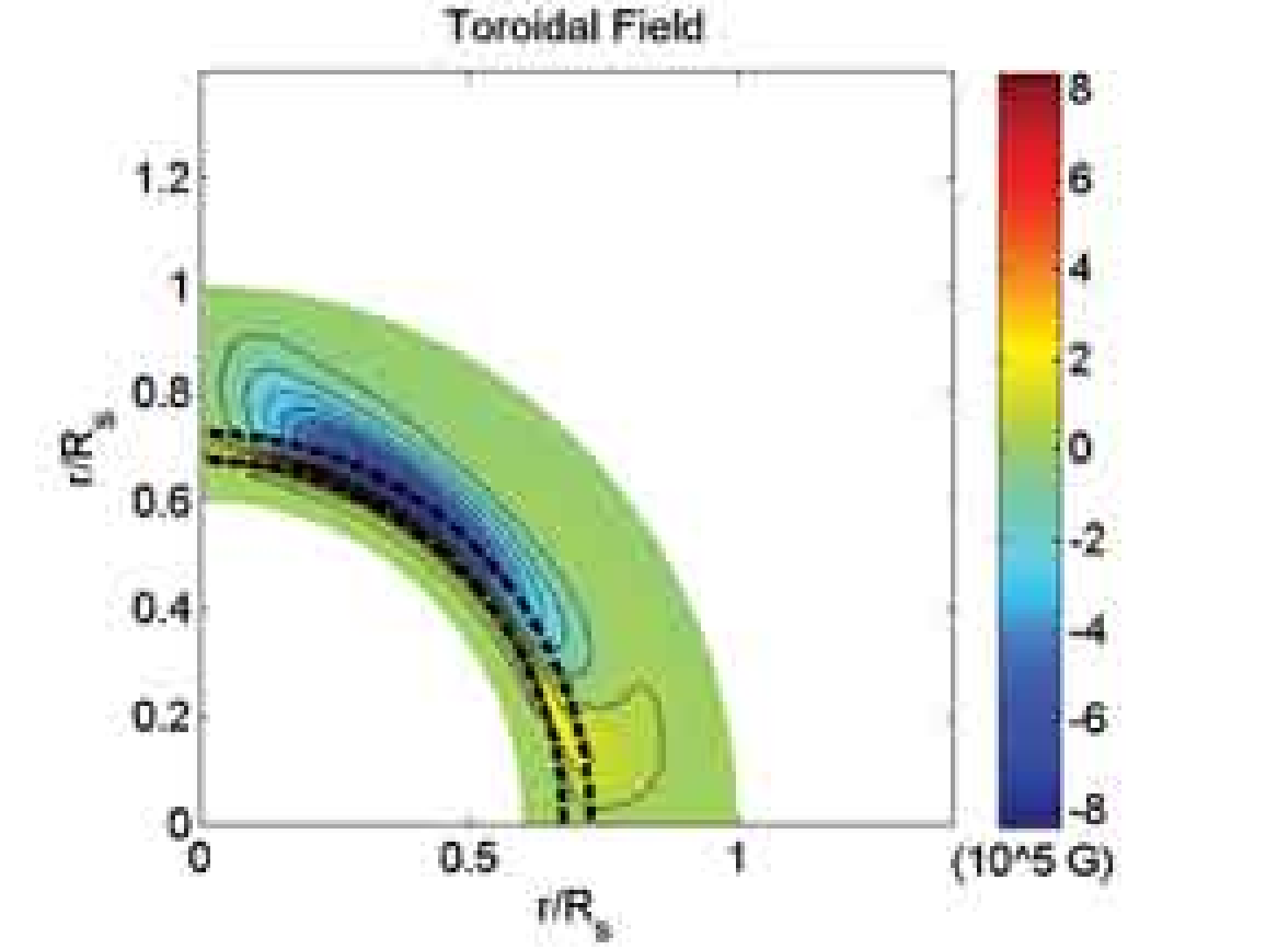} & \includegraphics[scale=0.25]{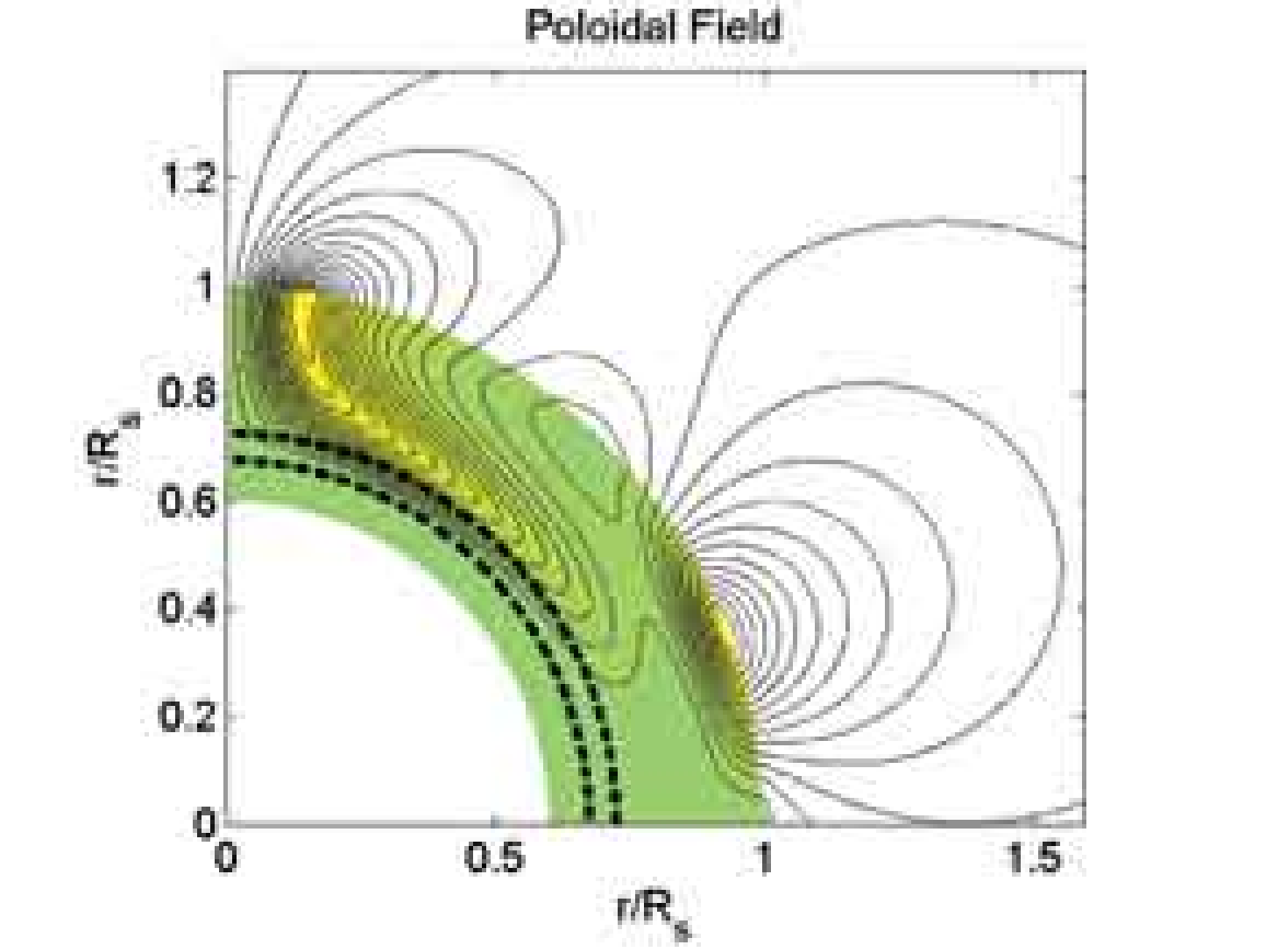} & \includegraphics[scale=0.25]{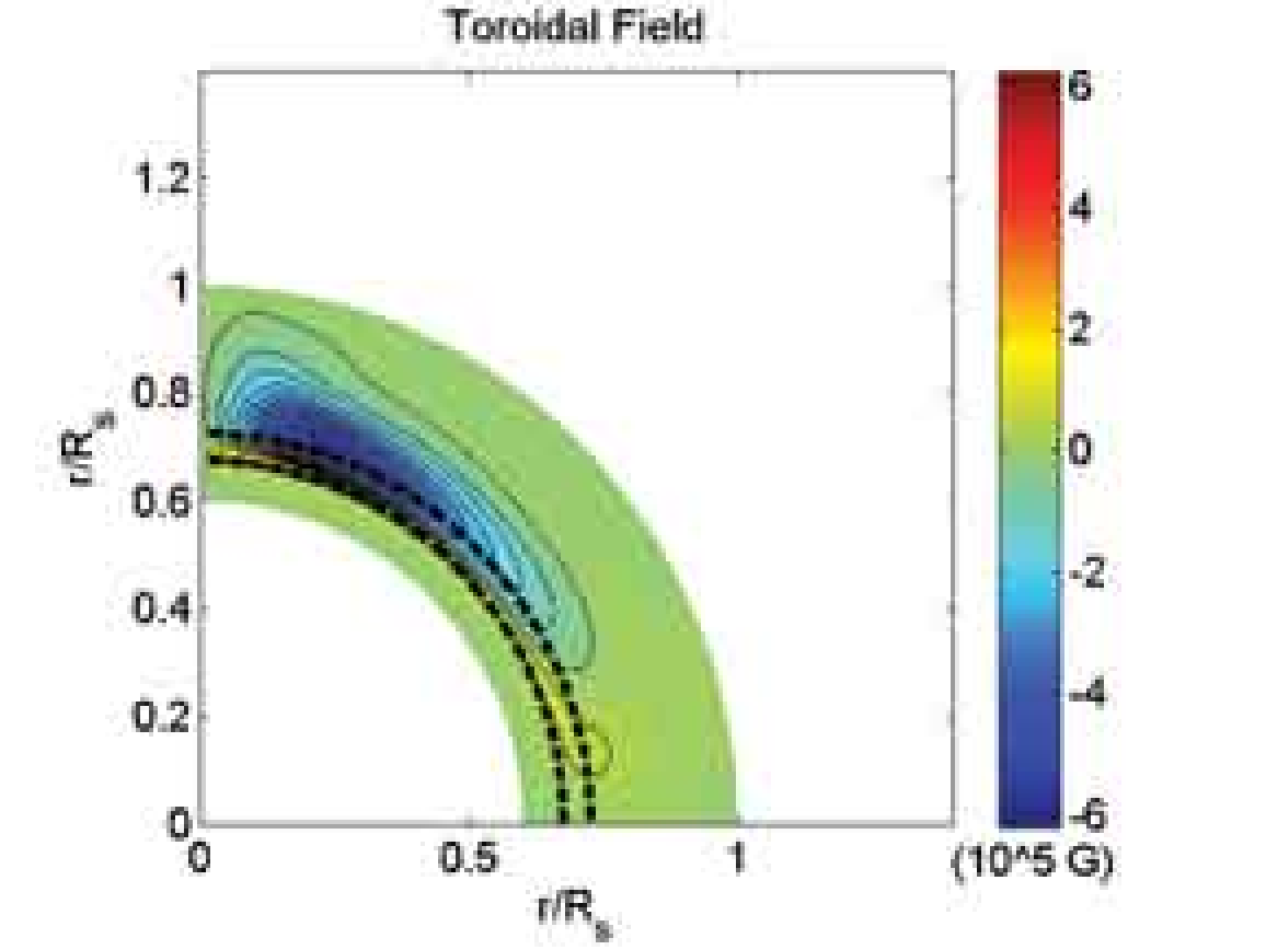} & \includegraphics[scale=0.25]{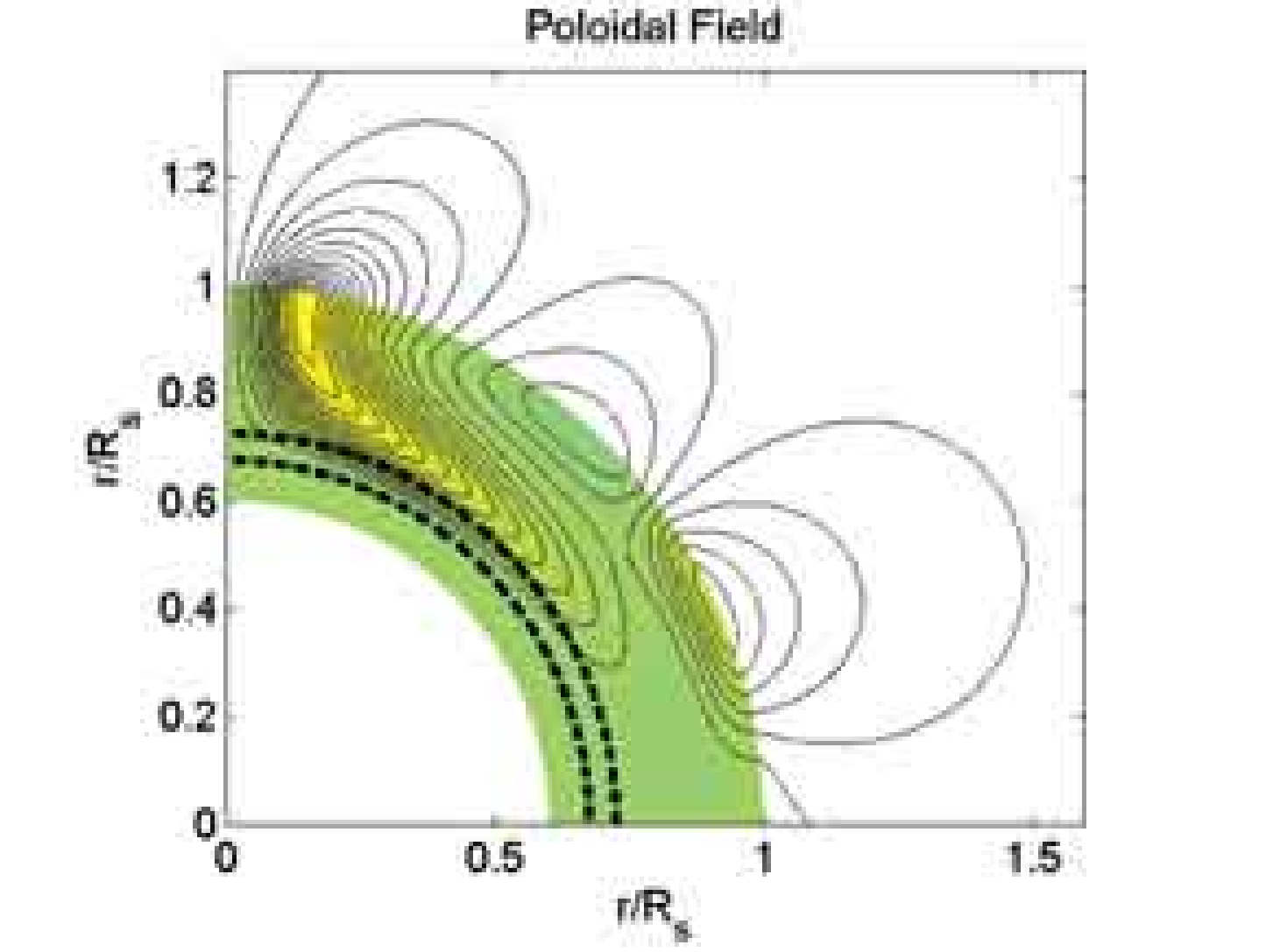}\\
  \includegraphics[scale=0.25]{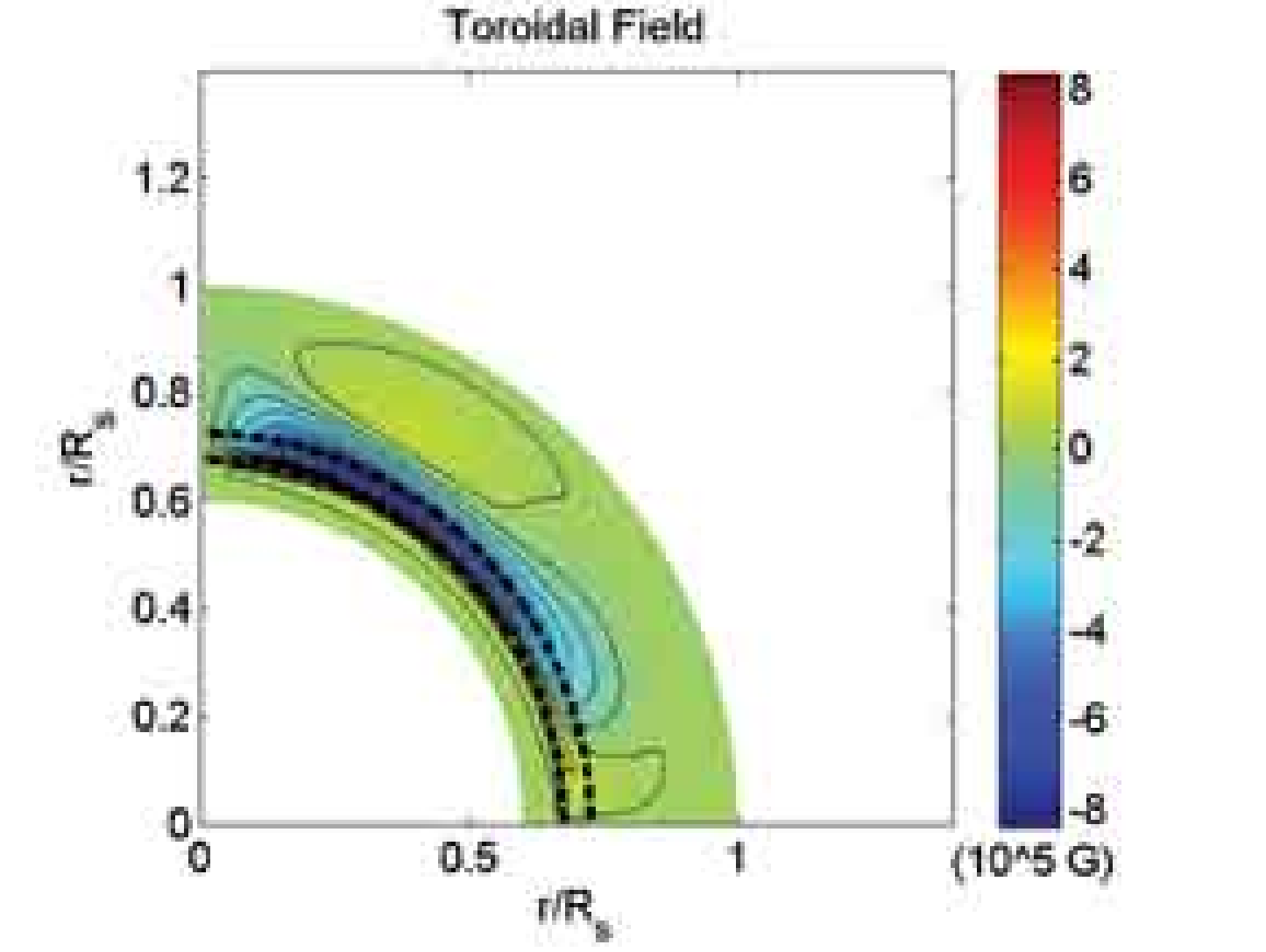} & \includegraphics[scale=0.25]{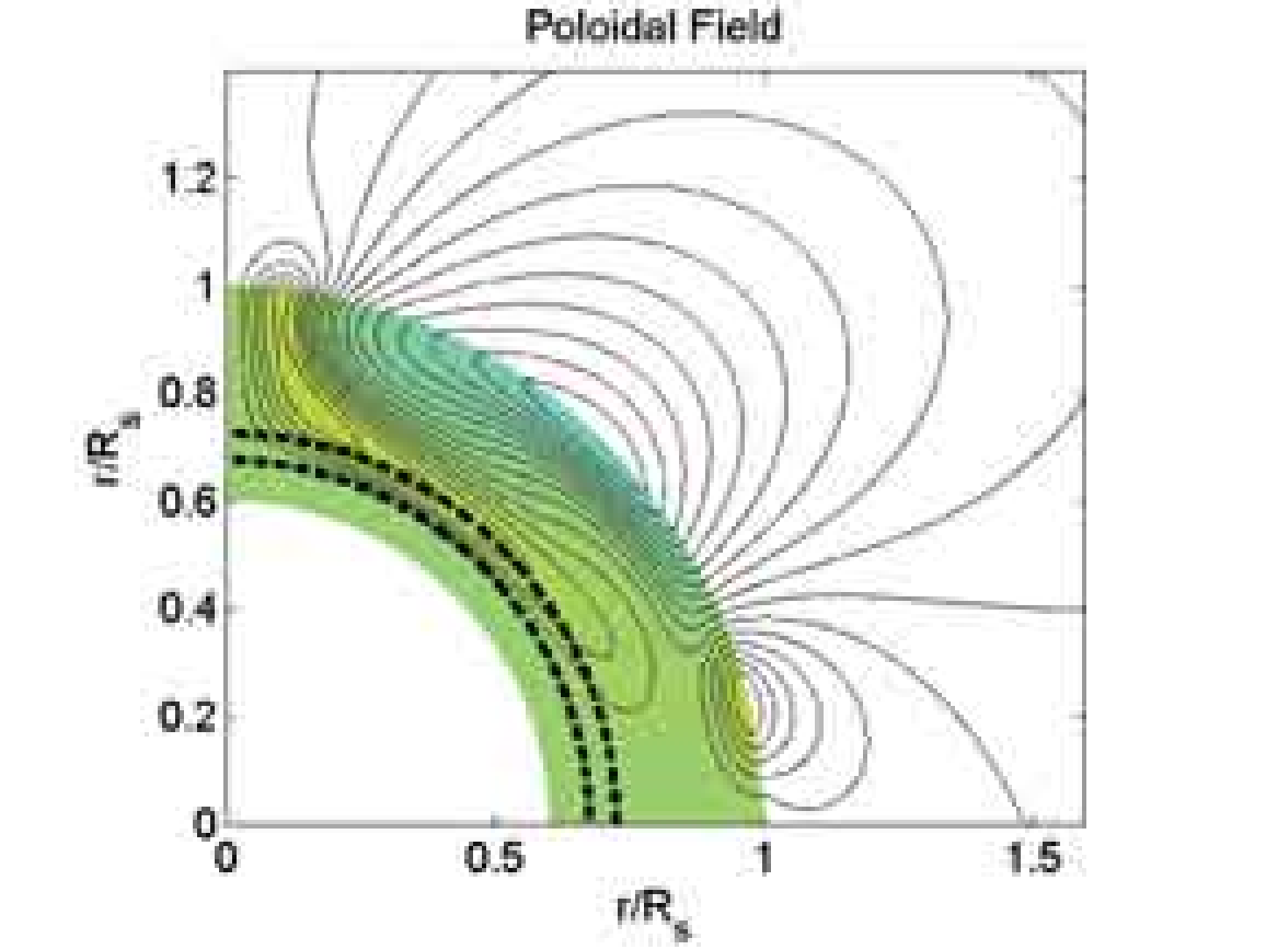} & \includegraphics[scale=0.25]{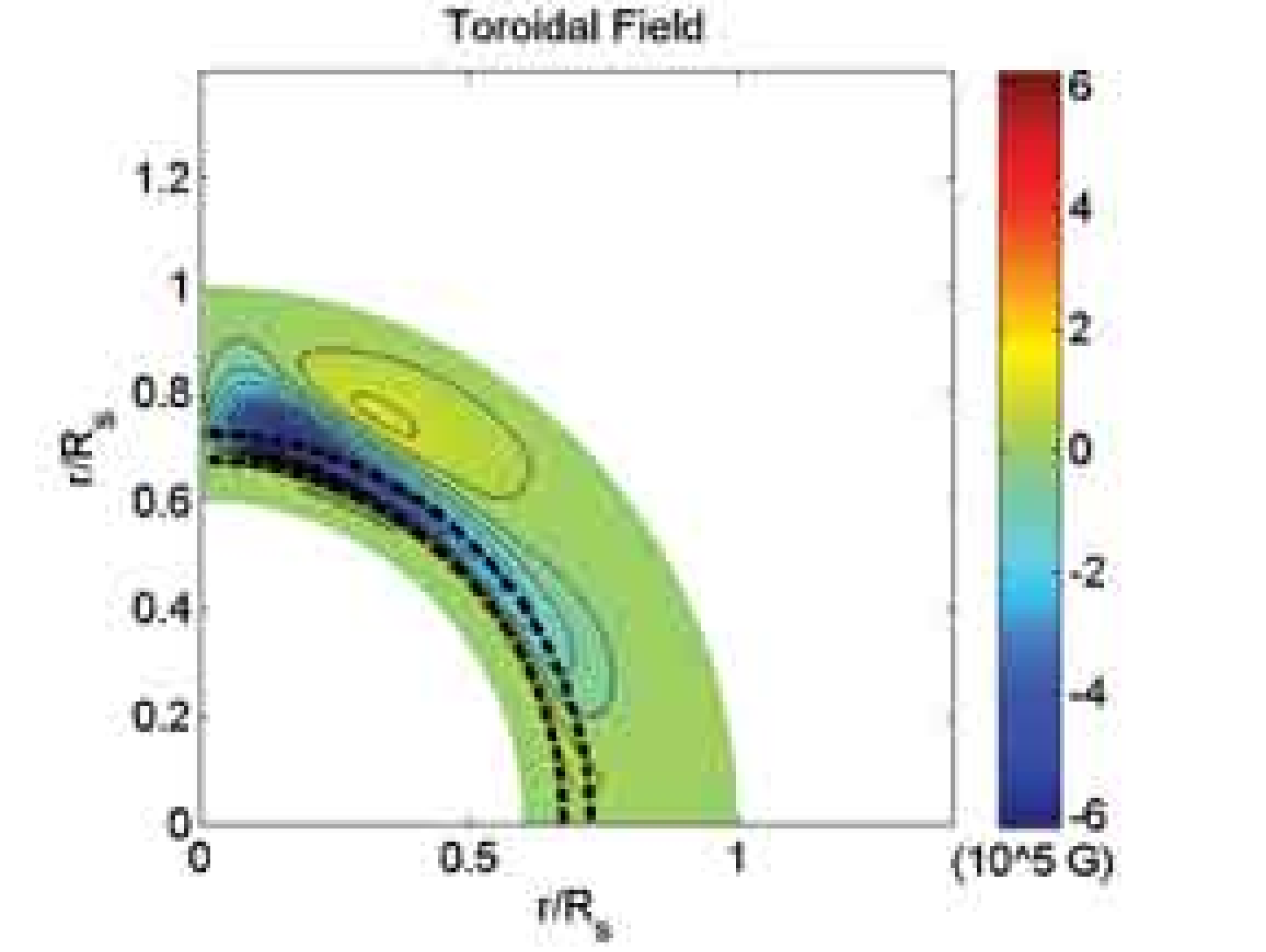} & \includegraphics[scale=0.25]{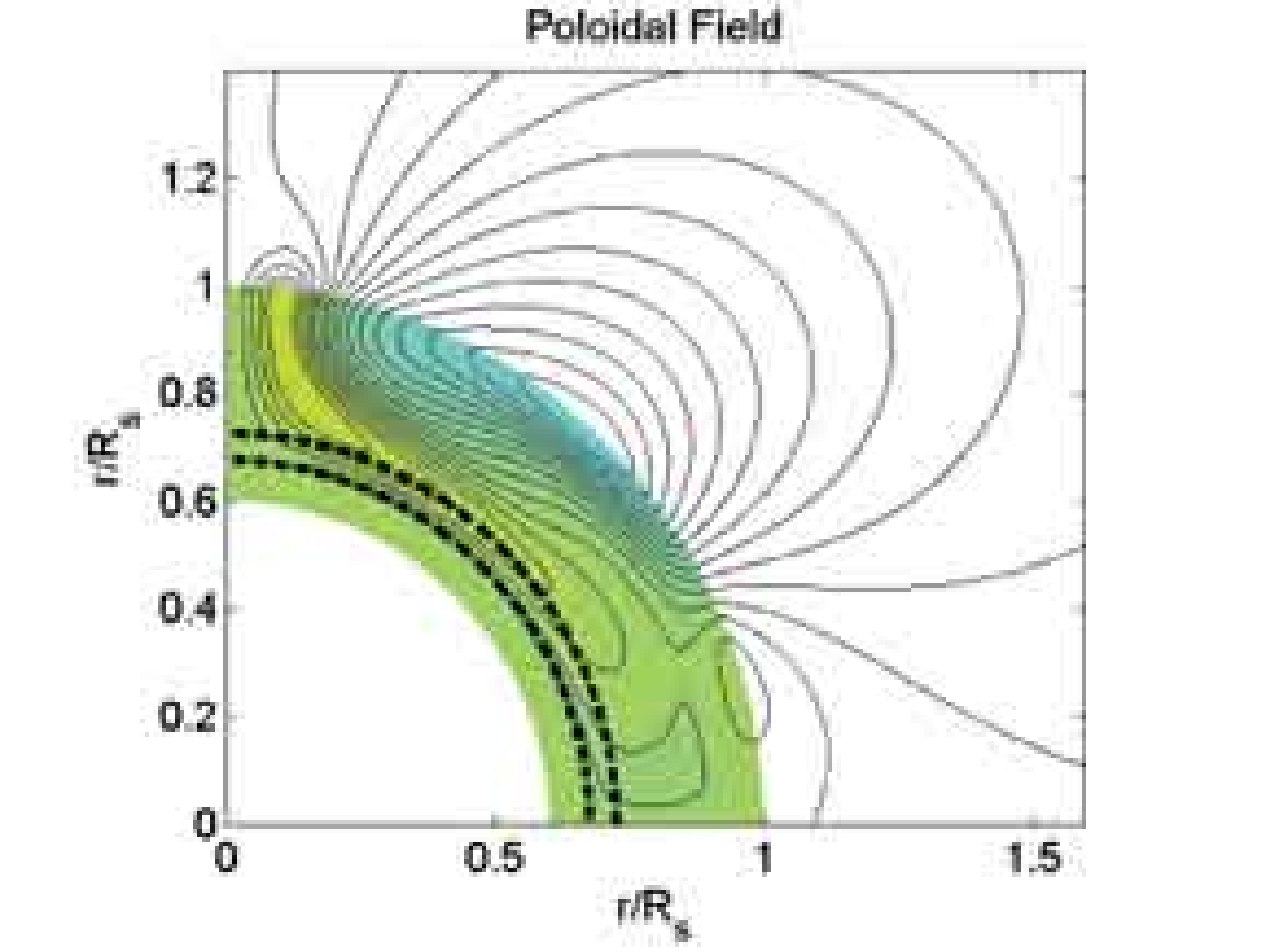}\\
  \includegraphics[scale=0.25]{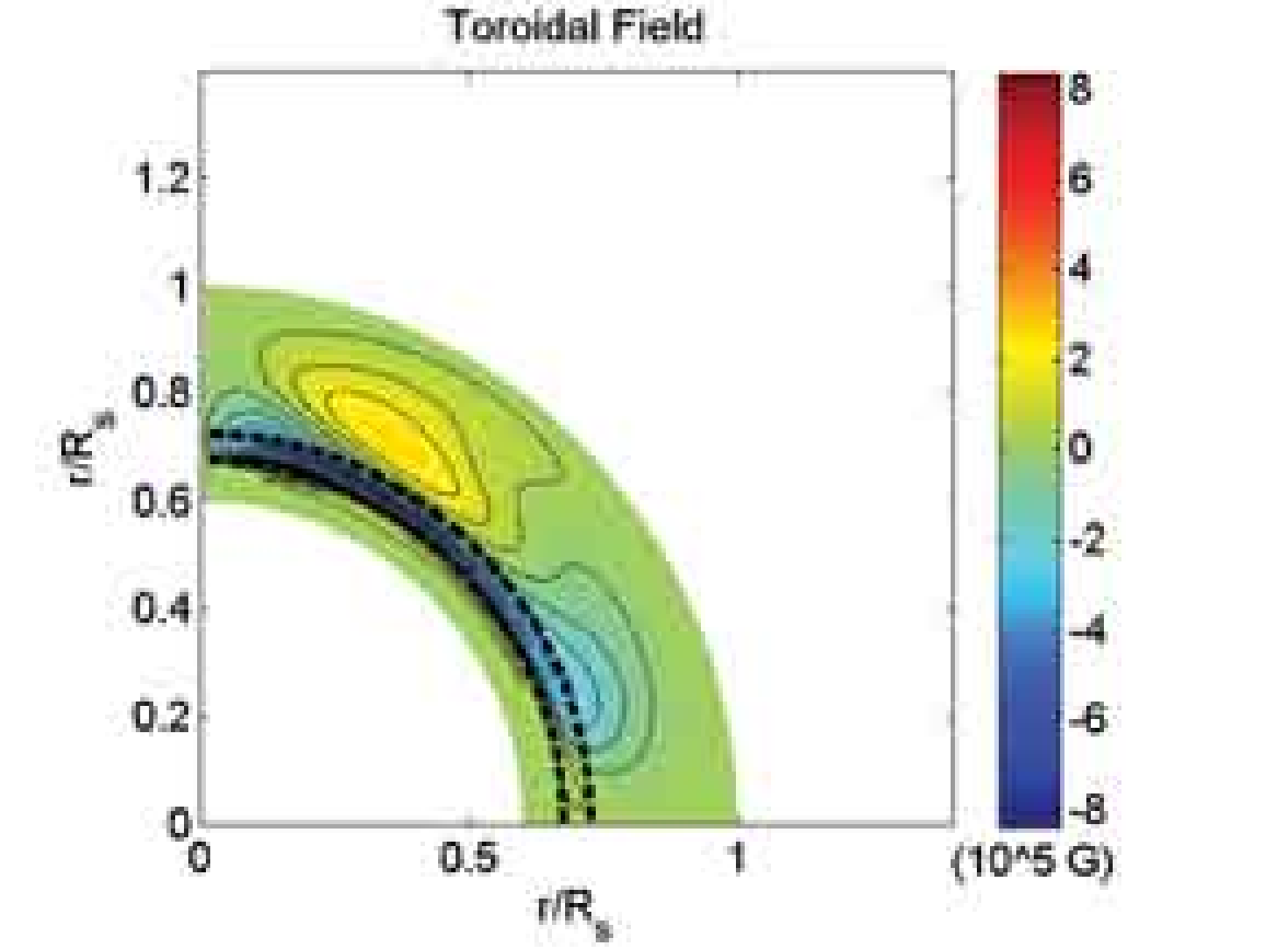} & \includegraphics[scale=0.25]{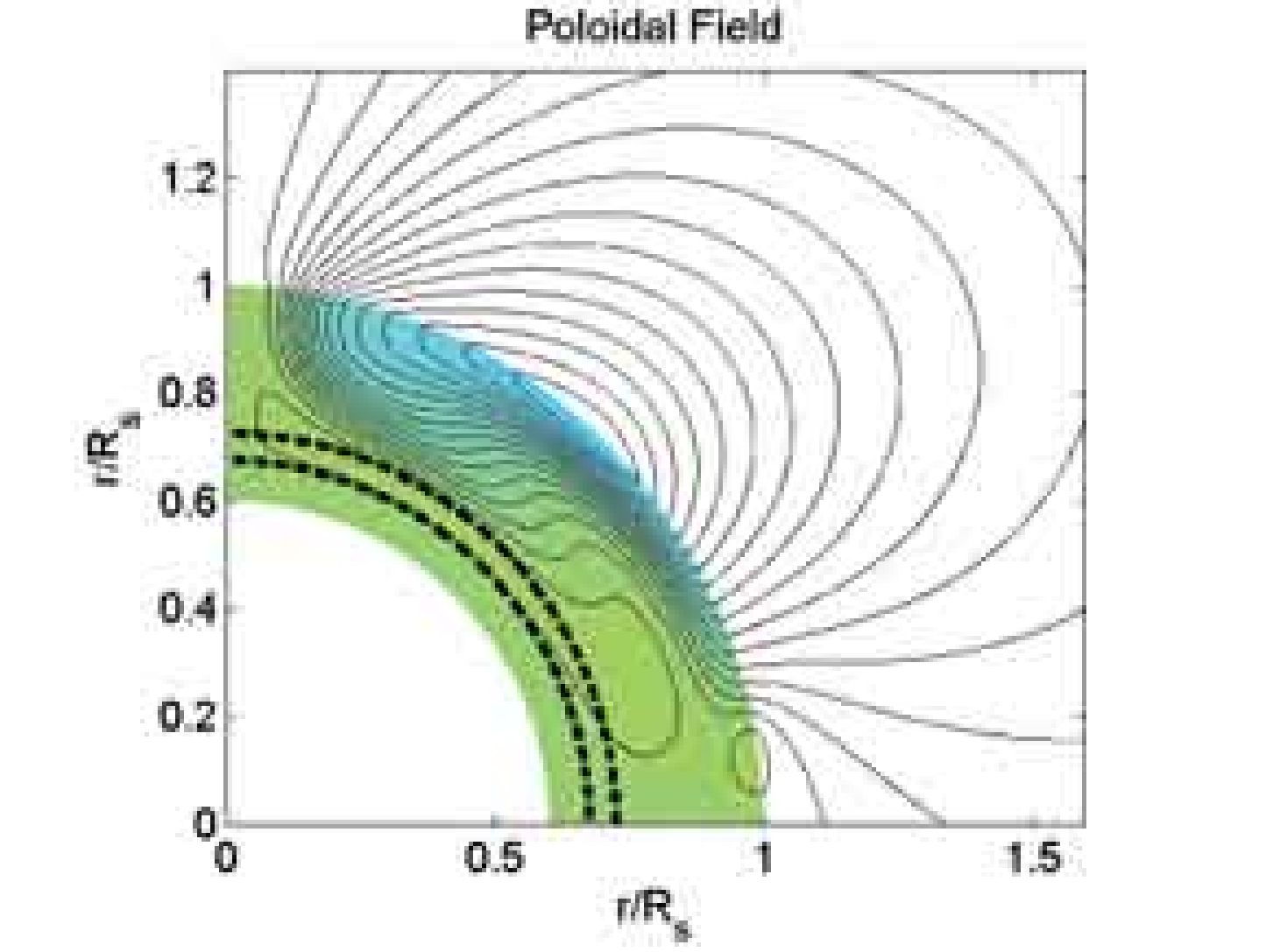} & \includegraphics[scale=0.25]{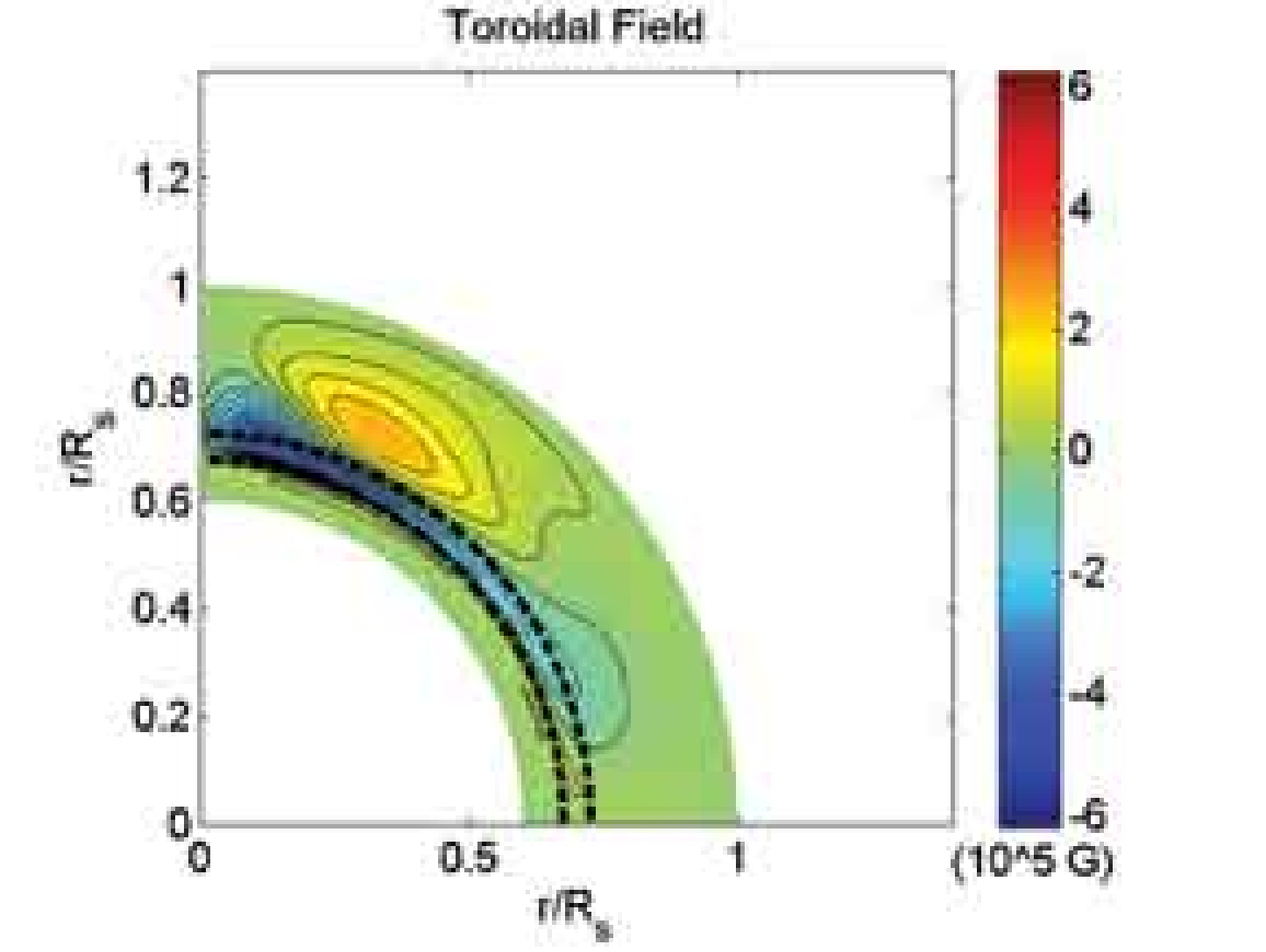} & \includegraphics[scale=0.25]{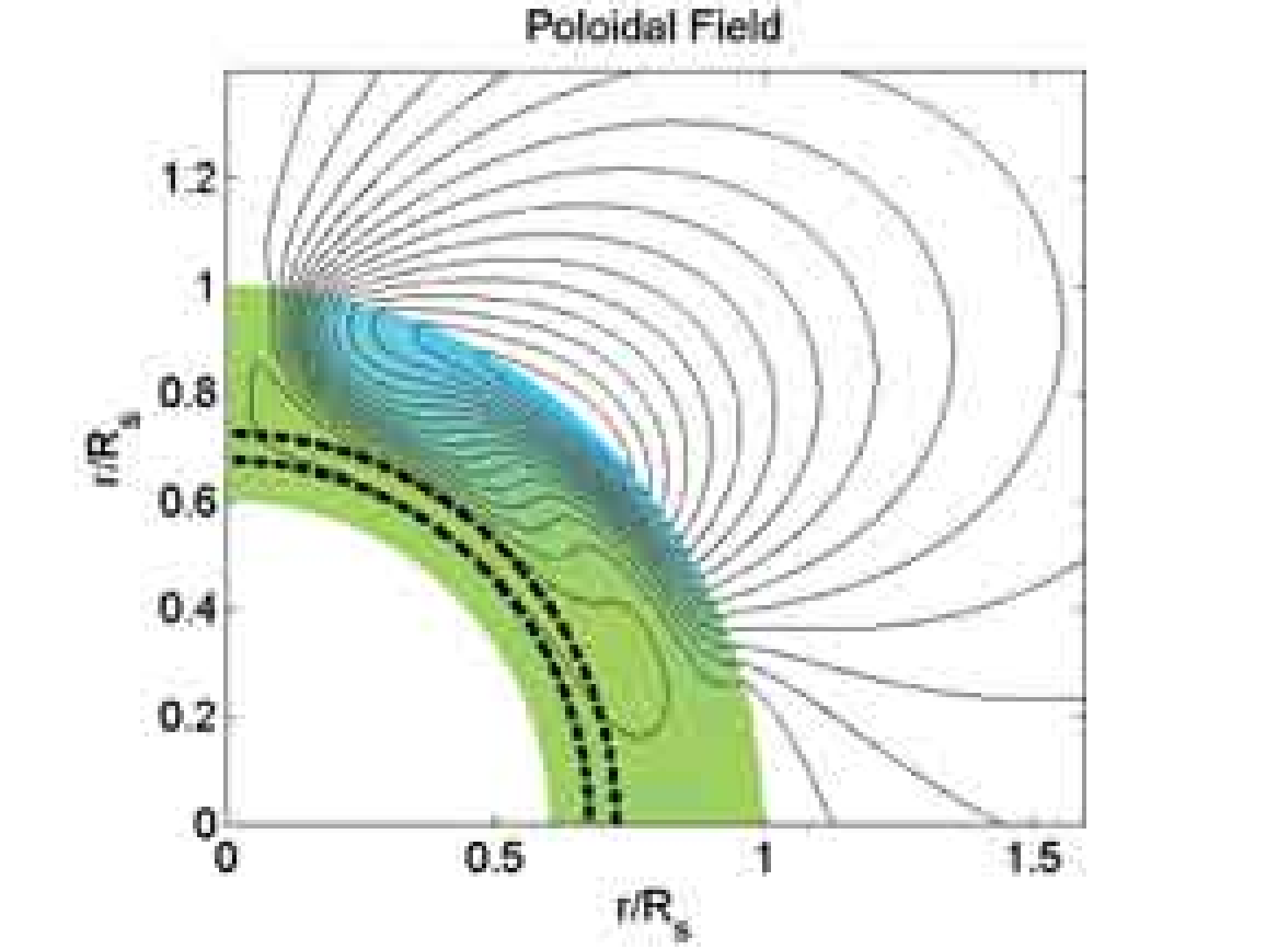}\\
  \includegraphics[scale=0.25]{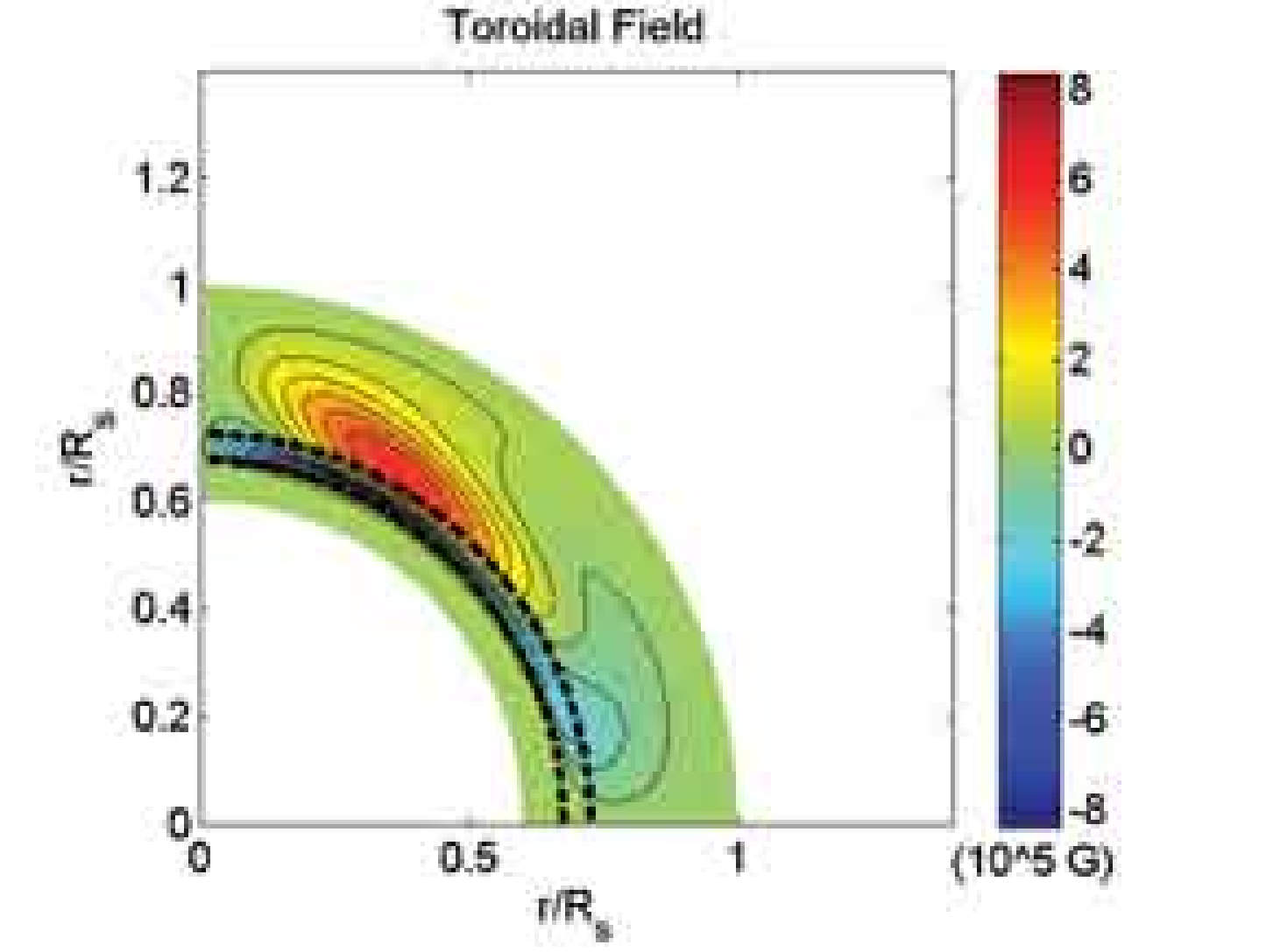} & \includegraphics[scale=0.25]{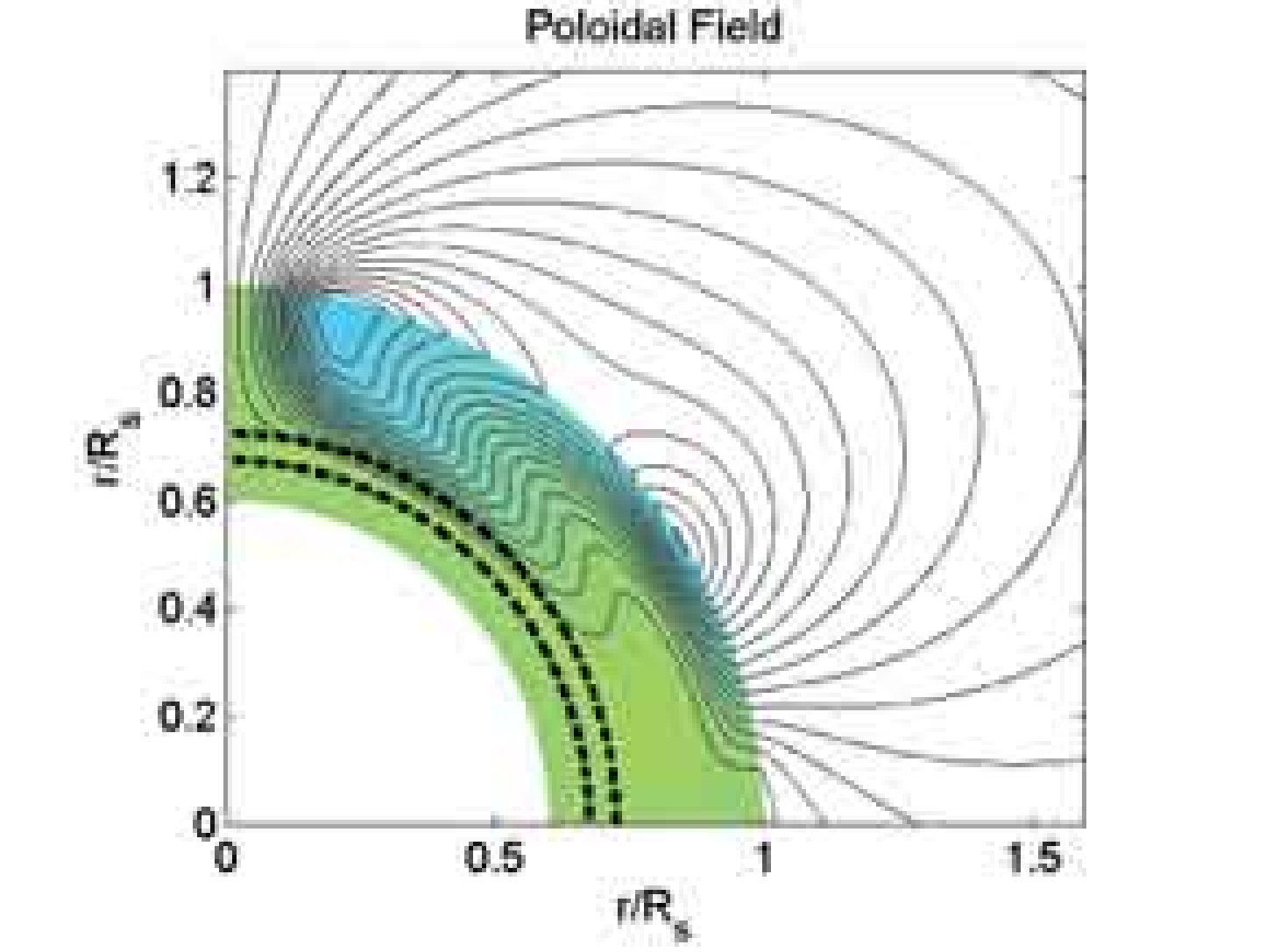} & \includegraphics[scale=0.25]{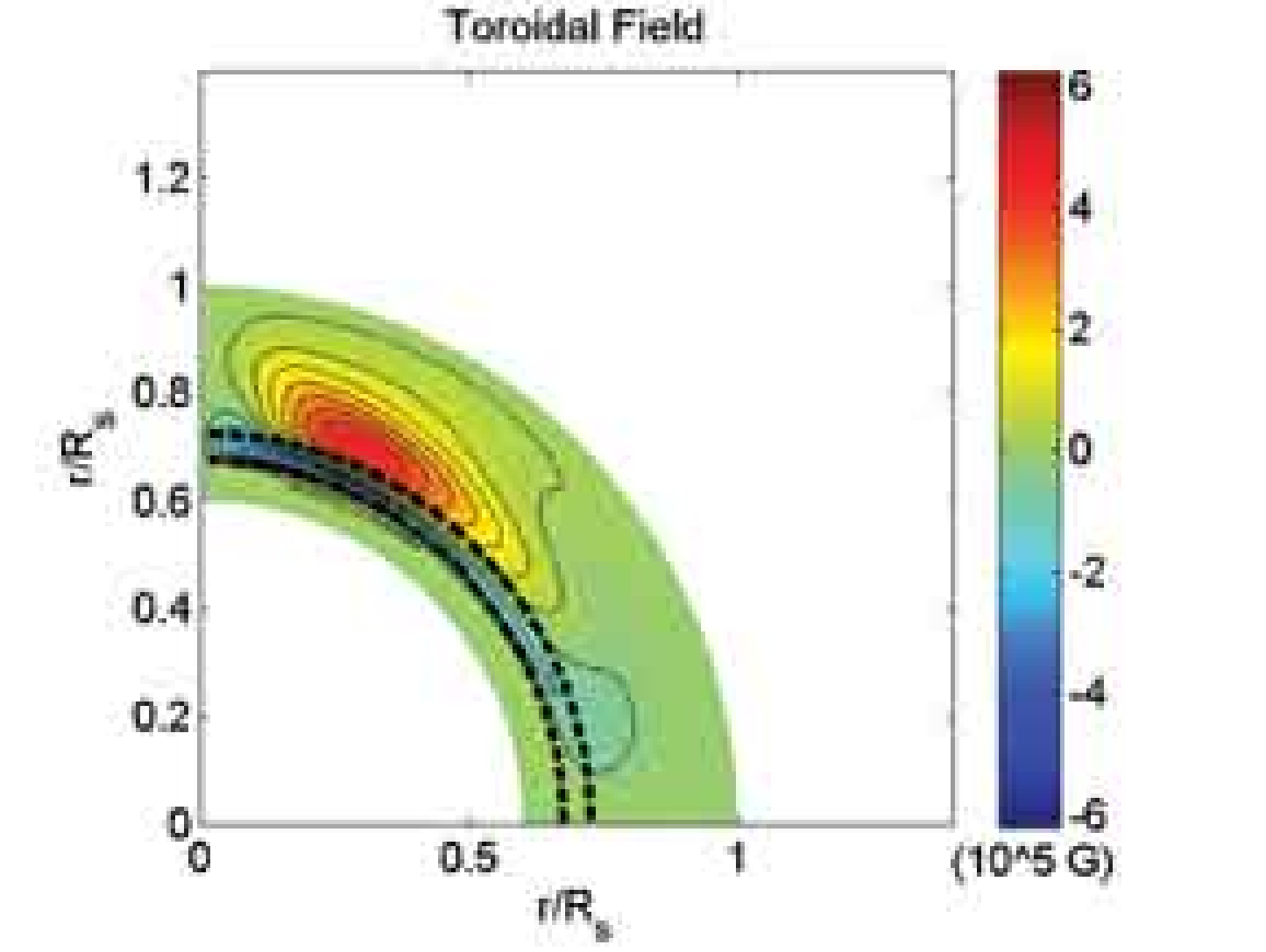} & \includegraphics[scale=0.25]{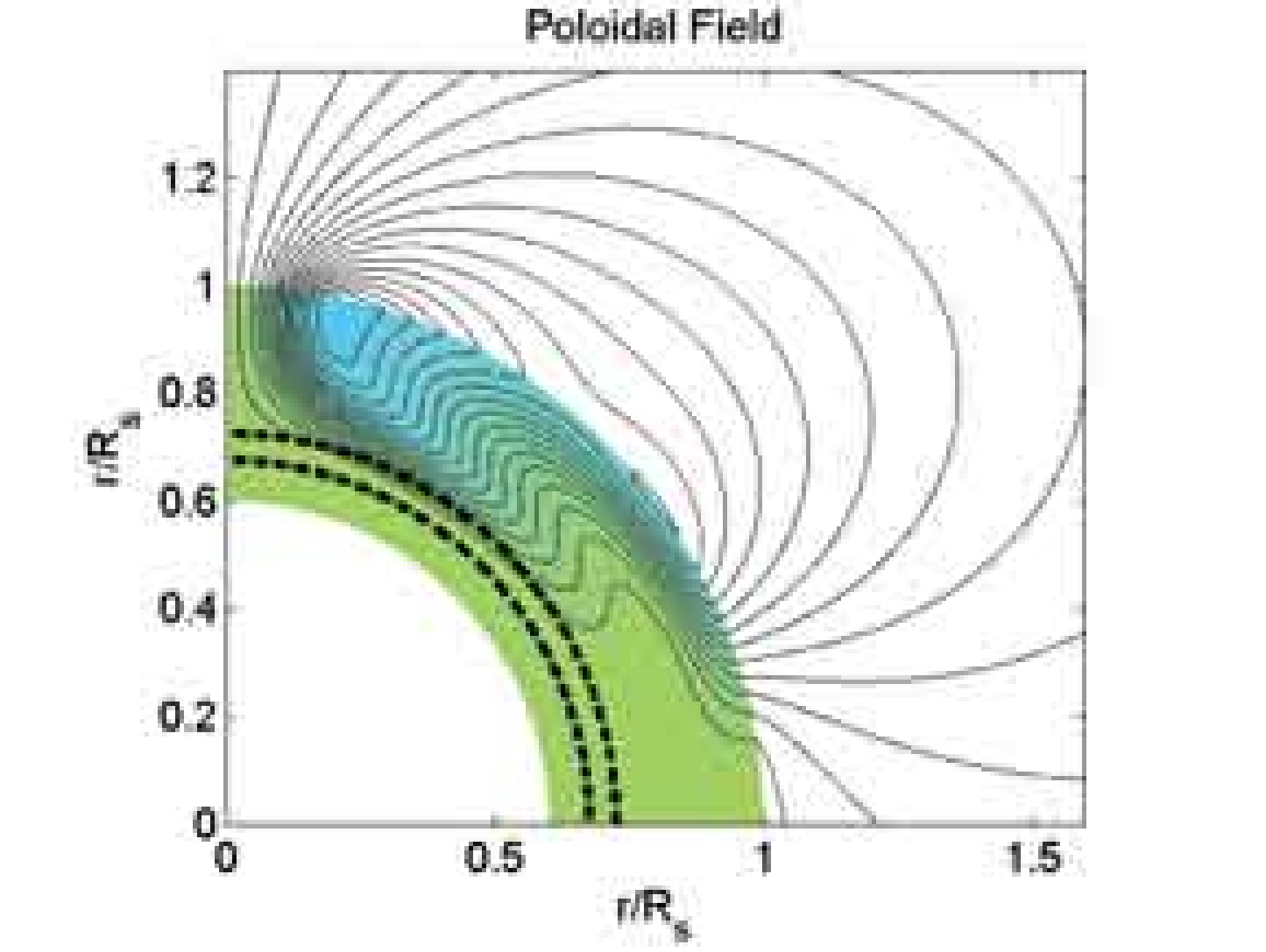}\\
  \end{tabular}
\caption{Snapshots of the magnetic field over half a dynamo cycle (a
sunspot cycle) when using a low super-granular diffusivity ($\eta_{cz}=10^{11} cm^2/s$). Each row is advanced by an eight of the dynamo cycle (a quarter of the sunspot
cycle) i.e., from top to bottom $t = 0, \tau/8, \tau/4$ and $3\tau/8$.  The solutions correspond to
the meridional flow Set 2 (deepest penetration with a peak
flow of 12 m/s) and analytic differential rotation (left) and composite data (right)}\label{S2C_ld}
\end{figure}

In the second parameter space experiment, we lower the super-granular diffusivity $\eta_{sg}$ from $10^{13} cm^2/s$ to $10^{11} cm^2/s$.  This was done to study the impact of the surface radial shear under low diffusivity conditions.  However, as can be seen in Figure~\ref{S2C_ld}, there is very little difference between the two solutions. This means that even after reducing the super-granular diffusivity by two orders of magnitude, the radial shear has very little impact on the solutions and the upper boundary conditions still play an important role in limiting the relative contribution from the near-surface shear layer.

\section{Conclusions}

In summary, we have presented here methods which can be used to better integrate helioseismic
data into kinematic dynamo models. In particular, we have demonstrated that using a composite
between helioseismic data and an analytical profile for the differential rotation, we can
directly use the helioseismic rotation data in the region of trust and substitute the suspect data by smoothly
matching it to the analytical profile where the data is noisy. This paves the way for including
the helioseismically inferred rotation profile directly in dynamo simulations. We have also shown
how mathematical properties of the commonly used analytic stream functions describing the
meridional flow can be fit to the available near-surface helioseismic data to entirely constrain
the latitudinal dependence of the meridional flow, as well as weakly constrain the radial (depth)
dependence.

In our simulations, comparing the helioseismic data for the differential rotation with the
analytical profile of Charbonneau et al.\ (1999), with four plausible meridional flow profiles,
we find that there is little difference between the solutions using the helioseismic composite
and the analytical differential rotation profile -- specially for shallow penetrations of the
meridional flow and even at reduced super-granular diffusivity. This is because the impact of the surface radial shear, which is present in the
helioseismic composite but not the analytic profile, is greatly reduced by the proximity of the upper boundary conditions. Also, for the shallow circulation, the toroidal field generation occurs in a region
located above the tachocline with mainly latitudinal shear, where the difference between the
composite data and the analytical profile is not significant.

The main result from this comparative analysis is that the latitudinal shear in the rotation is
the most dominant source of toroidal field generation in these type of models that are
characterized by high diffusivity at near surface layers, but lower diffusivity within the bulk
of the SCZ -- specially near the base where most of the toroidal field is being created. Since
this latitudinal shear exists throughout the convection zone, an interesting question is whether
toroidal fields can be stored there long enough to be amplified to high values by the shear in
the rotation, without being removed by magnetic buoyancy. If this were to be the case, i.e., the
latitudinal shear is indeed confirmed to be the dominant source of toroidal field induction, we
anticipate then that downward flux pumping (Tobias et al.\ 2001; see also Guerrero \& de Gouveia
Dal Pino 2008\nocite{guerrero-deouveiadalpino08}) -- which tends to act against buoyant removal of flux, may have an important role
to play in this context. This could also call into question the widely held view that the solar
tachocline is where most of the toroidal field is created and stored (see Brandenburg 2005\nocite{brandenburg05} for
arguments favoring a more distributed dynamo action throughout the SCZ).

Our attempts to integrate helioseismic meridional flow data into dynamo models and related
simulations have uncovered points that are both encouraging and discouraging.

On the discouraging side, we find that the currently available observational data are inadequate to
constrain the nature and exact profile of the deep meridional flow, especially the return flow.
Neither do the simulation results and their comparison with observed features of the solar cycle
clearly support or rule out any possibility. A recent analysis on light-element depletion due to
transport by meridional circulation indicates that solar light-element abundance observations
restrict the penetration to $0.62R_\odot$ (Charbonneau 2007\nocite{charbonneau07}); however this
analysis does not necessarily suggest that the flow {\it{does}} penetrate that deep. Also vexing
is the fact that different inversions, involving different helioseismic techniques such as
ring-diagram or time-distance analysis recovers different profiles and widely varying peak
meridional flow speeds (Giles et al.\ 1997\nocite{giles-etal97}; Braun \& Fan 1998; Gonz\'alez-Hern\'andez et al.\ 2006; Gizon \& Rempel 2008). In our analysis, we
chose to use the Gonz\'alez-Hern\'andez et al.\ data because at present, this provides the
(relatively) deepest full inversion of the flow within the SCZ.   Chou and
Ladenkov (2005)\nocite{chou-ladenkov05} reported time-distance diagrams reaching a depth of $0.79R_\odot$ but have not yet reported a full inversion that could be used on our simulations.

We point out that there is an important consequence of the presence of the flow speed maximum
inside the convection zone -- which is related to mass conservation:  If the maximum poleward flow
speed is found to be deeper inside the convection zone this would result in a stronger mass flux
poleward, which needs to be balanced by a deeper counterflow subject to mass conservation; the
density of the plasma increases rapidly as one goes deeper; e.g., the density at $0.97R_\odot$ is
ten thousand times larger than at the surface. Although that is not achieved currently, our
extensive efforts to fit the data point out that stronger constraints on the return flow may be
achieved even with data that does not necessarily go down to where this return flow is located, a
fact that may be usefully utilized when better depth-dependent helioseismic data on meridional
circulation becomes available.

Although the depth of penetration of the circulation is an important constraint on the flow
itself, our results indicate that the period of of the dynamo cycle does not in fact depend on
this depth. Rather, our simulations point out that the period of the dynamo cycle is more sensitive to changes on the speed of the counterflow than changes anywhere else in the transport circuit, as this is where the dynamo loop originates. An accurate determination of the average meridional flow speed over this loop closing at the SCZ base is very important in the context of the field transport
timescales. As shown by the analysis of Yeates, Nandy \& Mackay (2008), the relative timescales of
circulation and turbulent diffusion determines whether the dynamo operates in the advection or
diffusion dominated regime -- two regimes which have profoundly different flux transport dynamics
and cycle memory (the latter may lead to predictability of future cycle amplitudes). Getting a
firm handle on the average meridional flow speed is therefore very important and that is not
currently achieved from the diverging helioseismic inversion results on the meridional flow.

This suggests that a concerted effort using different helioseismic techniques on data for the
meridional flow over at least a complete solar cycle (over the same period of time) may be
necessary to generate a more coherent picture of the observational constraint on this flow
profile. It's important to note that even though we used time averaged data, nothing prevents one
from using the same methods to assimilate time dependent helioseismic data at different phases of the solar
circle, allowing us to study the impact of time varying velocity flows on solar cycle properties
and their predictability.

On the encouraging side, our dynamo simulations show that it is relatively straightforward to
use the available helioseismic data on the differential rotation (on which there is more
consensus and agreement across various groups) within dynamo models. Also encouraging is the fact
that the type of solar dynamo model presented here is able to handle the real helioseismic
differential rotation profile and generate solar-like solutions. Moreover, as evident from our
simulations, this dynamo model also generates plausible solar-like solutions over a wide range of
meridional flow profiles, both deep and shallow, and with fast and slow peak flow speeds. This
certainly bodes well for assimilating helioseismic data to construct better constrained
solar dynamo models -- building upon the techniques outlined here.

\acknowledgements

\section{Acknowledgements}

We are grateful to Irene Gonz\'alez-Hern\'andez and Rachel Howe at the National Solar
Observatory for providing us with helioseismic data and useful counsel regarding it's use. We also
thank J{\o}rgen Christensen-Dalsgaard of the Danish AsteroSeismology Center for data and
discussions related to the solar Model S.  Finally, we want to thank Paul Charbonneau and an anonymous
referee for useful comments and recommendations. The simulations presented here were performed at the
computing facilities of the Harvard Smithsonian Center for Astrophysics and this research was
funded by NASA Living With a Star Grant NNX08AW53G to the Smithsonian Astrophysical Observatory.

\appendix

\section{Numerical methods}

In order to use exponential propagation, we transform our system of Partial Differential
Equations(PDEs) in to a system of coupled Ordinary Differential Equations(ODEs) by discretizing
the spatial operators using the following finite difference operators:\\

For advective terms $\left(\frac{\partial A}{\partial t} = -v \frac{\partial A}{\partial x}  +
\chi(x) \right)$, where $v$ is the velocity, we use a third order upwind scheme:

\begin{equation}\label{ADV}
    v\frac{\partial A}{\partial x} = \left\{ \begin{array}{cc}
                                        \frac{v}{6\triangle x} ( -2 A_{i-1} - 3 A_i + 6 A_{i+1} - A_{i+2}) & \mbox{if $v < 0$}\\
                                        \frac{v}{6\triangle x} ( 2 A_{i+1} + 3 A_i - 6 A_{i-1} + A_{i-2}) & \mbox{if $v \geq 0$}\\
                                      \end{array} \right. + \mathcal{O}(\triangle x^3)
\end{equation}

For diffusive terms $\left(\frac{\partial A}{\partial t} = \eta \frac{\partial^2 A}{\partial x^2}
+ \chi(x) \right)$, where $\eta$ is the diffusion coefficient, we use a second order space
centered scheme:\\

\begin{equation}\label{DIF}
    \eta\frac{\partial^2 A}{\partial x^2} = \frac{\eta}{(\triangle x)^2} ( A_{i-1} - 2 A_i +
    A_{i+1}) + \mathcal{O}(\triangle x^2)
\end{equation}

For other first derivative terms $\left(\frac{\partial A}{\partial t} = \frac{\partial B}{\partial
x} + \chi(x) \right)$ we use a second order space centered scheme:

\begin{equation}\label{OFD}
    \frac{\partial B}{\partial x} = \frac{1}{2\triangle x} ( A_{i-1} - A_{i+1})
    + \mathcal{O}(\triangle x^2)
\end{equation}

Here, $\chi(x)$  corresponds to all the additional terms a PDE might have on the righthand side
besides the term under discussion and $A_i = A(x_0 + i\triangle x), i = 1,2,...,N_x$ is our
variable evaluated in a uniform grid of $N_x$ elements separated by a distance $\triangle x$.

\subsection{Exponential Propagation}

After discretization and inclusion of the boundary conditions we are left with an initial value
problem of ordinary differential equations:
\begin{eqnarray}
  \frac{\partial \textbf{U}(t)}{\partial t} &=& F(\textbf{U}(t)) \\
                            \textbf{U}(t_0) &=& \textbf{U}_0
\end{eqnarray}
where $\textbf{U}$ is the solution vector in $\mathbb{R}^N$.  Provided that the Jacobian $\partial
F(\textbf{U}(t))$ exists and is continuous in the interval $[t_0,t_0+\Delta t]$, we can linearize
$F(\textbf{U}(t_0+\Delta t))$ around the initial state to obtain
\begin{equation}\label{Lin}
    \frac{\partial \textbf{U}(t)}{\partial t} = F(\textbf{U}_0) + \partial F(\textbf{U}_0)(\textbf{U}(t_0+\Delta t) -
    \textbf{U}_0) + R(\textbf{U}(t_0+\Delta t))
\end{equation}
where $R(\textbf{U}(t_0+\Delta t)$ are the residual high order terms.   The solution to this
equation can be written as
\begin{equation}\label{In}
    \textbf{U}(t_0+\Delta t) = \textbf{U}_0 + \frac{e^{A\Delta t}-I}{A} + \mathcal{O}(\Delta t^2)
\end{equation}
where $A = \partial F(\textbf{U}_0)$. Neglecting higher order terms leaves us with a scheme that
is second order accurate in time and is an exact solution of the linear case.  However there is a
way of increasing the time accuracy of this method by following a generalization of Runge-Kutta
methods for non-linear time-advancement operators proposed by Rosenbrock
(1963\nocite{rosenbrock63}).  The combination of exponential propagation with Runge-Kutta methods
was first proposed by Hochbruck and Lubich (1997) and then generalized by Hochbruck, Lubich  and
Selhofer (1998\nocite{hochbruck-lubich-selhofer98}).  In this work we use a fourth order
algorithm which goes to the following intermediary steps to advance the solution vector between
timesteps ($\textbf{U}^n)\rightarrow\textbf{U}^{n+1})$):
\begin{eqnarray}
  k_1 &=& \Phi\left(\frac{1}{2}\Delta t A\right)F(\textbf{U}^n) \nonumber\\
  k_2 &=& \Phi\left(\Delta t A\right)F(\textbf{U}^n) \nonumber\\
  w_3 &=& \frac{3}{8}(k_1+k_2) \nonumber\\
  u_3 &=& \textbf{U}^n + \Delta t w_3 \nonumber\\
  d_3 &=& F(u_3) - F(\textbf{U}^n) - \Delta t A w_3 \nonumber\\
  k_3 &=& \Phi\left(\frac{1}{2}\Delta t A\right)d_3 \nonumber\\
  \textbf{U}^{n+1} &=& \textbf{U}^n + \Delta t \left(k_2 + \frac{16}{27}k_3 \right) \nonumber\\
  & & \nonumber \\
  A = F(\textbf{U}^n) &,& \Phi\left(\Delta t A\right) = \frac{e^{A\Delta t}-I}{A} \nonumber
\end{eqnarray}

\subsection{Krylov Approximation to the Exponential Operator}

Without any further approximation, this method is very expensive computationally due to the need of
continuously evaluating the matrix exponential.  However, it is possible to make a good
approximation by projecting the operator into a finite dimensional Krylov subspace
\begin{equation}\label{KR}
    S_{Kr} =
    \operatorname{span}\{\textbf{U}_0,A\textbf{U}_0,A^2\textbf{U}_0,...,A^{m-1}\textbf{U}_0\}.
\end{equation}

In order to do this we first compute an orthonormal basis for this subspace using the Arnoldi
algorithm (Arnoldi 1951\nocite{arnoldi51}):

\begin{enumerate}
    \item $v_1 = \textbf{U}_0/|\textbf{U}_0|^2$
    \item For $j=1,...,m$ do
    \begin{itemize}
        \item for $i=1,...,i$ compute $h_{i,j}=v_i^T*Av_j$
        \item calculate $w=Av_j-\displaystyle\sum_{i=0}^j h_{i,j}v_i$
        \item evaluate $h_(j+1,j)=|w|^2$
        \item if $h_{j+1,j}<\epsilon$ stop, else compute the next basis vector $v_{j+1}=w/h_{j+1,j}$
    \end{itemize}
\end{enumerate}
where $\epsilon$ is the parameter that sets the error tolerance in this approximation.  Once we have
finished computing the algorithm, we have the relationship
\begin{equation}\label{AV}
    AV_m \approx V_mH,
\end{equation}
where $V_m = [v_1,...,v_m]$ and H is a matrix whose elements are $h_{i,j}=v_i^T*Av_j$. The validity
of this approximation depends on the dimension of the Krylov subspace but numerical experiments
have found that 15-30 Krylov vectors are usually enough (see Hochbruck and Lubich 1997; Tokman
2001\nocite{tokman01}). Since $\{v_1,...,v_m\}$ is an orthonormal basis of the Krylov subspace
$S_{Kr}$ then $V_m^TV_M$ is a $m\times m$ identity matrix and $V_mV_m^T$ is a projector from
$\mathbb{R}^N$ onto $S_{Kr}$. Using this projector we can find an approximation to any matrix
vector multiplication by projecting them onto the Krylov subspace by calculating $Ab \approx
V_mV_m^TAV_mV_m^Tb$.  After using Equation \ref{AV} this becomes
\begin{equation}\label{Ab}
    Ab \approx V_mHV_m^Tb.
\end{equation}
In fact we can approximate the action of any operator $\Phi(A)$, that can be expanded on a Taylor
series, on the vector $\textbf{U}_0$ using the Krylov subspace projection:
\begin{equation}\label{Operator}
    \Phi(A)\textbf{U}_0 \approx V_m\Phi(H)V_m^T\textbf{U}_0 = |\textbf{U}_0|^2V_m\Phi(H)e_1
\end{equation}
where $e_1 =$ $[1$ $0$ ... $0]$ and we used $v_1 = \textbf{U}_0/|\textbf{U}_0|^2$ and $V_m^Tv_1 =
e_1$.  As we can see in Equation \ref{Operator}, the use of the Krylov approximation effectively
reduces the size of the matrix operator; this makes the use of the exponential operator relatively
inexpensive computationally.

These two algorithms, in combination with a robust error control and an adaptative time-step
mechanism strategy, form the core of the SD-Exp4 integrator (for more details see Hochbruck and
Lubich 1997; Hochbruck, Lubich  and Selhofer 1998 and Tokman 2001).

In order to verify the performance standards of the SD-Exp4 code we ran case C' of the dynamo benchmark by
Jouve et al.\ (2008)\nocite{Jouve-etal08}.  This case is similar in nature to the simulations done in this
work with the following differences:

\begin{itemize}
  \item The meridional flow profile has different radial and latitudinal dependence.
  \item The poloidal source term has no quenching term and a different radial and latitudinal dependence.
  \item The turbulent diffusivity profile consists of only one step and reaches a peak value of $10^{11}$ $cm^2/s$.
  \item The analytic differential rotation has no $cos^4(\theta)$ dependence and uses a thinner tachocline.
\end{itemize}

\begin{figure}[c]
  \includegraphics[scale=0.6]{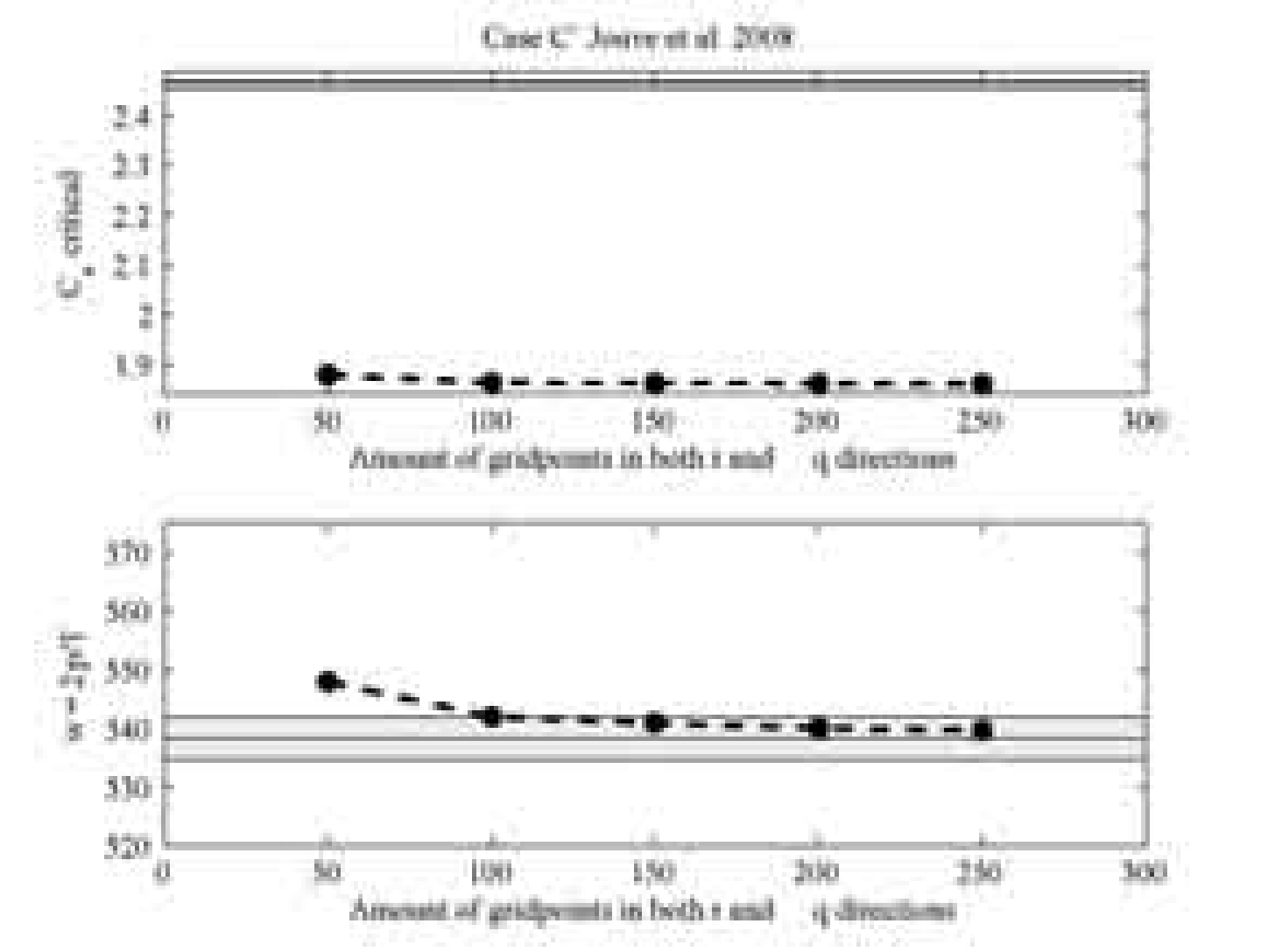}\\
  \caption{Simulation results (dots) of running case C' of the dynamo benchmark by Jouve et al.\ (2008).  The top figure shows how the minimum value of $C_\alpha=\alpha_0R_\odot/\eta_{cz}$ for which the dynamo has stable oscillations depends on resolution.  The bottom figure shows how $\omega = 2\pi R_\odot^2/(T\eta_{cz})$ depends on resolution.  In order to make the plot comparable to the benchmark, we include the mean value found by the different codes and a shaded area corresponding to one standard deviation around this mean.  This run was made using the same meridional flow and poloidal source profiles as in the benchmark study (for details see Jouve et al.\ 2008).}\label{BMRK}
\end{figure}

In order to compare the performance of the different codes in the benchmark study, two quantities are used: $C^{crit}_s = \alpha^{crit}_0R_\odot/\eta_{cz}$ which quantifies the minimum strength of the source-relative-to-dissipation that yields stable oscillations and $\omega = 2\pi R_\odot^2/(T\eta_{cz})$ which quantifies the frequency of the magnetic cycle relative to the diffusion timescale.  The dependence of this quantities is then plotted versus resolution for all the different codes (see Figure 11 of Jouve et al.\ 2008).  In order to compare our code, we perform simulations with the same parameters and model ingredients as in case C' of Jouve et al.\ (2008) and plot this two quantities versus resolution.  We also plot the location of the mean found by the other codes and a shaded area encompassing one standard deviation around the mean. As can be seen in Figure \ref{BMRK}, we find lower values of $C^{crit}_s$ than the values obtained in Jouve et al.\ 2008 (1.86 as opposed to an average 2.46) for oscillatory solutions; this may be due to a lower amount of numerical diffusion in our code.  Moreover, we find values of $\omega$ in very good agreement with those found in Jouve et al.\ 2008 (540 as opposed to an average of 538.2).  We also find that for our code this quantities are less sensitive to resolution when compared to other codes used in the benchmark.  This is likely due to the fact that we try to avoid numerical differentiation whenever we can do it analytically.

As a final remark, Table 4 of Jouve et al.\ 2008 specifies the different time steps that are used by the different codes, which range from values of $10^{-8}$ to $10^{-5}$ code time units.  Thanks to the use of Krylov approximations we are able to use time steps as large as $10^{-2}$ in code time units while solving Case C'.  However, this does not translate to a performance improvement of several orders of magnitude (due to the added computations in the Krylov approximation).  In order to quantify the relative performance improvement, we also made comparisons with the Surya code, which has been studied extensively in different
contexts (e.g. Nandy \& Choudhuri 2002, Chatterjee, Nandy \& Choudhuri 2004, Choudhuri, Chatterjee \& Jiang 2007),
and is made available to the public on request.  In comparison to the Surya code, we find that the SD-Exp4 code achieves a performance improvement that reduces runtime from a half to a tenth of the total runtime depending on the particularities of the simulation.



\end{document}